\documentclass[10pt,
							 twocolumn,
							 superscriptaddress,
							 english,
							 prl,
							 showpacs,
							 floatfix,
							 aps
							]{revtex4-1}
\usepackage[utf8]{inputenc}
\usepackage{amsmath}
\usepackage{amssymb}
\usepackage{graphicx}
\usepackage{xspace}

\makeatletter
\@ifundefined{textcolor}{}
{%
 \definecolor{BLACK}{gray}{0}
 \definecolor{WHITE}{gray}{1}
 \definecolor{RED}{rgb}{1,0,0}
 \definecolor{GREEN}{rgb}{0,1,0}
 \definecolor{BLUE}{rgb}{0,0,1}
 \definecolor{CYAN}{cmyk}{1,0,0,0}
 \definecolor{MAGENTA}{cmyk}{0,1,0,0}
 \definecolor{YELLOW}{cmyk}{0,0,1,0}
}

\usepackage{soul}
\usepackage{braket}
\usepackage[backref=none,
bookmarksnumbered=true,
bookmarks=true,
bookmarksopen=true,
colorlinks=true,
citecolor=blue,
linkcolor=blue,
anchorcolor=green,
urlcolor=blue,unicode=false]{hyperref}
\let\baraccent=\= % rename builtin command \= to \baraccent
\renewcommand{\=}[1]{\stackrel{#1}{=}} % for putting numbers above =

\newcommand{\didv}{\ensuremath{\mathrm{d}I/\mathrm{d}V}\xspace}
\newcommand{\etal}{\textit{et al.}\xspace}

\newcommand{\Fig}[1]{Fig.\unitspace\ref{fig:#1}}
\newcommand{\Figure}[1]{Figure\unitspace\ref{fig:#1}}

\newcommand{\twoH}{\ensuremath{2H}-NbSe\ensuremath{_2}}
\newcommand{\nbse}{NbSe\ensuremath{_2}}
\newcommand{\spat}{\ensuremath{4\umV, 200\upA}}
\newcommand{\sptop}{\ensuremath{500\umV, 80\upA}}

\DeclareMathOperator{\unm}{\,\mathrm{nm}}
\DeclareMathOperator{\upm}{\,\mathrm{pm}}

\DeclareMathOperator{\umV}{\,\mathrm{mV}}

\DeclareMathOperator{\umeV}{\,\mathrm{meV}}
\DeclareMathOperator{\umueV}{\,\mathrm{\mu eV}}

\DeclareMathOperator{\upA}{\,\mathrm{pA}}

\DeclareMathOperator{\umuS}{\,\mathrm{\mu S}}

\DeclareMathOperator{\umbar}{\,\mathrm{mbar}}

\makeatother

\usepackage{babel}
\begin{document}

\title{Yu-Shiba-Rusinov states in the charge-density modulated superconductor \nbse}

\author{Eva Liebhaber}
\affiliation{\mbox{Fachbereich Physik, Freie Universit\"at Berlin, 14195 Berlin, Germany}}

\author{Sergio Acero Gonz\'alez}
\affiliation{\mbox{Dahlem Center for Complex Quantum Systems and Fachbereich Physik, Freie Universit\"at Berlin, 14195 Berlin, Germany}}

\author{Rojhat Baba}
\affiliation{\mbox{Fachbereich Physik, Freie Universit\"at Berlin, 14195 Berlin, Germany}}

\author{Ga\"el Reecht}
\affiliation{\mbox{Fachbereich Physik, Freie Universit\"at Berlin, 14195 Berlin, Germany}}

\author{Benjamin W. Heinrich}
\affiliation{\mbox{Fachbereich Physik, Freie Universit\"at Berlin, 14195 Berlin, Germany}}

\author{Sebastian Rohlf}
\affiliation{\mbox{Institut f\"ur Experimentelle und Angewandte Physik, Christian-Albrechts-Universit\"at zu Kiel, 24098 Kiel, Germany}}
\affiliation{\mbox{Ruprecht-Haensel-Labor, Christian-Albrechts-Universit\"at zu Kiel, 24098 Kiel, Germany}}

\author{Kai Rossnagel}
\affiliation{\mbox{Institut f\"ur Experimentelle und Angewandte Physik, Christian-Albrechts-Universit\"at zu Kiel, 24098 Kiel, Germany}}
\affiliation{\mbox{Ruprecht-Haensel-Labor, Christian-Albrechts-Universit\"at zu Kiel, 24098 Kiel, Germany}}
\affiliation{\mbox{Deutsches Elektronen-Synchrotron DESY, 22607 Hamburg, Germany}}

\author{Felix von Oppen}
\affiliation{\mbox{Dahlem Center for Complex Quantum Systems and Fachbereich Physik, Freie Universit\"at Berlin, 14195 Berlin, Germany}}

\author{Katharina J. Franke}
\affiliation{\mbox{Fachbereich Physik, Freie Universit\"at Berlin, 14195 Berlin, Germany}}
\date{\today}

\begin{abstract}

\nbse\ is a remarkable superconductor in which charge-density order coexists with pairing correlations at low temperatures. Here, we study the interplay of magnetic adatoms and their Yu-Shiba-Rusinov (YSR) bound states with the charge density order. Exploiting the incommensurate nature of the charge-density wave (CDW), our measurements provide a thorough picture of how the CDW affects both the energies and the wavefunctions of the YSR states. Key features of the dependence of the YSR states on adsorption site relative to the CDW are explained by model calculations. Several properties make \nbse\ a promising substrate for realizing topological nanostructures. Our results will be important in designing such systems.

\end{abstract}

\pacs{%
      %75.70.Rf 	Surface magnetism
			%74.25.Jb, % Superconcuctivity: Electronic structure (photoemission, etc.)
			%74.55.+v % SC: Tunneling phenomena: single particle tunneling and STM
			} %  73.22.-f 	Electronic structure of nanoscale materials and related systems
\maketitle 
%\maketitle must follow title, authors, abstract and \pacs

%%############################ INTRODUCTION ###############################

NbSe$_2$ is a fascinating and in part still enigmatic superconductor. There is evidence for an anisotropic gap as well as multiband superconductivity \cite{Hess1990, Yokoya2001, Noat2010, Noat2015, Rodrigo2004a, Boaknin2003, Fletcher2007, Guillamon2008, Guillamon2008a}, and the spin physics is highly nontrivial due to strong spin-orbit coupling, most prominently in monolayers of NbSe$_2$ \cite{Xi2016, Ugeda2016}. Already far above the superconducting transition, NbSe$_2$ undergoes a transition into a charge-density-wave (CDW) ordered state ($T_{\mathrm{CDW}}\approx 33\,$K) which coexists with superconductivity ($T_{\mathrm{c}}\approx 7.2\,$K) at low temperatures \cite{Rossnagel2001, Johannes2006, Kiss2007, Johannes2008, Rahn2012, Malliakas2013, Dai2014, Arguello2014, Arguello2015, Flicker2015, Soumyanarayanan2013, Weber2011}. The precise interplay between CDW ordering and superconductivity remains only partially understood at present \cite{Kiss2007, Suderow2005, Borisenko2009, Cho2018}. 

At the same time, recent experiments suggest that NbSe$_2$ has attractive properties in the context of engineering topological superconducting phases \cite{Sticlet2019, Glodzik2019}. Following the seminal work of Nadj-Perge \etal \cite{Nadj-Perge2014}, a topological phase and Majorana bound states can appear in chains of magnetic adatoms on superconducting substrates. One way to understand these remarkable phases is based on coupling  Yu-Shiba-Rusinov (YSR) bound states \cite{NadjPerge2013, Pientka2013} in the superconducting gap which are induced by individual adatoms \cite{Yu1965, Shiba1968, Rusinov1969,Yazdani1997}. Due to its layered structure, YSR states in NbSe$_2$ fall off laterally much more slowly than in three-dimensional superconductors \cite{Menard2015}. This enhances the coupling between the YSR states of neighboring magnetic impurities \cite{Kezilebieke2018}, potentially increasing the energy scale and thus the stability of possible topological phases. NbSe$_2$ is also a relatively robust material which may enable one to build nontrivial adatom structures from the bottom up by manipulation using a scanning tunneling microscope tip. 

The YSR states of different adatoms hybridize particularly strongly when the YSR energies of the isolated adatoms are identical and their wavefunctions extend sufficiently far along their connecting line \cite{Ruby2018, Choi2018, Kezilebieke2018, Kamlapure2018, Kim2019}. For NbSe$_2$, this is nontrivial to achieve due to the CDW ordering. Indeed, the CDW is incommensurate with the underlying atomic lattice so that adatoms which are nominally identical in terms of their atomic adsorption site, will in general be positioned differently with respect to the CDW. Here, we observe that both the energy and the wavefunction of YSR states depend sensitively on the adatom position relative to the CDW, and provide model calculations to understand these effects qualitatively. The dependence on the CDW imposes requirements which are crucial to take into account when designing adatom structures to realize topological superconducting phases.

To isolate the effect of the CDW on the YSR states, we focus on adatoms on clean \twoH\ surfaces instead of buried impurities as used in previous experiments \cite{Menard2015, Senkpiel2018}. This gives us full control over the adsorption site and allows us to compare adatoms which differ only in their locations relative to the CDW. Indeed, numerous previous studies \cite{Yazdani1997, Ji2008, Ruby2015, Ruby2016, Choi2017, Cornils2017, Kamlapure2018} on other substrates have shown that in the absence of CDW ordering, YSR spectra and wavefunctions are fully reproducible for a particular adsorption configuration of the adatom on the substrate.

\begin{figure}[ht]\centering
\includegraphics[width=8cm]{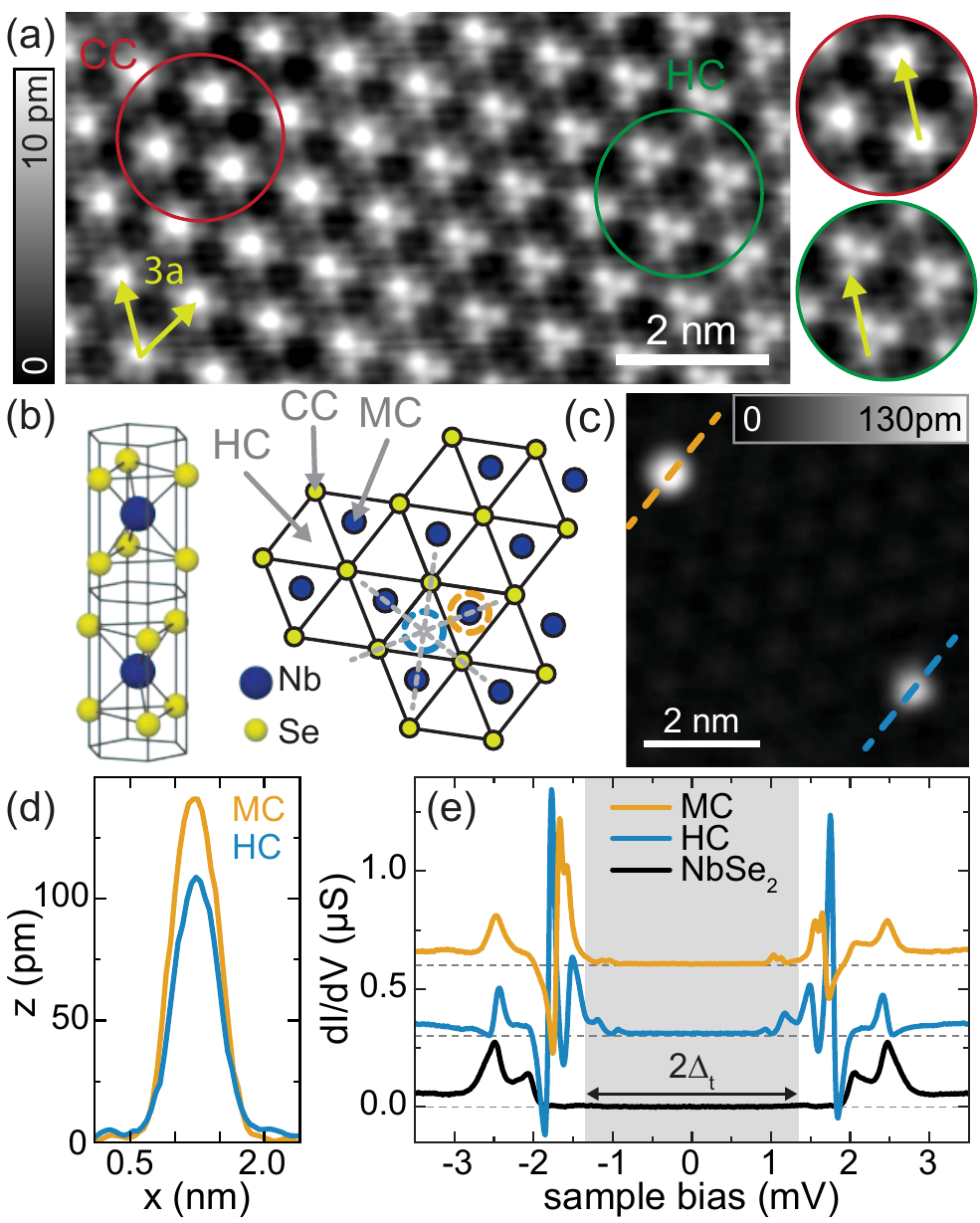}
\caption{
(a) Atomic-resolution STM images showing the incommensurate CDW modulation (constant-current set point: \spat) with close-ups of different regions. (b) Unit cell and top view of \twoH, with different lattice sites labeled as HC, CC and MC. Grey dashed lines indicate mirror axes. (c) Topography of HC and MC adatoms. (d) Line profiles across the atoms shown in (c). (e) Constant-height \didv spectra taken on the substrate (black) and on the atoms shown in (c) (set point: \spat; $V_{\mathrm{rms}}=15\umueV$, spectra offset by $0.3\umuS$). $2\Delta_{\mathrm{t}}$ is indicated by the shaded area.  
}
\label{Fig:figure1}
\end{figure}

%%############################EXPERIMANTAL RESULTS AND DISCUSSION##############################
%
%

A clean \twoH\ surface is prepared by cleaving a bulk crystal (grown by iodine vapor transport \cite{Rahn2012}) in ultra-high vacuum. Scanning tunneling spectroscopy (performed by standard lock-in techniques at a temperature of $1.1\,$K) shows that the superconducting energy gap of the substrate is framed by characteristic coherence peaks in an energy range of $2.0-2.7\umeV$ (black \didv spectrum in Fig.\ \ref{Fig:figure1}e). The origin of this peculiar quasiparticle density of states (DOS) has been discussed in terms of multiband or anisotropic superconductivity \cite{Hess1990, Yokoya2001, Noat2010, Noat2015, Rodrigo2004a, Boaknin2003, Fletcher2007, Guillamon2008, Guillamon2008a}. Notice that we use superconducting Pb tips to improve the energy resolution beyond the Fermi-Dirac limit. The convolution of the DOS of tip and substrate shifts all spectral features by the tip's energy gap $\Delta_{\mathrm{t}}\approx 1.35\umeV$ (for details see Supporting Information (SI) \cite{SM}). 

Scanning tunneling microscopy (STM) images of the clean surface clearly reveal the modulation of the local DOS induced by the CDW superimposed on the atomic corrugation (Fig.\ \ref{Fig:figure1}a). The CDW has a lattice constant $a_{\mathrm{cdw}}\gtrsim 3a$, making it incommensurate with the underlying atomic lattice. Thus, the phase of the CDW relative to the atomic lattice varies smoothly across the surface \cite{Gye2019, Guster2019}. The close-ups show two extremal cases for which the maxima coincide either with a Se atom (chalcogen-centered, CC, red circle in Fig.\ \ref{Fig:figure1}a and Fig.\ \ref{Fig:figure1}b) or a hollow site (hollow-centered, HC, green circle in Fig.\ \ref{Fig:figure1}a and Fig.\ \ref{Fig:figure1}b \cite{SC1}). 

\begin{figure}[tp]\centering
\includegraphics[width=8cm]{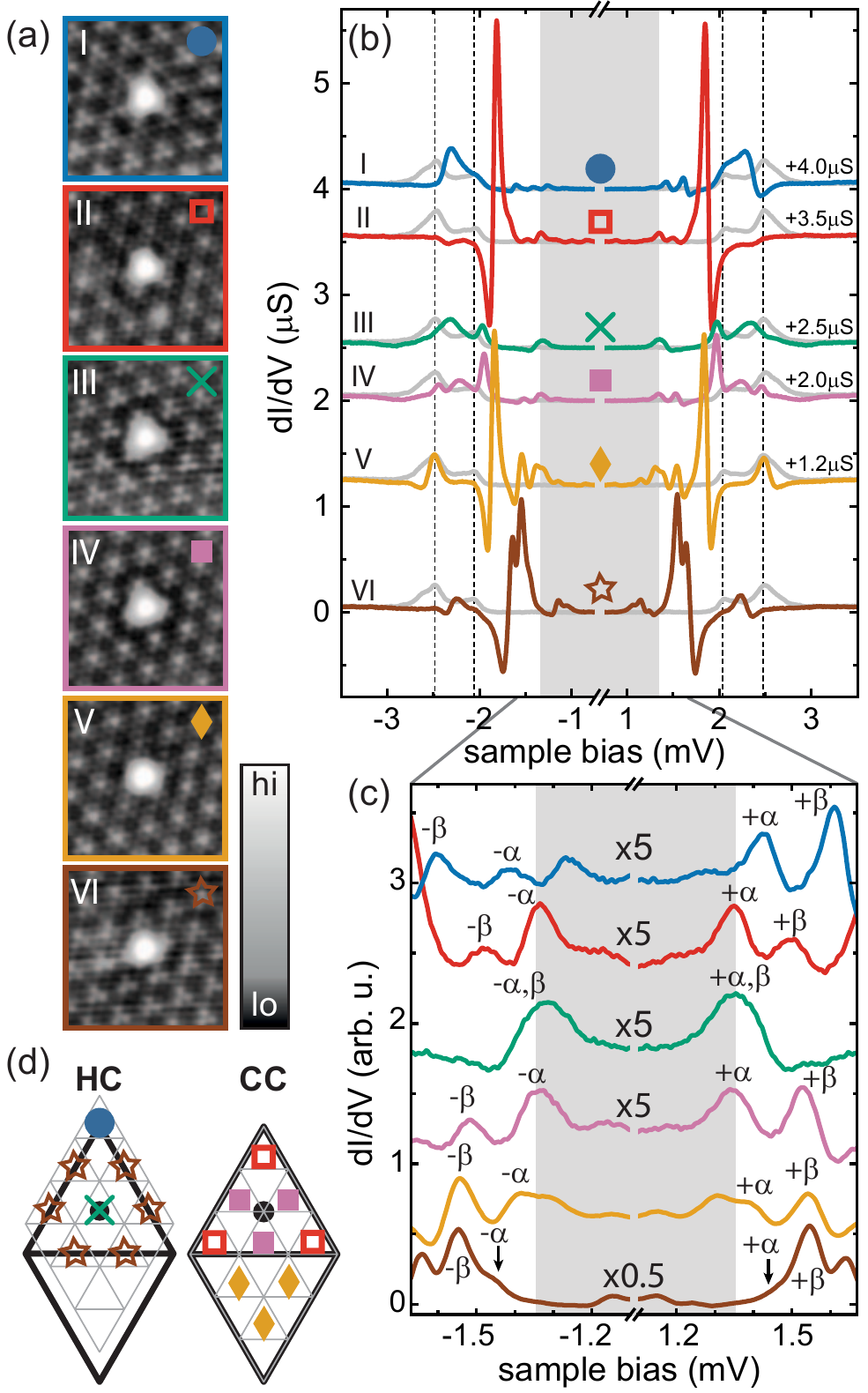}
\caption{
(a) Constant-current STM images ($5\times5\unm^2$) of several HC Fe atoms labeled by I-VI. A non-linear color code is used to resolve the atomic background (set point: \spat). (b) Constant-height \didv spectra taken at the center of atoms I-VI (color) and on the substrate (grey). Spectra offset for clarity (set point: \spat; $V_{\mathrm{rms}}=15\umueV$). 
(c) Close-ups of the spectra in (b). YSR resonances are labeled by $\pm\alpha,\pm\beta$. (d) Superimposed atomic (grey) and CDW lattices (black) for two CDW structures (HC and CC). The Se atoms are located at vertices of the atomic grids, so that the HC (MC) adsorption sites correspond to triangles pointing up (down). In the CDW lattice, vertices (dots) are maxima (minima) of the CDW. Colored symbols indicate the positions of atoms I-VI. 
}
\label{Fig:figure2}
\end{figure}

Fe atoms (deposited at temperatures below 12\,K into the STM) appear as two stable species with different apparent heights (Fig.\ \ref{Fig:figure1}c,d). By atomic-resolution imaging (for details see SI \cite{SM}), we assign adatoms with small and large apparent height to adsorption in the two inequivalent hollow sites of the terminating Se layer, identified as hollow-centered (HC) and metal-centered (MC) sites, respectively (Fig.\ \ref{Fig:figure1}b) \cite{SC3}. \didv spectra taken above the centers of HC and MC adatoms (blue and orange in Fig.\ \ref{Fig:figure1}e) show several YSR states inside the superconducting gap as well as in the energy range of the substrate's coherence peaks as has been observed previously for buried impurities \cite{Senkpiel2018}. The energy and intensity of the YSR states differ between the two species. Presumably, the splitting of the adatom $d$-levels is sensitive to the different local environments of the adsorption sites, which in turn affects the potential- and exchange-scattering strengths \cite{Ruby2016}.

To isolate the influence of the CDW, we focus on Fe atoms which are all sitting in the same atomic adsorption site (HC) in the following. Six different atoms (labeled by I-VI) are shown in Fig.\ \ref{Fig:figure2}a. The corresponding \didv spectra (Fig.\ \ref{Fig:figure2}b) reveal that the energy and intensity of their YSR states differ strongly even though their atomic adsorption sites are identical with respect to the unperturbed lattice. 

\begin{figure*}[t]
\includegraphics[width=16cm]{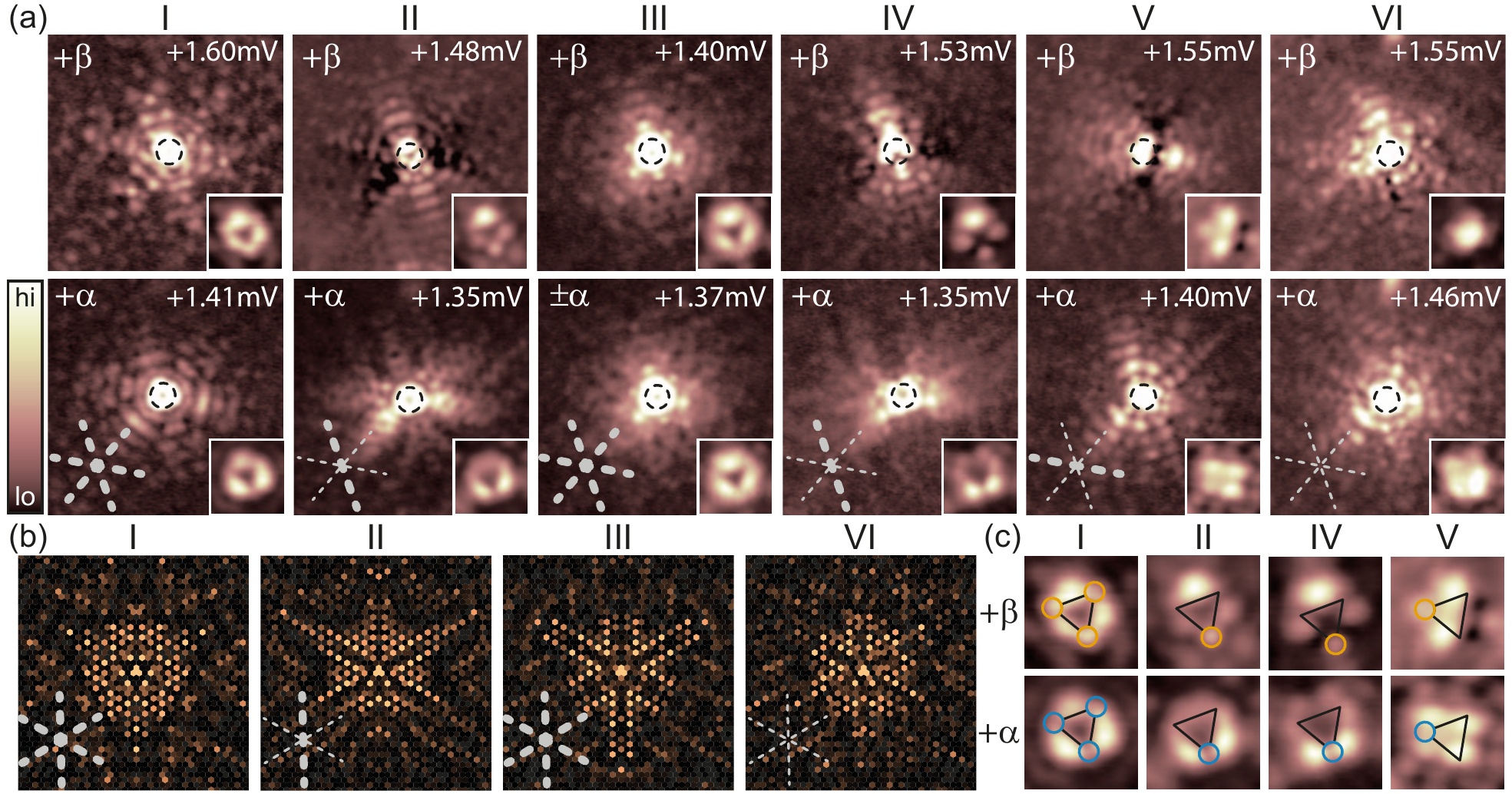}
\caption{
(a) Constant-contour \didv maps ($9.5\times9.5\unm^2$) of the YSR $(\alpha,\beta)$ states of HC atoms (I-VI) (set point: \spat; $V_{\mathrm{rms}}=15$-$25\umueV$; $V_{\mathrm{bias}}$ as indicated). Black dashed circles (diameter $1\unm$) outline the atoms' position. The insets show a $2\times2\unm^2$ close-up view around the center of the atoms. (b) Numerical results for YSR states within a phenomenological description of the CDW (field of view: $40a\times 40a$ around adatom). For details of the model and parameters, see SI \cite{SM}, specifically Fig.\ S2 where these plots are complemented by additional adsorption sites. Thick (thin) grey lines in the maps indicate presence (absence) of mirror axes. (c) Close-up view around the center of the atoms I, II, IV, and V with superimposed triangles illustrating the presence (orange circles) and absence (blue circles) of intensity at their vertices. On atom III, states $\alpha$ and $\beta$ overlap, on atom VI there is no symmetry.
}
\label{Fig:figure3}	
\end{figure*}

YSR states in the energy range of the superconducting coherence peaks are difficult to disentangle from the background. To avoid this complication, we focus on deep-lying YSR states, specifically the two lowest YSR pairs labeled as $\pm\alpha$ and $\pm\beta$ in the close-up view presented in Fig.\ \ref{Fig:figure2}c. \didv maps recorded at the corresponding bias voltages $V_{\alpha}$ and $V_{\beta}$ are shown in Fig.\ \ref{Fig:figure3}a for adatoms I-VI. The main panels show the extended patterns while the insets focus on the immediate vicinity of the adatom. All extended maps show patterns with oscillating intensity \cite{SC2}. The overall symmetries of the patterns clearly differ between adatoms I-VI. 

The symmetry of the patterns associated with YSR states is expected to originate from the anisotropy of the Fermi surface \cite{Salkola1997,Flatte1997PRL} and the local crystal field \cite{Ruby2016}. The threefold-symmetric atomic adsorption site together with the sixfold symmetry of the Fermi surface should therefore lead to $D_3$ symmetry (threefold rotation as well as three mirror axes, {\em cf.} dashed lines in Fig.\ \ref{Fig:figure1}b). Interestingly, this $D_3$ symmetry is observed for both YSR states $\alpha$ and $\beta$ of adatoms I and III, which reside at a maximum and a minimum of the CDW, respectively (Figs.\ \ref{Fig:figure2}d, \ref{Fig:figure3}a). However, this symmetry is lost in both, the long-range and the immediate vicinity of the adatoms for the other atoms shown in Fig.\ \ref{Fig:figure3}a. For adatoms II, IV, and V, the symmetry is reduced to $D_1$ symmetry (single mirror axis) while no symmetry axis can be discerned for adatom VI. 

The symmetry reductions of the YSR patterns coincide with the reductions of the local symmetry of the adsorption sites by the CDW (Fig.\ \ref{Fig:figure2}d). Adatoms I and III are positioned at a maximum or a minimum of the CDW, so that the CDW respects the atomic $D_3$ symmetry (HC structure of the CDW).  Atoms II, IV, and V are located on one of the three equivalent symmetry axes connecting the high-symmetry positions, but unlike adatoms I and III do not directly fall on an extremum (CDW in the CC domain). Thus, the CDW breaks the atomic $D_3$ symmetry, leaving only $D_1$ symmetry consistent with the observed YSR patterns. Finally, the position of adatom VI is totally asymmetric with respect to the CDW, which is reflected in the absence of any symmetry in the corresponding \didv maps. 

The microscopic physics behind these symmetry reductions can be understood theoretically within a phenomenological mean-field description of the CDW combined with a tight-binding model of the \twoH\ band structure. Foregoing a realistic description of the Fe $d$-orbital physics, we model the adatom as a classical impurity with isotropic potential and exchange couplings to the substrate (see SI \cite{SM} for details). Due to the electron-phonon interaction, the CDW acts on the electrons as a weakly incommensurate, static periodic potential. This potential reduces the symmetry of the YSR wavefunctions in a manner consistent with the experimental results. Corresponding numerical results are shown in Fig.\ \ref{Fig:figure3}b for a subset of the adsorption sites investigated experimentally (see SI \cite{SM} for the remaining sites and further details). We find that the wavefunctions (and the number) of the YSR states are quite sensitive to details of the band structure and the neglected $d$-orbital physics, so that one expects agreement only for qualitative aspects of symmetry.

So far, the assignments $\alpha$ and $\beta$ for the YSR states simply refer to the two deepest resonances. 
Analyzing the observed YSR patterns allows us to track these two YSR states as a function of position relative to the CDW. For the YSR state $+\beta$ of adatom I, the short-range pattern exhibits a triangular shape (Fig.\ \ref{Fig:figure3}c) with high intensity on its sides and slightly smaller intensity at its vertices (orange circles). In contrast, the YSR state $+\alpha$ for adatom I lacks intensity at the vertices (blue circles) and thus exhibits pronounced nodes. Signatures of this distinction persist at the lower-symmetry positions II, IV, and V. While the intensity distributions change for both states, we still observe distinct intensities at one of the vertices which distinguish between the $+\alpha$ and $+\beta$ resonances (while present, the distinction is somewhat less pronounced for adatom V, see also SI \cite{SM}). A similar distinction cannot be made for adatom III. We attribute this to the fact that in this case, both the $\pm\alpha$- and the $\pm\beta$-resonances lie very close to the Fermi level. This prevents us from individually resolving them within our energy resolution (Fig.\ \ref{Fig:figure2}c, green trace).

\begin{figure}[t]
\centering
\includegraphics[width=8cm]{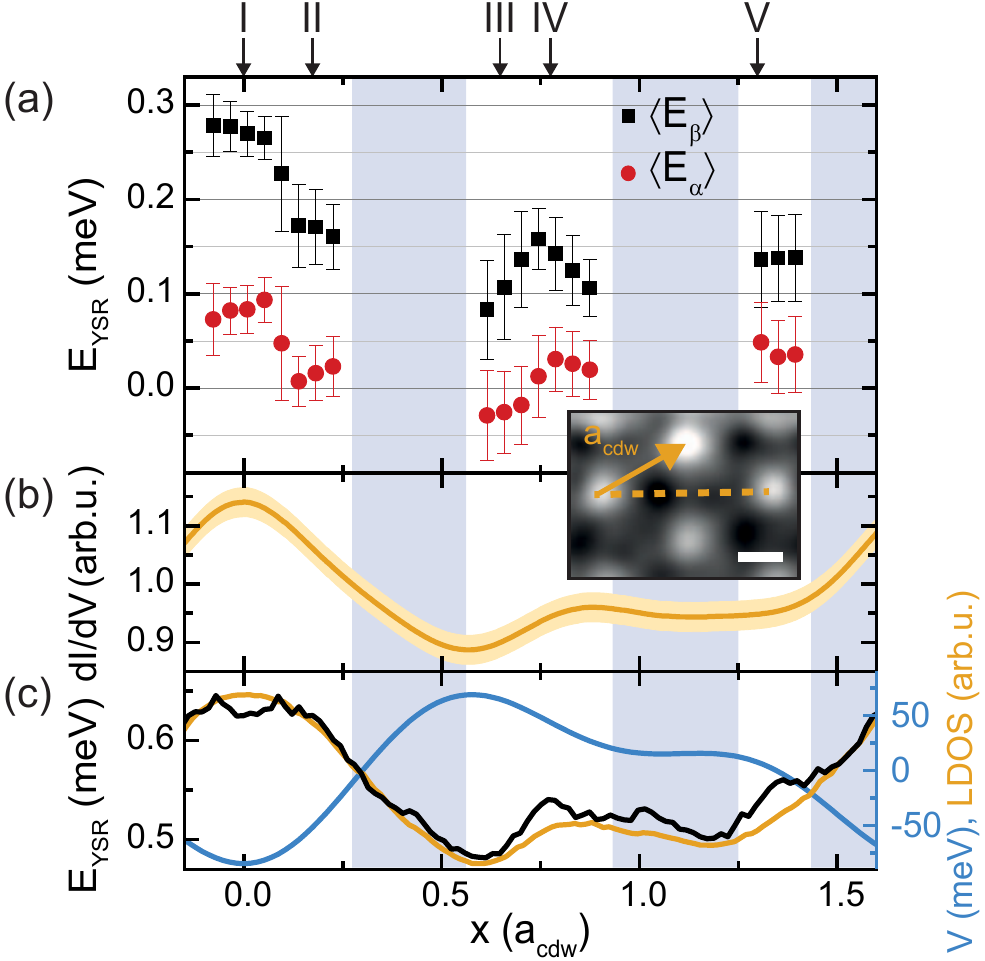}
\caption{
(a) Evolution of the $\alpha$-,$\beta$-YSR energy with position relative to the CDW. Displayed points are obtained from averaging the energies of various atoms within intervals of $x\pm0.05$\,a$_{\mathrm{cdw}}$. Error bars include standard deviation from the averaging method and experimental errors (see SI \cite{SM}). (b) Linecut along the orange-dashed line through a FFT-filtered constant-height \didv map [taken at $T=8\,$K and $V_{\mathrm{bias}}=0$ (set point: \spat; $V_{\mathrm{rms}}=100\umueV$)] shown as inset (scale bar is 4\,\r A) representing the variation of the local DOS (the non-filtered data can be found in Fig.\ S17a). (c) CDW potential (blue), local DOS (orange) and YSR energies (black) obtained from the theoretical model for $JS/2=360\umeV , i.e.$\ in the strong-coupling regime. 
}
\label{Fig:figure4}
\end{figure}

With these characteristics, we can track the resonances as a function of position relative to the CDW and probe its effect on the YSR energies. As the CDW transforms smoothly between HC and CC domains, we can study numerous positions along its symmetry axis (such as positions II, IV, and V in Fig.\ \ref{Fig:figure2}d, but also in between). Altogether, we analyze approximately $90$ adatoms adsorbed close to one of the three equivalent CDW symmetry axes, combining data from several samples and Pb tips. For all adatoms, we extract the energies of the YSR states by deconvolving the \didv spectra to remove the influence of the Pb tip (see SI \cite{SM}). Figure \ref{Fig:figure4}a plots the energies of both the $\alpha$ and $\beta$ states as a function of adatom position along a high-symmetry axis of the CDW. Here, we assume that the energy dependences of both resonances follow the same trend which implies that the energy of the $\alpha$ resonance changes sign and crosses the quantum phase transition from a screened-spin to a free-spin ground state \cite{Choi2017, Franke2011, Farinacci2018, Malavolti2018} as a function of adatom position. Note that some regions (shaded in Fig.\ \ref{Fig:figure4}a) of adatom positions remain inaccessible for Fe atoms in HC sites, because the CDW maxima avoid MC domains \cite{SC1}. 

A qualitatively similar dependence of the energy of the YSR states is found in our model calculations (Fig.\ 
\ref{Fig:figure4}c and SI \cite{SM}). The YSR energies are (anti)correlated with the CDW potential and the local DOS, with the type of correlation depending on whether the impurity is in the weak- or strong-coupling regime and on the strength of potential scattering by the impurity \cite{SM}. (We find that the CDW potential is anticorrelated with the local DOS.) Except for very strong potential scattering (see SI \cite{SM}), our results are consistent with a simple model for the energy of YSR states \cite{Yu1965, Shiba1968, Rusinov1969},  
\begin{align}
E_{\text{YSR}}=\pm\Delta\frac{1-\left(\pi J S \nu_0 /2\right)^2}{1+\left(\pi J S \nu_0 /2\right)^2}\, .
\label{Eq:YSR}
\end{align}
Here, $\Delta$ is the superconducting gap, $S$ the (classical) spin, $J$ the exchange coupling, and $\nu_0$ the local DOS of the superconductor. This predicts that YSR energies are (anti)correlated with the local DOS for impurities in the strong (weak) coupling limit. Experimentally, we record a constant-height \didv map at zero bias above $T_{\mathrm{c}}$ and remove the atomic corrugation by an FFT filter ({\em cf.} inset Fig.\ \ref{Fig:figure4}b). The remaining signal is proportional to the local DOS due to the CDW at the Fermi level. We find that the YSR energies correlate with the local DOS as seen in Fig.\ \ref{Fig:figure4}a,b. This suggests that the YSR states are in most of the adsorption sites in the strong-coupling regime. Only when the Fe atoms are located in the minima of the CDW, resonance $\alpha$ undergoes the quantum phase transition to the free-spin ground state.

In conclusion, we find that the properties of YSR states are strongly affected by CDW ordering. The CDW 
shifts the energy of YSR states and reduces the symmetry of YSR wavefunctions both in the immediate vicinity of and further from the adatoms. These observations impose stringent constraints on the design of strongly coupled magnetic adatom structures on \nbse\ substrates. Neverthless, \nbse\ remains a highly attractive substrate in this context, owing to the long-range nature of the YSR states and its suitability for adatom manipulation. Our results extend to truly two-dimensional \nbse\ substrates with their promise of device applications. Indeed, single-layer \nbse\ also exhibits an incommensurate CDW \cite{Guster2019, Ugeda2016}. Importantly, pairing correlations in single-layer \nbse\ are remarkably robust against in-plane fields as a consequence of spin-valley locking (Ising superconductivity) \cite{Xi2016}. This may enable measurements of YSR states at magnetic fields which polarize the adatom spin without suppressing superconductivity. The interaction of YSR states with superconductivity, charge-density order, and external magnetic fields entails substantial flexibility to tune into or control a topologically non-trivial superconducting phase. Finally, our study suggests that YSR states provide a sensitive probe of these competing interactions which should extend to other systems.

\acknowledgements
Financial support by Deutsche Forschungsgemeinschaft through grant CRC 183 (FvO), as well as by the European Research Council through the consolidator grant ``NanoSpin" (KJF) is gratefully acknowledged. 

\bibliographystyle{apsrev4-1}
%\bibliography{library}
%

\clearpage

\setcounter{figure}{0}
\setcounter{section}{0}
\setcounter{equation}{0}
\setcounter{table}{0}
\renewcommand{\theequation}{S\arabic{equation}}
\renewcommand{\thefigure}{S\arabic{figure}}
	\renewcommand{\thetable}{S\arabic{table}}%
	\setcounter{section}{0}
	\renewcommand{\thesection}{S\arabic{section}}%

\onecolumngrid

\renewcommand{\Fig}[1]{\mbox{Fig.\unitspace\ref{Sfig:#1}}}
\renewcommand{\Figure}[1]{\mbox{Figure\unitspace\ref{Sfig:#1}}}

\newcommand{\vsigma}{\mbox{\boldmath $\sigma$}}

\section*{\Large{Supporting Information}}

\section{Theoretical considerations}

\subsection{Mean-field description of the charge-density wave}

We resort to a phenomenological mean-field description. We start with a generic Fr\"ohlich-type Hamiltonian describing electrons coupled to phonons \cite{SGruner2018},
\begin{equation}
   H=\sum_{\bf k} \epsilon_{\bf k} c^\dagger_{\bf k} c_{\bf k} + \sum_{\bf q} \omega_{\bf q} 
   b^\dagger_{\bf q} b_{\bf q} + \sum_{\bf k,q} g_{\bf q} (b_{\bf q}+b^\dagger_{-\bf q})c^\dagger_{\bf k+q}c_{\bf k}.
\end{equation}
Here, $c_{\bf k}$ annihilates electrons with momentum ${\bf k}$ and energy $\epsilon_{\bf k}$, $b_{\bf q}$ annihilates a phonon with wavevector ${\bf q}$ and frequency $\omega_{\bf q}$, and $g_{\bf q}$ denotes the strength of the electron-phonon coupling. Within mean-field theory, we assume that certain phonon modes (denoted by ${\bf Q}$) go soft due to their coupling to the electronic system and develop a finite expectation value. Restricting to the lowest Fourier components and neglecting phonon dynamics beyond the static charge-density wave (CDW) distortions, we find the mean-field Hamiltonian
\begin{equation}
   H=\sum_{\bf k} \epsilon_{\bf k} c^\dagger_{\bf k} c_{\bf k} + \sum_{\bf k}\sum_{\bf Q} g_{\bf Q} \langle b_{\bf Q}+b^\dagger_{-\bf Q}\rangle c^\dagger_{\bf k+Q}c_{\bf k}
\end{equation}
for the electronic degrees of freedom. Within this approach, the CDW acts on the electrons as a periodic potential $V({\bf r})$. Indeed, the electron-phonon interaction term can also be expressed in real space as 
\begin{equation}
   H_{\rm el-ph}= \int {\rm d}{\bf r} V({\bf r}) \psi^\dagger({\bf r})\psi({\bf r}) ,
\end{equation}
where the mean-field CDW potential takes the form
\begin{equation}
  V({\bf r})= \sum_{\bf Q} e^{-i{\bf Q}\cdot {\bf r}} g_{\bf Q} \langle b_{\bf Q}+b^\dagger_{-\bf Q}\rangle.
\end{equation}

To find the vectors \textbf{Q}, we consider a triangular lattice of Nb atoms with lattice vectors
\begin{align}
{\bf a}_1 & =a(1,0), \\
{\bf a}_2 & =a(1/2,\sqrt{3}/2)
\end{align}
and bond length $a=3.445\,$\r A. The corresponding reciprocal lattice vectors ${\bf b}_i$ satisfying ${\bf a}_i \cdot {\bf b}_j=2\pi \delta_{ij}$ are
\begin{align}
{\bf b}_1 & =\frac{4\pi}{\sqrt{3}a}(\sqrt{3}/2,-1/2), \\
{\bf b}_2 & =\frac{4\pi}{\sqrt{3}a}(0,1).
\end{align}
The unit cell of the CDW has a linear dimension which is approximately three times larger than the unit cell of the atomic lattice, ${\bf a}^{\rm cdw} \approx 3{\bf a}_i$ (for $i=1,2$). Correspondingly, the reciprocal lattice vectors of the CDW are approximately a factor of three smaller, ${\bf b}^{\rm cdw}_i\approx {\bf b}_i/3$. Then, the first harmonics of the CDW have wavevectors
\begin{align}
{\bf Q}_1 & =q(1-\delta) (\sqrt{3}/2,-1/2), \nonumber \\
{\bf Q}_2 & =q(1-\delta) (0,1) ,  \label{eq:FourierComps} \\
{\bf Q}_3 & =q(1-\delta)(-\sqrt{3}/2,-1/2)=-({\bf Q}_1+{\bf Q}_2), \nonumber 
\end{align}
where $q=\frac{4\pi}{3\sqrt{3}a}$, and $\delta\ll 1$ accounts for the fact that the CDW is not exactly commensurate with the lattice. 

\subsection{Tight-binding model}

\begin{figure}[t]
\centering
\includegraphics[width=\textwidth]{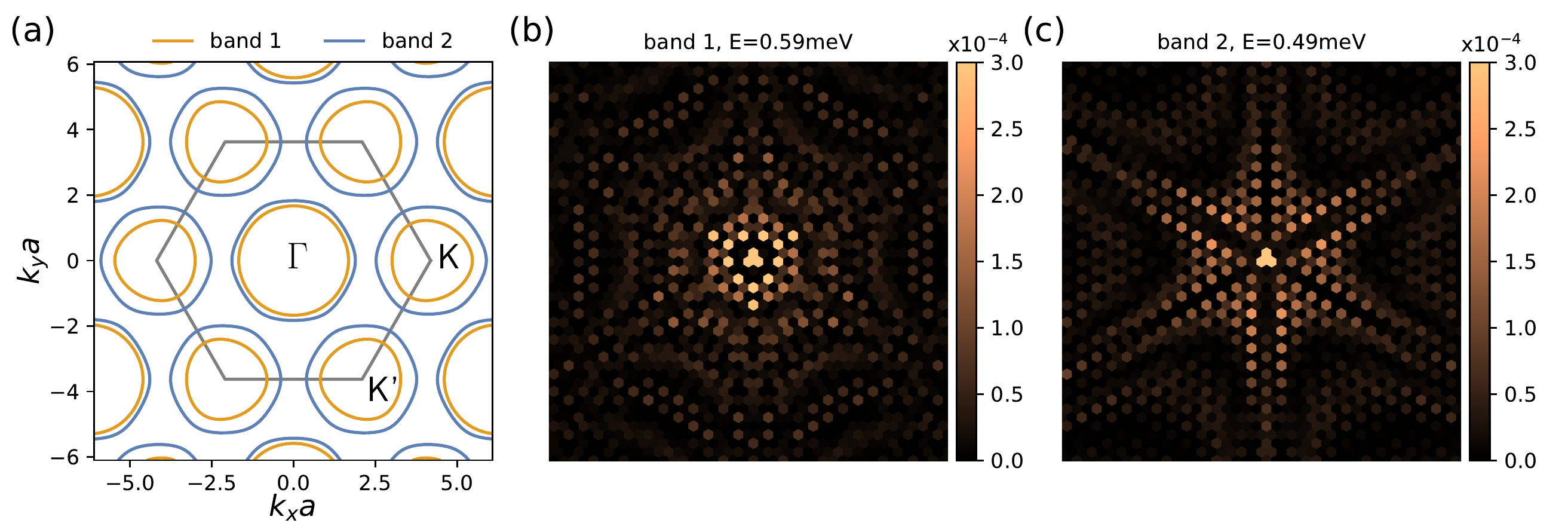}
\caption{(a) Fermi surfaces of bands 1 and 2 (see Eq.\ (\ref{eq:band12}) and Table \ref{table1}). (b,c) Electronic probability density $|u({\bf r})|^2$ of the YSR state in the absence of the CDW for $JS/2=120$\,meV, $K=0$ and a lattice of size $504 \times 504$ with periodic boundary conditions for band 1 (b) and band 2 (c). The region shown includes $40\times 40$ lattice spacings.} \label{Fig:S1}
\end{figure}

To illustrate the effect of the CDW on the YSR state wavefunction, we perform model calculations within an effective tight-binding description. Specifically, we implement a model for the NbSe$_2$ band structure first introduced in Ref.\ \cite{SSmith1985} and later used in Refs. \cite{SRossnagel2005,SInosov2008,SInosov2009,SRahn2012,SMenard2015}$ $. 
The model uses that the relevant states near the Fermi energy have mostly Nb character and therefore
focuses on one atomic $d$-orbital per niobium atom. (The model neglects an additional band centered at the $\Gamma$ point which predominantly derives from Se orbitals and has strongly three-dimensional character \cite{SBawden2016}.) 
Reducing the band-structure problem to the triangular Nb lattice makes it necessary to include hopping up to fifth nearest neighbors to reproduce the NbSe$_2$ band structure. The resulting band structure
\begin{equation}
\begin{split}
E({\bf k}) &= t_0 + t_1 (2 \cos \xi \cos \eta + \cos 2\xi)+t_2(2\cos 3\xi\cos\eta+\cos 2\eta) +t_3(2\cos 2\xi\cos 2\eta +\cos 4\xi) \\ &+ t_4( \cos\xi \cos 3\eta + \cos 5\xi\cos \eta+\cos 4\xi\cos 2\eta) + t_5(2 \cos 3\xi\cos 3\eta + \cos 6\xi)  \label{eq:band12}
\end{split}
\end{equation}
\begin{table}
\begin{center}
\begin{tabular}{ ccccccc } 
\hline
       & $t_0$ & $t_1$ & $t_2$ & $t_3$ & $t_4$ & $t_5$    \\
\hline
band 1 & 10.9  & 86.8  & 139.9 & 29.6  & 3.5   &  3.3   \\
band 2 & 203.0 & 46.0  & 257.5 & 4.4   & -15.0 &  6.0    \\ 
\hline
\end{tabular}
\caption{Values of the fitting parameters $t_n$ in Eq.\ (\ref{eq:band12}) in meV.} \label{table1}
\end{center}
\end{table}is sixfold symmetric about the $\Gamma$ point. Here, we defined $\xi=k_x/2$ and $\eta=\sqrt{3}k_y/2$. The hopping strengths $t_n$ are used as fitting parameters to reproduce the NbSe$_2$ band structure. The values of the parameters $t_n$ are given in Table \ref{table1} \cite{SRahn2012} for the two bands. The parameters for band 1 reproduce the inner cylindrical bands of NbSe$_2$, while the parameters for band 2 reproduce the outer bands. Figure\ \ref{Fig:S1}a shows the Fermi surfaces of the two bands, with the Brillouin zone outlined in gray.

While this model accounts for the symmetries of the band structure and the approximate shapes of the Fermi surfaces, it is limited in other ways. The model neglects interlayer coupling which, however, is expected to be weak compared to the intralayer couplings. In this approximation, it suffices to consider a single (tri)layer of \twoH, consisting of a Nb layer sandwiched between two Se layers with equal orientations of the triangles. The model also neglects spin-orbit coupling \cite{SXi2016}. This precludes a one-to-one identification of the two sets of bands in Fig.\ \ref{Fig:S1}a with specific spin directions. 

The elementary layer of bulk \twoH\ consists of two trilayers with the orientations of the Se lattices rotated by $180^\circ$ with respect to one another. Neglecting interlayer coupling, the band structure of the second trilayer is identical to the first (but differs in the spin assignments). Thus, the bulk crystal has twofold degenerate bands, consistent with the fact that it has an inversion center located between the trilayers. The adatom predominantly couples to one of the trilayers and the band structure in Eq.\ (\ref{eq:band12}) should be a good starting point. In the following, we follow Ref.\  \cite{SMenard2015} in neglecting the splitting between the two sets of bands and model the system either by band 1 or band 2, taking these bands as spin degenerate. 

Superconductivity is included by incorporating the tight-binding Hamiltonian into a Bogoliubov-deGennes Hamiltonian with conventional (isotropic) $s$-wave pairing.  We choose the pairing strength to be  $\Delta=1$\,meV. In experiment, one finds multiple coherence peaks for \twoH. Since we focus on YSR states far from the gap edge, we choose a representative value for $\Delta$ which falls into the range of the peak distribution ($0.7-1.4$\,meV) found in experiment. 

The multiple coherence peaks originate from multiband or anisotropic superconductivity, or both. Our modeling does not account for these effects. In principle, an anisotropic gap function affects the wavefunction patterns of the YSR states. However, the (observable) spatial extent of YSR states is smaller than the coherence length through which effects of the pairing function would enter. We thus expect that the effect of an anisotropic pairing function on the YSR wavefunctions is weak and therefore account for superconductivity via an isotropic pairing function. 

\subsection{Coupling to the magnetic impurity}

While the origin of the magnetic moment and the resulting exchange coupling can be understood starting with an Anderson model and its mapping to a Kondo Hamiltonian in the absence of valence fluctuations (see Ref.\ \cite{SHewson1993} for an extensive pedagogical discussion), YSR states are typically well described within a model which treats the impurity spin in the Kondo model as classical \cite{SYu1965, SShiba1968, SRusinov1969}. For YSR states in NbSe$_2$, this model was previously used in Ref.\ \cite{SMenard2015}$ $. Following this reference, we model a magnetic impurity as a classical spin $S$ which interacts with the substrate electrons through the exchange interaction $JS\sigma_z/2$ and accompanying potential scattering $K$. In our experiment, we focus on magnetic impurities located above the center of a Nb triangle (hollow-centered, see Fig.\ 1b of the main text). The impurity induces not only separate exchange and potential couplings to its three neighbors, but also nonlocal exchange couplings which scatter electrons between the three sites via the impurity. It is essential to retain the latter to ensure that the impurity induces only one pair of YSR resonances. If we assume identical hopping amplitudes between the three sites and the impurity and assume that the impurity spin is aligned along the $z$-direction, we obtain the exchange coupling 
\begin{equation}
  H_{\rm exch} = -\frac{JS/2}{3}\sum_{i,j}\sum_{s,s'} c_{is}^\dagger \sigma^z_{ss'}c_{js'}
\end{equation}
and potential scattering terms  
\begin{equation}
  H_{\rm pot} = \frac{K}{3} \sum_{i,j}\sum_s c_{is}^\dagger c_{js}.
\end{equation}
Here, the sums over $i$ and $j$ run over the three nearest-neighbor Nb sites of the impurity, $\sigma^z$ denotes a Pauli matrix, and $c_{is}$ annihilates an electron on site $i$ with spin $s$. 

Figures\ \ref{Fig:S1}b and c show the probability densities of the YSR states for (a spin-degenerate version of) bands 1 and 2, respectively, for a specific choice of impurity parameters. The spatial distribution of the probability density is threefold symmetric in both cases. Due to the more isotropic character of band 1, the corresponding YSR state is also more spatially isotropic than its band 2 counterpart, which exhibits a more pronounced anisotropy. In a more realistic modeling including the effects of spin-orbit coupling on the band structure (but within the approximation that the impurity is coupled only to a single trilayer), the YSR states would be simultaneously coupled to spin-textured, but polarized versions of  both bands. While this implies that our results for the YSR wavefunctions are only qualitative, the symmetry of the YSR wavefunctions should not be affected by this approximation. Similarly, we expect that the qualitative variation of the YSR energies with the location of the adsorption site relative to the CDW would also be unaffected. 

We note that in our approximation, there is only a single pair of  YSR states for a classical magnetic impurity. In fact, this conclusion would remain valid even when spin-orbit splitting of the bands was taken into account. On the one hand, it is known that even in multiband situations, a classical magnetic impurity induces only a single pair of YSR states. In fact, if we write the matrix of exchange couplings $J_{\alpha\alpha'}$ between bands $\alpha$ and $\alpha'$, it rather robustly satisfies the relation $J_{11}J_{22}=J_{12}J_{21}$.  With this relation, a classical magnetic impurity generates only a single pair of YSR states. For instance, this was shown explicitly for a generic model motivated by the multiband superconductor MgB$_2$ \cite{SMoca2008} and for a quantum dot coupled to multiple superconducting leads \cite{SKirsanskas2015}. In the presence of spin-orbit coupling, the same conclusions can be shown to remain valid. This can be seen by expressing the  Hamiltonian in terms of creation and annihilation operators for Kramers pairs of electron states rather than conventional spin states. 

\subsection{Coupling to the charge-density wave}

As explained above, the effect of the CDW on the electrons of the substrate can be described through a periodic potential. Within the tight-binding model, we include the CDW as a modulation of the on-site potential,
\begin{equation}
H_{\rm cdw} = \sum_{{\bf r},\sigma}  c_{\mathbf{r}\sigma}^{\dagger} V(\mathbf{r}) c_{\mathbf{r}\sigma},
\end{equation}
where the sum runs over the lattice sites of the triangular Nb lattice. The CDW potential  $V(\mathbf{r})$ is constructed from the main Fourier components of the CDW $[$see Eq.\ (\ref{eq:FourierComps})$]$,
\begin{equation}
V(\mathbf{r})= V_0 \left[ \cos (\mathbf{Q}_1\cdot \mathbf{r} + \phi_1 ) + \cos (\mathbf{Q}_2\cdot \mathbf{r} + \phi_2 ) +  \cos (\mathbf{Q}_3\cdot \mathbf{r} + \phi_3 ) \right],
\end{equation}
where $V_0$ denotes the amplitude of the CDW potential. Except for $\phi_1=\phi_2=\phi_3$ and similar fine-tuned cases, typical choices of the phases $\phi_i$ yield CDW potentials with the desired symmetry and shape, exhibiting an absolute maximum, an absolute minimum, and a local minimum/saddle point. For definiteness, we choose $\phi_1=\phi_2=0$. 

Due to the small deviations from commensurability, the CDW shifts slowly as a function of position relative to the underlying atomic lattice. On the scale of the YSR states, these shifts can be considered constant so that the CDW potential can be written as
\begin{equation}
V(\mathbf{r})= V_0\left[ 2\cos \left(\sqrt{3}\frac{q}{2}(x-x_0)\right)\cos \left(\frac{q}{2}(y-y_0)\right) + \cos \left(q(y-y_0)+\phi_3\right) \right],
\end{equation}
where $q=4\pi/3\sqrt{3}$. The offsets $x_0$ and $y_0$ originate from the deviation $\delta$ from commensurability, and describe the local shift of the CDW relative to the atomic lattice. In the experiment, the measured maxima of the CDW correspond to regions where the lattice deformation  compresses the ions, creating an attractive potential for the electrons. For this reason we choose the potential $V(\mathbf{r})$ to have minima where the measured tunneling density of states has maxima, \textit{i.e.}, position I of the adatom corresponds to a minimum of $V(\mathbf{r})$ and position III corresponds to a maximum. In principle, the CDW may also modulate the hopping parameters of the tight-binding model, but the effective long-range hopping processes of the model make this cumbersome to include. 

We note that this model also includes configurations of the CDW relative to the atomic lattice which are not observed experimentally (see Fig.\ \ref{Fig:S18}). Theoretical results for these unobserved configurations are only included for completeness.  The results for observed configurations are not affected since the Moire-like variations occur on relatively large scales.  

\subsection{Numerical results}

Within this model, we can study how the wavefunction and the energy of YSR states depend on the adatom position relative to the CDW. As discussed above, the model cannot reproduce the experimental results quantitatively as it neglects various relevant effects. Nevertheless, the model is expected to capture the symmetries of the YSR wavefunctions and to provide insights into the qualitative dependence of the YSR energy on the adsorption site relative to the CDW. 
\begin{figure}[t]
\centering
\includegraphics[width=0.9\textwidth]{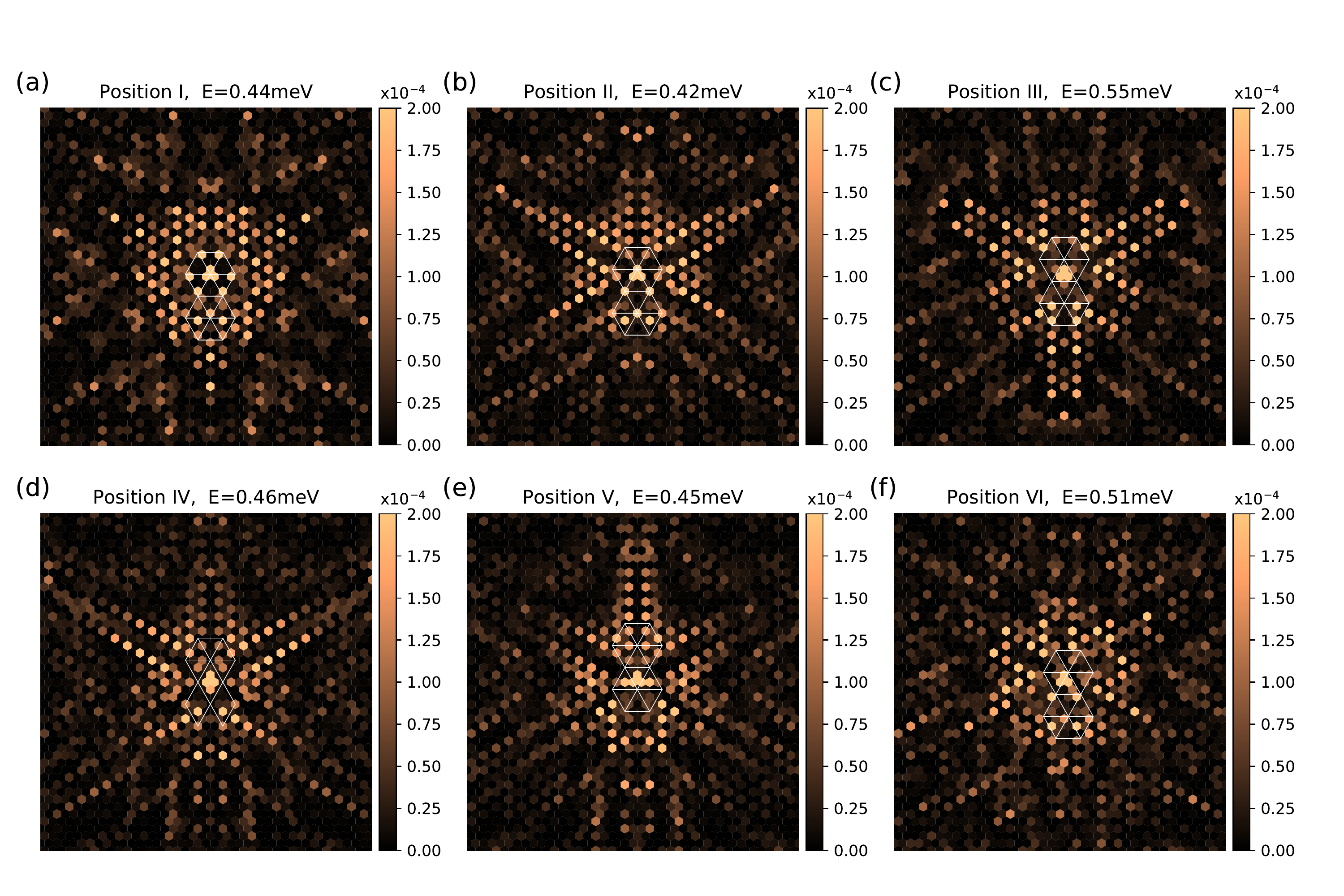}
\caption{Electronic probability density $|u({\bf r})|^2$ of the YSR state in the presence of the CDW potential for band 2 and for adatoms located at positions equivalent to those studied experimentally as presented in the main text. For all plots, the parameters are $JS/2=120$\,meV, $K=0$, $V_0=-30$\,meV, $\phi=\pi/3$ and a lattice of size $504 \times 504$ with periodic boundary conditions. The region shown includes $40\times 40$ lattice spacings. The white lines outline the CDW, with crossing lines indicating a maximum of the CDW.} \label{Fig:S2}
\end{figure}
Figure\ \ref{Fig:S2} shows the probability density of the YSR wavefunction for the six different positions of the adatom relative to the CDW, as discussed in the main text. The results are obtained for the parameters of band 2. The adatoms in positions I and III are at the maximum and the minimum of the CDW, respectively, \textit{i.e.}, at points with a threefold symmetric environment. Consequently, the corresponding density plots  (Fig.\ \ref{Fig:S2}a and c) present threefold symmetry. Positions II, IV, and V exhibit reflection symmetry about an axis that passes through the adatom. For the corresponding plots shown in Fig.\ \ref{Fig:S2}b, d, and e, this axis is aligned along the vertical direction. Finally, position VI has no symmetries with respect to the CDW and consequently, the state shown in Fig.\ \ref{Fig:S2}f also exhibits no symmetries. Interestingly, the YSR state seems to retain the original threefold symmetry (without CDW) to some degree (compare with Fig.\ \ref{Fig:S1}c). 

\begin{figure}[t]
\centering
\includegraphics[width=0.9\textwidth]{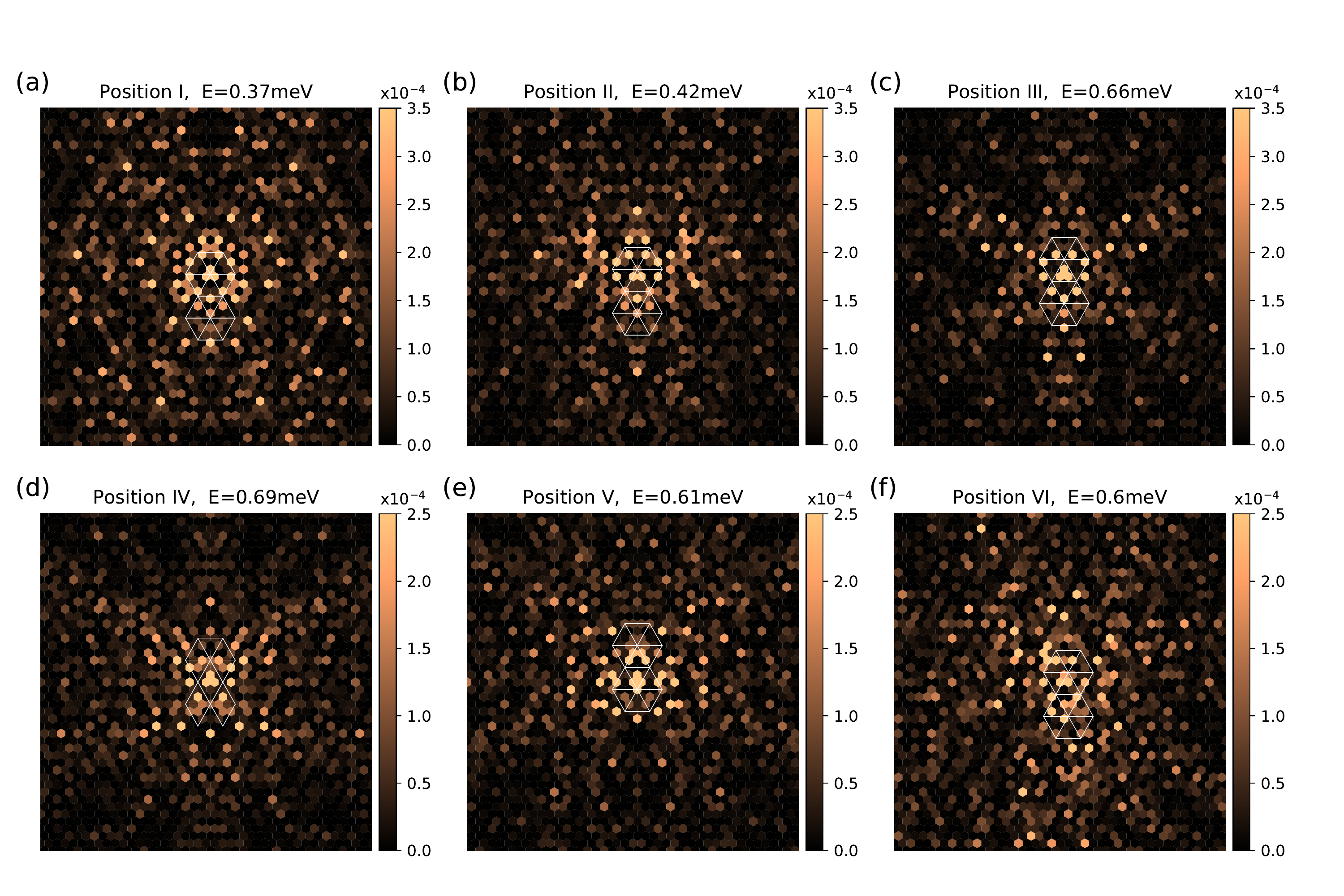}
\caption{Analogous plots to those in Fig.\ \ref{Fig:S2} for band 1. Parameters: $JS/2=120$\,meV, $K=0$, $V_0=-30$\,meV, $\phi=\pi/3$ and a lattice of size $504 \times 504$ with periodic boundary conditions. The region shown includes $40\times 40$ lattice spacings.} \label{Fig:S3}
\end{figure}

Figure\ \ref{Fig:S3} shows equivalent plots for band 1. Due to the more isotropic character of band 1, the threefold symmetry is less pronounced in these plots. Remarkably, position VI (Fig.\ \ref{Fig:S3}f) does not preserve any of the original threefold symmetry in the vicinity of the adatom. More generally, a comparison between Figs.\ \ref{Fig:S2} and \ref{Fig:S3} shows that the detailed YSR wavefunctions depend quite sensitively on the band structure parameters.

Finally, we discuss the correlation of the energy of the YSR state with the modulation of the local density of states by the CDW, as shown in Fig.\ 4 in the main text. Within a simple model of a classical magnetic impurity, the energy of YSR states depends on the strength of the exchange coupling as measured by the dimensionless parameter $\alpha=\pi\nu_0 JS/2$, where $\nu_0$ is the density of states, $J$ the exchange coupling, and $S$ the impurity spin. Thus, one indeed expects the energy of the YSR state to depend on the density of states $\nu_0$. The direction of the energy shift depends on whether the impurity is weakly or strongly coupled, corresponding to unscreened or (partially) screened impurity spins, respectively. The energy of the positive-energy YSR state decreases (increases) with the density of states, depending on whether the impurity spin is unscreened (screened). Variations in the local density of states due to the CDW should thus be reflected in the energy of the YSR states. Within this picture, the observed correlations suggest that in our experiment the impurity is in the strong-coupling limit. 

\begin{figure}[t]
\centering
\includegraphics[width=0.9\textwidth]{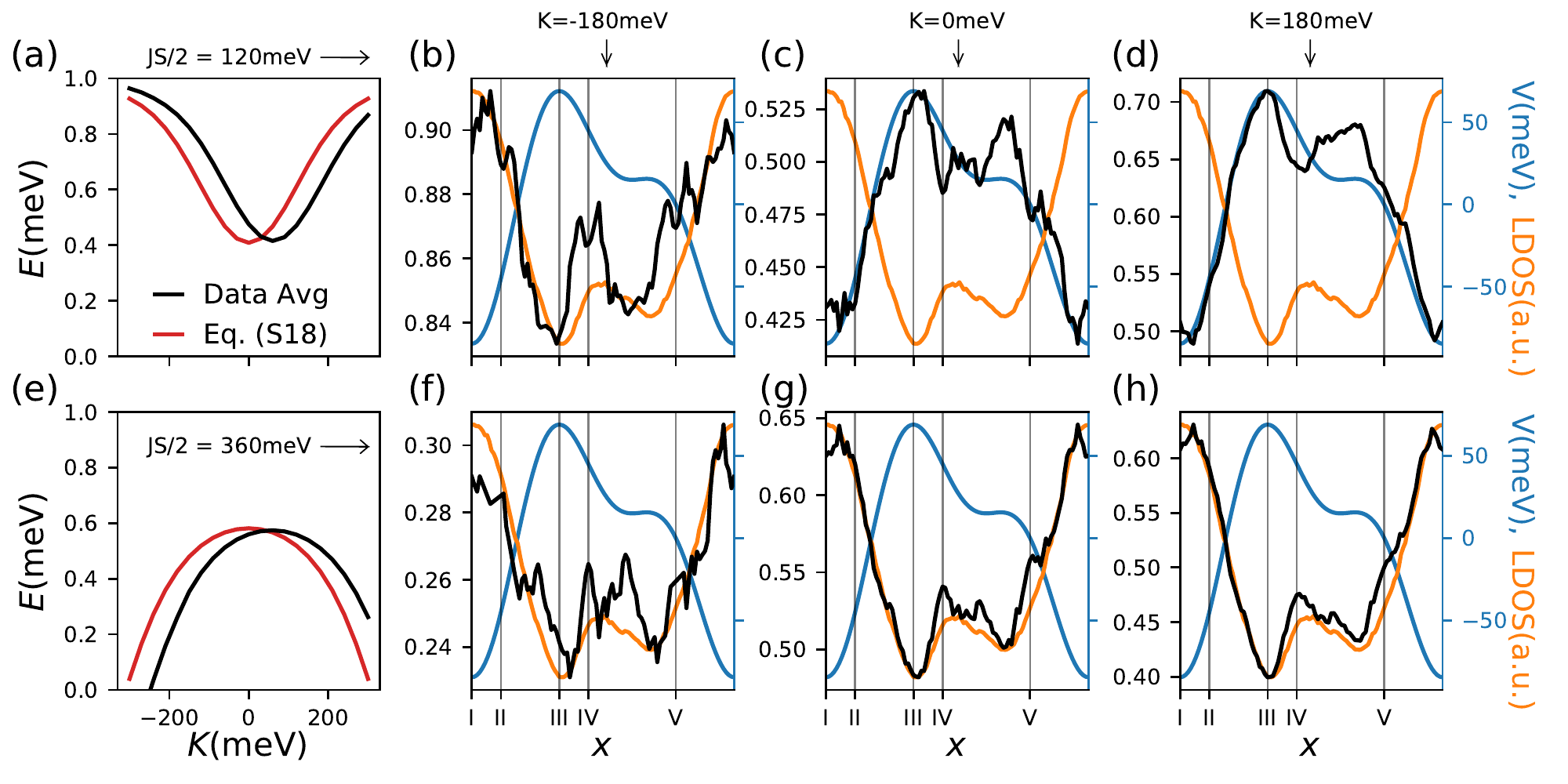}
\caption{Tight-binding calculations for the energy of the YSR state as a function of  the position of the adatom relative to the CDW,  for various potential scattering strengths $K$. Panels (a) and (e) show the energy of the YSR state $E$ vs. $K$ for the weakly and strongly coupled regimes, respectively [as obtained by averaging over different positions of the adatom relative to the CDW (black) and as given by Eq.\ (\ref{eq:E_YSR}) for an appropriately chosen $\nu_0$ (red)]. Panels (b)-(d) show $E$ vs.\ adatom position in the weakly coupled regime for three different $K$. The plots show the local density of states (LDOS, orange) to anticorrelate with the CDW potential $V$ (blue). Panels (c) and (d) show the energy to correlate with the CDW potential $V$ (anticorrelate with the LDOS). For $K=-180$\,meV shown in panel (b) the correlation inverts due to the shift of $K$ to $K+V(\mathbf{r}_{0})$ (see text). Panels (f)-(h) show equivalent plots for the strongly coupled regime, where the effect of $K$ is less relevant. The energy $E$ is correlated with the LDOS as expected from Eq.\ (\ref{eq:E_YSR}). For all plots, the parameters are $V_0=-30$\,meV, $\phi=\pi/3$, and a lattice size of $750\times 750$. } \label{Fig:S4}
\end{figure}

This interpretation is consistent with our theoretical results. However, we find that in general the CDW affects the energy of the YSR state through two mechanisms. In addition to the density of states effect, the CDW potential has a second effect which can be roughly rationalized as the CDW affecting the strength of potential scattering from the impurity and thereby shifting the YSR energy. Including potential scattering within the simple model, the energy of a YSR state is given by \cite{SYu1965,SShiba1968a,SRusinov1969}
\begin{equation}
E = \pm \Delta\frac{1-\alpha^2+\beta^2}{\sqrt{4\alpha^2+(1-\alpha^2+\beta^2)^2}}, \label{eq:E_YSR}
\end{equation}
where $\beta=\pi \nu_0 K$ is a dimensionless measure of the strength of the potential scattering $K$. The transition between weak and strong coupling takes place at $\alpha^2=1+\beta^2$ where the energy of the YSR state passes through zero. Roughly, the CDW potential $V$ seems to shift $K$ to $K+V(\mathbf{r}_{0})$, where $\mathbf{r}_{0}$ denotes the adsorption site of the impurity. For comparison with experiment, it is important to note that an increase in the CDW potential {\em decreases} the density of states. This can be seen in {\em e.g.} Fig.\ \ref{Fig:S4}g, which shows numerical results for the average local density of states of the three sites coupled to the impurity (orange) in the presence of the CDW potential (blue) as obtained from a straight-forward band structure calculation. Thus, the experiment implies an anticorrelation between the CDW potential $V$ and the energy of the YSR states. 

To illustrate these two effects of the CDW potential, we calculate the YRS energy for different adatom positions and various values of $K$. Consider first the results for weak coupling. We find that depending on the strength $K$ of potential scattering, the CDW potential is correlated or anticorrelated with the energy of the YSR state. For $K=180$\,meV and $K=0$, the energy is correlated with the CDW potential, see Fig.\ \ref{Fig:S4}c and d. In contrast, we find anticorrelations for $K=-180$\,meV, see Fig.\ \ref{Fig:S4}b. The correlations observed at $K=0$ are consistent with the effect of the CDW potential on the density of states. At this $K$, the energy shift of the YSR state due to the change in potential scattering, $K$ to $K+V(\mathbf{r}_{0})$, is weak since the energy of the YSR states is only weakly dependent on $K$ near the minimum at $K=0$. This is illustrated in Fig.\ \ref{Fig:S4}a which shows the dependence of the energy of the YSR state on $K$ in the weak-coupling regime. Specifically, the black curve averages data for the energy of the YSR state over different adatom positions. The red curve corresponds to Eq.\ (\ref{eq:E_YSR}) with the same values for $JS$ and $K$ and an appropriately chosen density of states $\nu_0$. Both curves exhibit similar `parabolic' behaviors.

For the strongly coupled regime, we find anticorrelations between the CDW potential and the energy of the YSR state -- and thus correlations between the local density of states and the YSR energy -- for all values of $K$, see Fig.\ \ref{Fig:S4}f,g and h. These correlations between the density of states and the YSR energy are consistent with experiment. To start with, the anticorrelations at $K=0$ are consistent with the expected density of states effect according to Eq.\ (\ref{eq:E_YSR}). For strong coupling, the dependence of the YSR energy on $K$ is generally weaker compared to the weak coupling case, see Fig.\ \ref{Fig:S4}e. For this reason, the shift of $K$ by the CDW potential no longer overcomes the density-of-states effect. Thus, unless the impurity is a surprisingly strong potential scatterer, we conclude that in experiment the impurity is in the strong coupling limit.  

\section{Experimental details}
\subsection{Superconducting tip}
In order to increase the energy resolution beyond the Fermi-Dirac limit, all measurements presented in the manuscript were performed using a superconducting Pb tip fabricated by indenting a W tip into a clean Pb(111) surface until the tip exhibits a bulk-like superconducting gap. Details of the preparation procedure for tip and sample are described in Ref.\ \cite{SRuby2015}$ $. The tunneling current is proportional to the convolution of the density of states of the sample $\rho_{\text{S}}$ and tip $\rho_{\text{t}}$ and can be expressed as \cite{STersoff1985}:
\begin{align}
I(V)\propto\int_{-\infty}^{+\infty} \text{d}E\,\, \rho_{\text{t}}(E-eV,T) \rho_{\text{s}}(E,T)[f(E-eV,T)-f(E,T)]|M_{\mu ,\nu}|^2 \,\, .
\label{Eq:current}
\end{align}
Here, $f(E,T)$ is the Fermi-Dirac-distribution function accounting for the thermal occupation of the states in the tip and the sample, and $|M_{\mu ,\nu}|$ is the tunneling matrix element between the initial state $\mu$ and final state $\nu$.

\begin{figure}[htbp]\centering
\includegraphics[width=\textwidth]{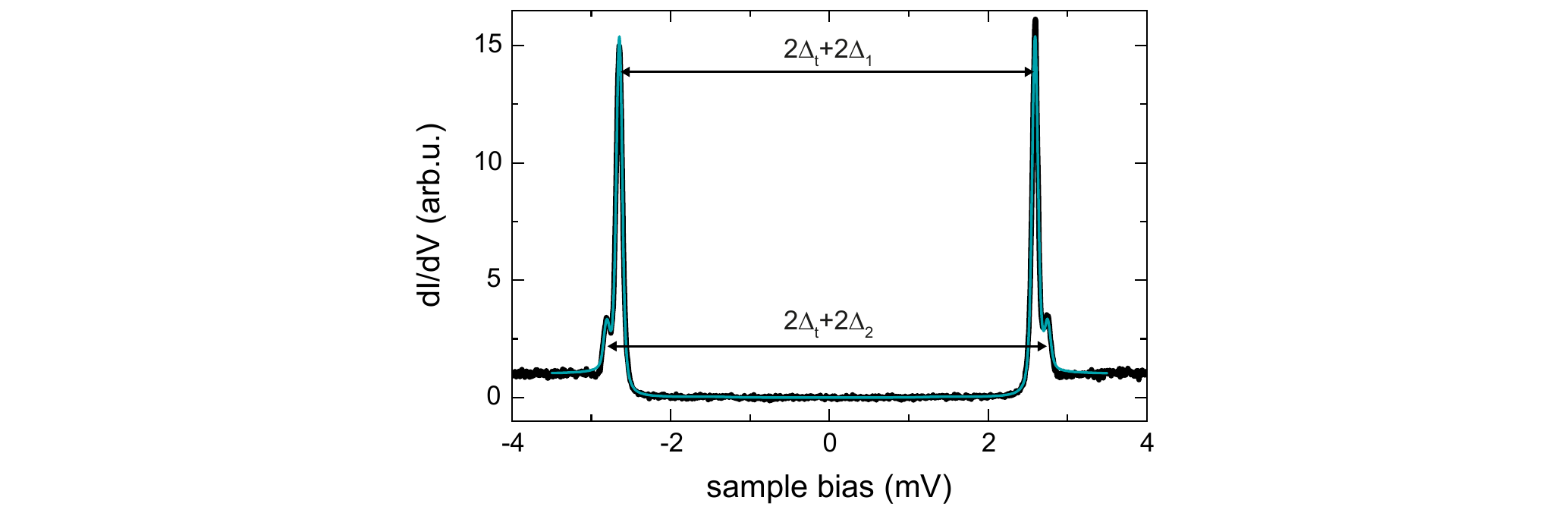}
\caption{
Constant-height spectrum taken with a Pb tip on a bare Pb(111) crystal (set point:\ \spat ; $T=1.1$\,K). The fit (shown in green) is performed as described in Ref.\ \cite{SRubyPb15}$ $.
}
\label{Fig:S5}
\end{figure}

This leads to the differential conductance:
\begin{align}
\frac{\partial I}{\partial V}\propto\int_{-\infty}^{+\infty} \text{d}E\,\, \frac{\partial \rho_{\text{t}}(E-eV)}{\partial V} \rho_{\text{s}}(E,T)[f(E-eV,T)-f(E,T)]  \nonumber \\  
+\int_{-\infty}^{+\infty} \text{d}E\,\,  \rho_{\text{t}}(E-eV) \rho_{\text{s}}(E,T) \frac{\partial f(E-eV,T)}{\partial V} \,\,\,.
\label{Eq:didv}
\end{align}
Before investigating the \nbse\ sample, the tip was characterized on a clean Pb(111) sample. A spectrum taken on bare Pb(111) after preparation of the Pb tip is shown in Fig.\ \ref{Fig:S5}. The two superconducting energy gaps of the substrate are well resolved \cite{SRubyPb15}. The Pb tip was characterized by the fit procedure described in detail in Ref.\ \cite{SRubyPb15} where a BCS-like density of states is assumed for the tip:
\begin{align}
\rho_{\text{t}}(E)= \text{sgn}(E)\,\Re\left(\frac{E-i\Gamma}{\sqrt{(E-i\Gamma)^2-\Delta_{\text{t}}^2}}\right) .
\label{Eq:rho_tip}
\end{align}
From this fitting procedure we determine the  values for the superconducting energy gap $\Delta_{\text{t}}$ and the depairing factor $\Gamma$ of the tip. The size of the energy gap $\Delta_{\text{t}}\approx 1.35\umeV$ does not vary for bulk-like tips, whereas the depairing factor was found to vary between $\Gamma\approx 10$-$20\umueV$ for different tips.

\subsection{Numerical deconvolution of the \didv spectra}

\begin{figure}[htbp]\centering
\includegraphics[width=\textwidth]{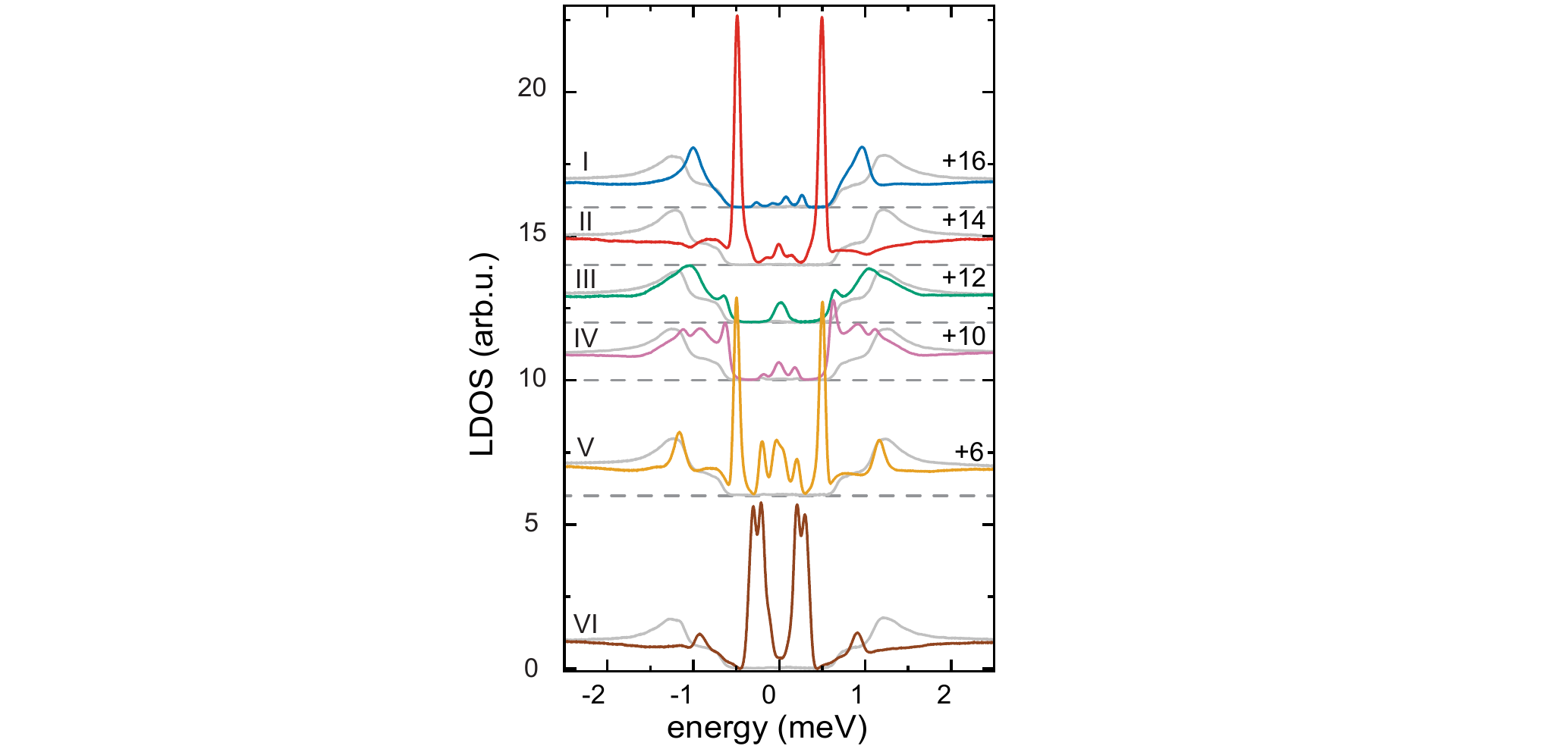}
\caption{
Deconvolved data of atom I-VI (substrate data is shown in grey). Offsets are indicated by the numbers on the right of each trace.
}
\label{Fig:S6}
\end{figure}

In order find the exact position of the YSR energies on the \nbse\ sample, all data was numerically deconvolved as described in the following (similar to the procedure described in Ref.\ \cite{SChoi2017}). Equation\ (\ref{Eq:didv}) can be discretized into matrix form and then reads:
\begin{align}
\overrightarrow{\frac{\partial I}{\partial V}}(V) \propto \mathbf{K}(E,V,T) \,\overrightarrow{\rho_{\text{s}}}(E)\, .
\label{Eq:matrix}
\end{align}
The vector on the left hand side contains the differential conductance data, $\overrightarrow{\rho_{\text{s}}}(E)$ is the local density of states of the sample. The matrix $\mathbf{K}$ can be determined by comparing Eq.\ (\ref{Eq:didv}) and Eq.\ (\ref{Eq:matrix}) to
\begin{align}
\mathbf{K}_{ij}(E_j, V_i, T)= \text{d}E\,\, \frac{\partial \rho_{\text{t}}(E_j-eV_i)}{\partial V} [f(E_j-eV_i,T)-f(E_j,T)] \nonumber \\
+ \text{d}E\,\,  \rho_{\text{t}}(E_j-eV_i) \frac{\partial f(E_j-eV_i,T)}{\partial V} \,\,\,.
\label{Eq:matrixcomponents}
\end{align}  
Thus, with the knowledge of the depairing factor and the energy gap of the tip, $\rho_{\text{s}}\left(E\right)$ can be calculated by finding the pseudoinverse of $\mathbf{K}$. 
In order to assure that the tip properties did not change during sample exchange from Pb to \nbse, we checked the accurateness of the tip's energy gap by deconvolution and recalculation of spectra on several Fe adatoms. If an inaccurate value of the energy gap is assumed for the tip, the energies of the thermally excited YSR resonances are found in a wrong position in the recalculated trace. Before deconvolving all spectra are normalized and smoothed using a Savitzky-Golay filter. Deconvolved spectra of atoms I-VI (\textit{cf.}\ Fig.\ 2 in the main text) are shown in Fig.\ \ref{Fig:S6}.

\newpage
\subsection{Identification of adsorption sites}
\begin{figure}[htbp]\centering
\includegraphics[width=\textwidth]{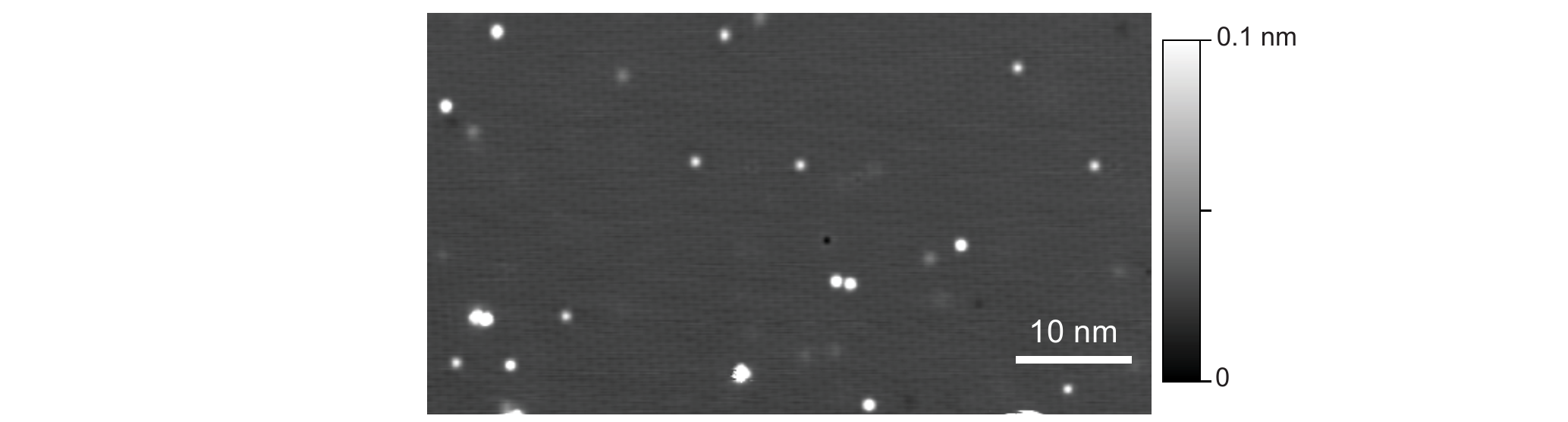}
\caption{
Overview image of the \nbse\ surface after deposition of Fe adatoms (set point:\ \sptop).
}
\label{Fig:S7}
\end{figure}

\begin{figure}[htbp]\centering
\includegraphics[width=\textwidth]{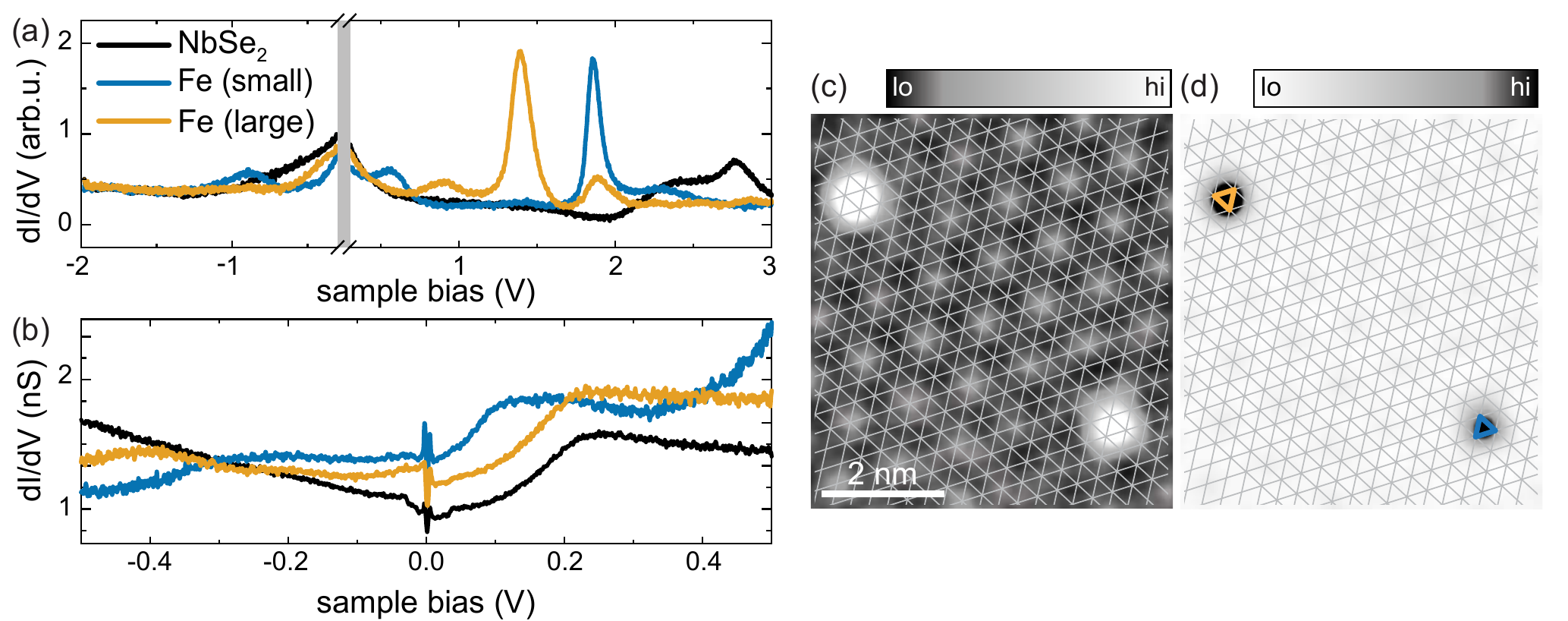}
\caption{
(a)  Constant-current ($I=100\upA$; $V_{\text{rms}}=5\umV$) spectra of \nbse\, (black) and on Fe adatoms in different hollow sites (blue, orange). (b) Constant-height spectra (set point:\ $I=500\upA, V=-500\umV$; $V_{\text{rms}}=2\umV$) of \nbse\, (black) and on Fe adatoms in different hollow sites (blue, orange). (c) Atomic-resolution STM image showing one adatom of each species (set point:\ \spat). The atomic lattice of the top Se layer is overlaid in grey. (d) The same topography and atomic lattice as shown in (c) with an inverted color code such that only the center of the atoms is visible. The orange and blue triangles point out the two different adsorption sites. 
}
\label{Fig:S8}
\end{figure}

A dilute coverage of Fe atoms ($\approx 50\, \mathrm{atoms}/(100\times 100\unm^2)$) was deposited by evaporation on the \nbse\ sample in the STM at temperatures below $12$\,K and a base pressure of $1\cdot10^{-10}\umbar$. Two distinct apparent heights of the Fe atoms are observed (Fig.\ \ref{Fig:S7} and Fig.\ 1c,d in the main text). Furthermore, the two adatom types differ in the $d$-level resonances (see blue and orange data in Fig.\ \ref{Fig:S8}a). Here, only minor differences arise within different atoms of one species.

To identify the precise adsorption site of the two species, we recorded atomic-resolution images as shown in Fig.\ \ref{Fig:S8}c,d. With the atomic lattice superposed on the STM images we find that all Fe atoms reside in hollow sites of the terminating Se layer. The Fe atoms with large apparent height solely occupy hollow sites, where the neighboring Se atoms form a triangle pointing in one direction, while the atoms with low apparent height are enclosed by a Se triangle rotated by 180$^\circ$.
Inspection of the atomic structure of \nbse\ shows that there are two distinct hollow sites which differ by the presence or absence of a Nb atom beneath (metal-centered, MC, or hollow-centered, HC). 

\begin{figure}[htbp]\centering
\includegraphics[width=\textwidth]{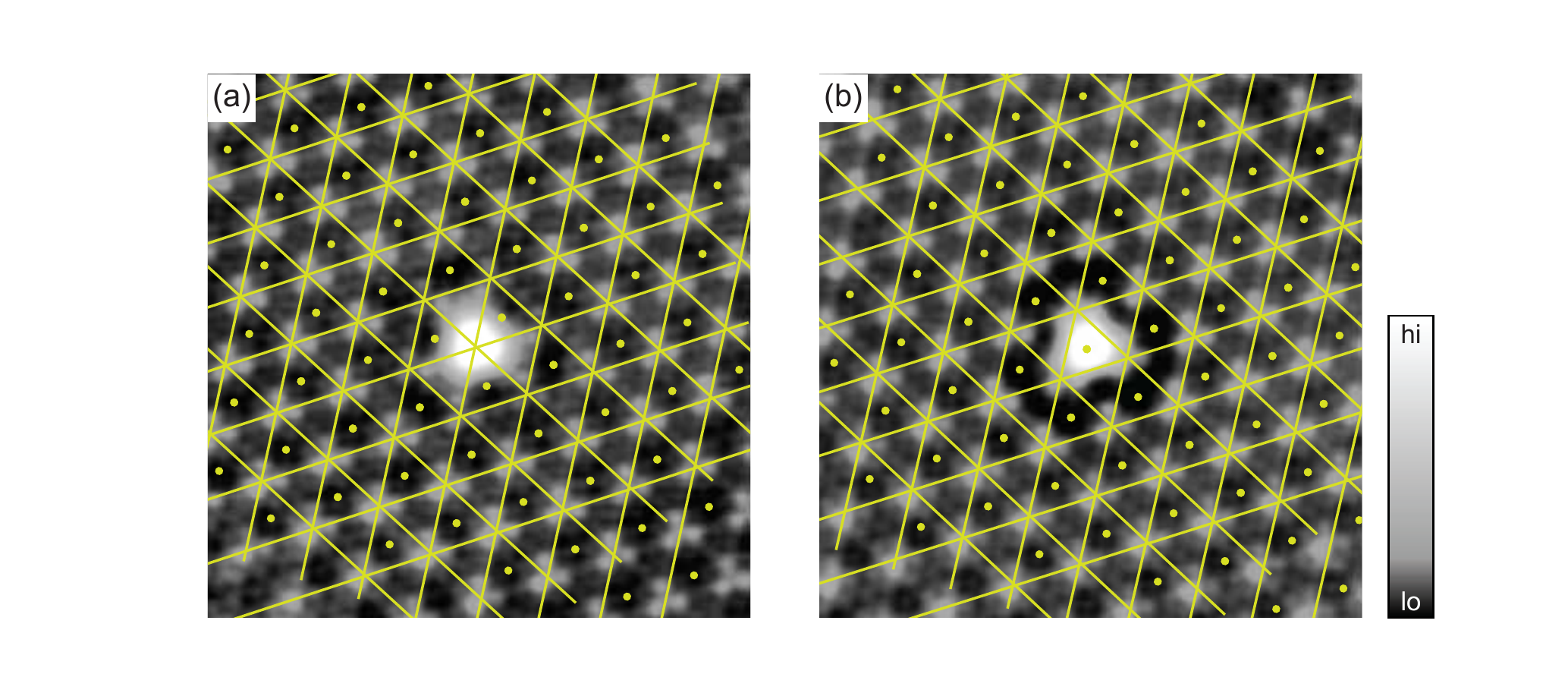}
\caption{
Atomic-resolution STM images of two adatoms with small apparent height, located on the maximum (a) and minimum (b) of the CDW (set point:\ \spat, $8\times 8\ \mathrm{nm}^2$). The intersections of the yellow grid correspond to maxima, the dots to the minima of the CDW.
}
\label{Fig:S9}
\end{figure}

As discussed in the main text, the CDW is incommensurate with the lattice ($a_{\text{cdw}}  \simeq 3.05 a$ \cite{SSoumyanarayanan2013}) which implies that the CDW smoothly shifts along the atomic lattice of the surface (see Fig.\ 1a of the main text and Fig.\ \ref{Fig:S11}). Hence, the maxima of the CDW can be found at different sites with respect to the atomic lattice. Some striking motifs occur when the CDW maximum coincides with a hollow site (HC) or with a chalcogen site (CC) as illustrated in Fig.\ 1a of the main manuscript. Interestingly, the configuration with a CDW maximum on a Nb atom (MC) is not observed \cite{SGye2019, SGuster2019}. This local configuration is energetically unfavored \cite{SGye2019, SZheng2018, SGuster2019, SLian2018, SCossu2018}. To avoid the MC structure smooth grain boundaries are built into the CDW modulation. 
As illustrated in Fig.\ \ref{Fig:S9}a,b Fe atoms with small apparent height can be found exactly in the minimum and maximum of the CDW. Hence, this species is assigned to adsorption in the HC position. Consequently, atoms with large apparent height are adsorbed in the MC sites.

\newpage
\subsection{Determination of Fe-atom adsorption sites relative to the charge-density wave}

\begin{figure}[htbp]\centering
\includegraphics[width=\textwidth]{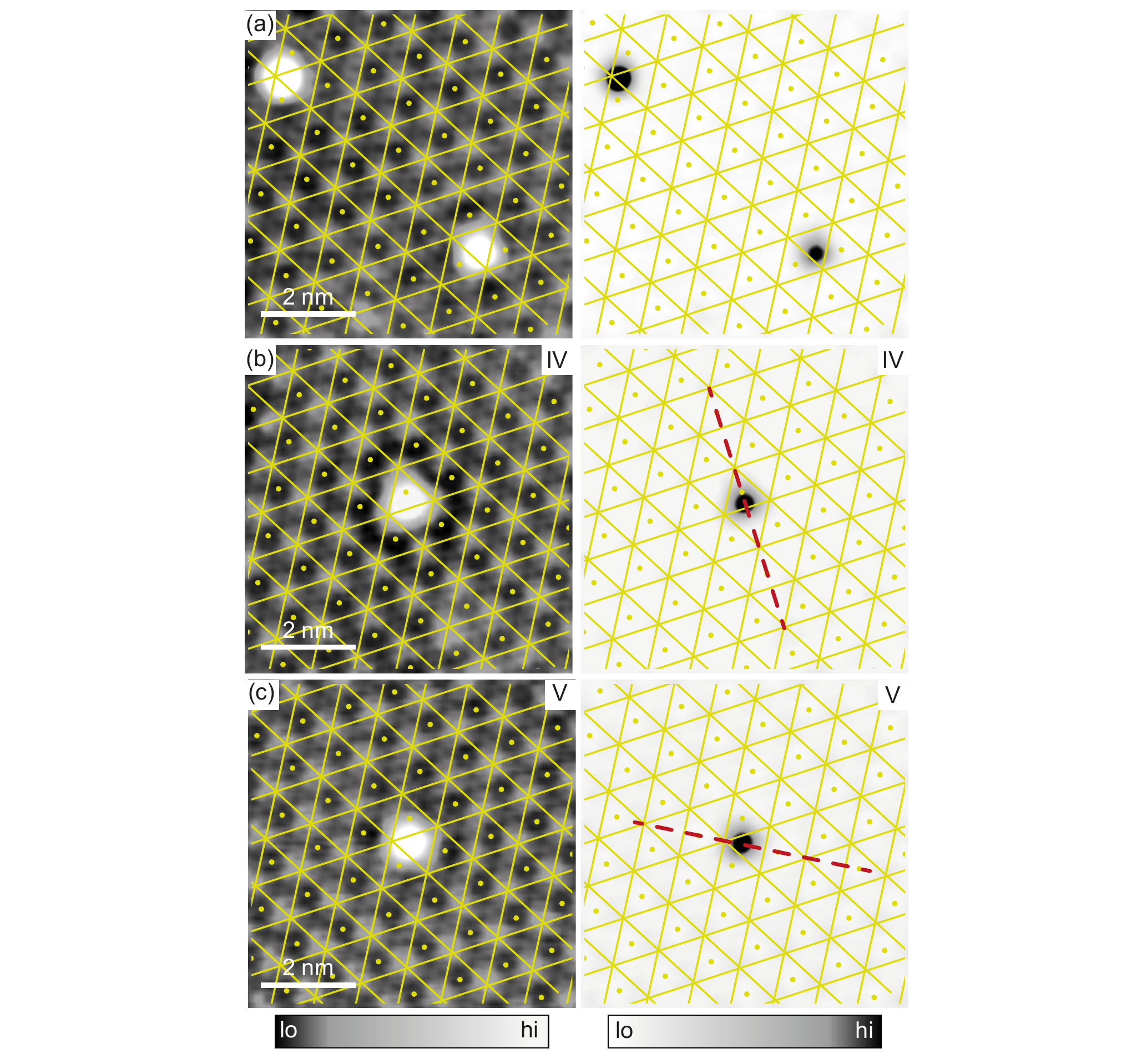}
\caption{
(a-c) The left panels show atomic-resolution topographies overlaid with the CDW lattice (set point:\ \spat). The intersections of the yellow grid correspond to maxima of the CDW, the dots correspond to minima. In the right panel the center of the atoms can be identified (same images as in the left panel with reversed color code). (a) shows both atoms of Fig.\ 1c. (b) and (c) show atoms IV and V of the main manuscript. Red dashed lines in (b) and (c) indicate the symmetry axis (\textit{cf.}\ Fig.\ 3 in the main text).
}
\label{Fig:S10}
\end{figure}

The adsorption site of Fe atoms relative to the CDW was determined for a set of approximately 90 atoms (used for the data compilation in Fig.\ 4a of the main manuscript). The center of the Fe atoms as well as the location of the CDW can be determined from the STM images (see superposed CDW grid on the STM topographies in Fig.\ \ref{Fig:S10}). The error in placing the CDW grids is estimated as $\delta x=\pm\,0.05\, \text{a}_{\text{cdw}}$, which accounts for the inaccuracy in placing the CDW grid, possible drift effects and for the inaccuracy in finding the center of the atom.

\newpage
\subsection{Pinning of the charge-density wave}

\begin{figure}[htbp]\centering
\includegraphics[width=\textwidth]{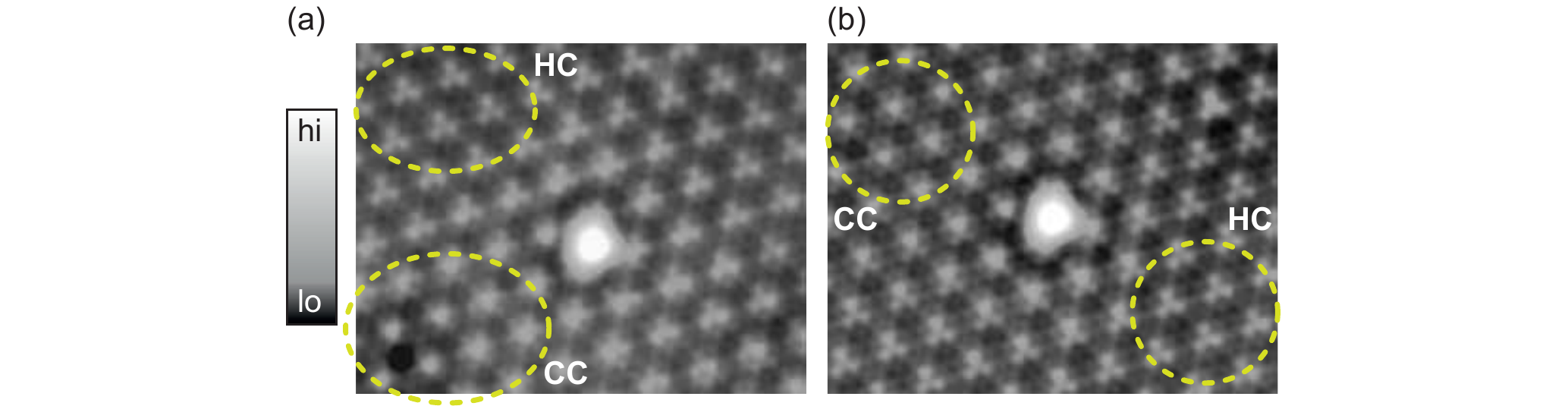}
\caption{
(a,b) Constant-current STM images of two HC adatoms (set point:\ \spat; $9\times 7\,\text{nm}^2$). A stretched color code is used to resolve the atomic corrugation. Dashed circles indicate pure HC and CC domains. In the bottom left of (a) there is a defect pinning the CC structure.
}
\label{Fig:S11}
\end{figure}

It has been shown that the CDW can be pinned by crystal defects \cite{SCossu2018, SLanger2014, SFlicker2015, SChatterjee2015, SArguello2014}. Therefore, the phase of the CDW is dictated by the complex interplay of lattice defects and long-range order of the CDW. One such example of a local modification of the CDW in the presence of a defect can be seen in Fig.\ \ref{Fig:S11}a in the bottom left, where the CC structure of the CDW is pinned around a defect (visible as depression).
Hence, one may argue that the Fe atoms can affect the CDW. 
However, a pinning of the CDW by the adatoms is not observed in the experiment. The adatoms seem not to favor a specific CDW domain as can be seen in Fig.\ \ref{Fig:S11}a and b. Here, pure HC and CC domains are marked. The adatoms are located in the smooth transition region.

\subsection{Spatial variation of \didv spectra}

\begin{figure}[htbp]\centering
\includegraphics[width=0.95\textwidth]{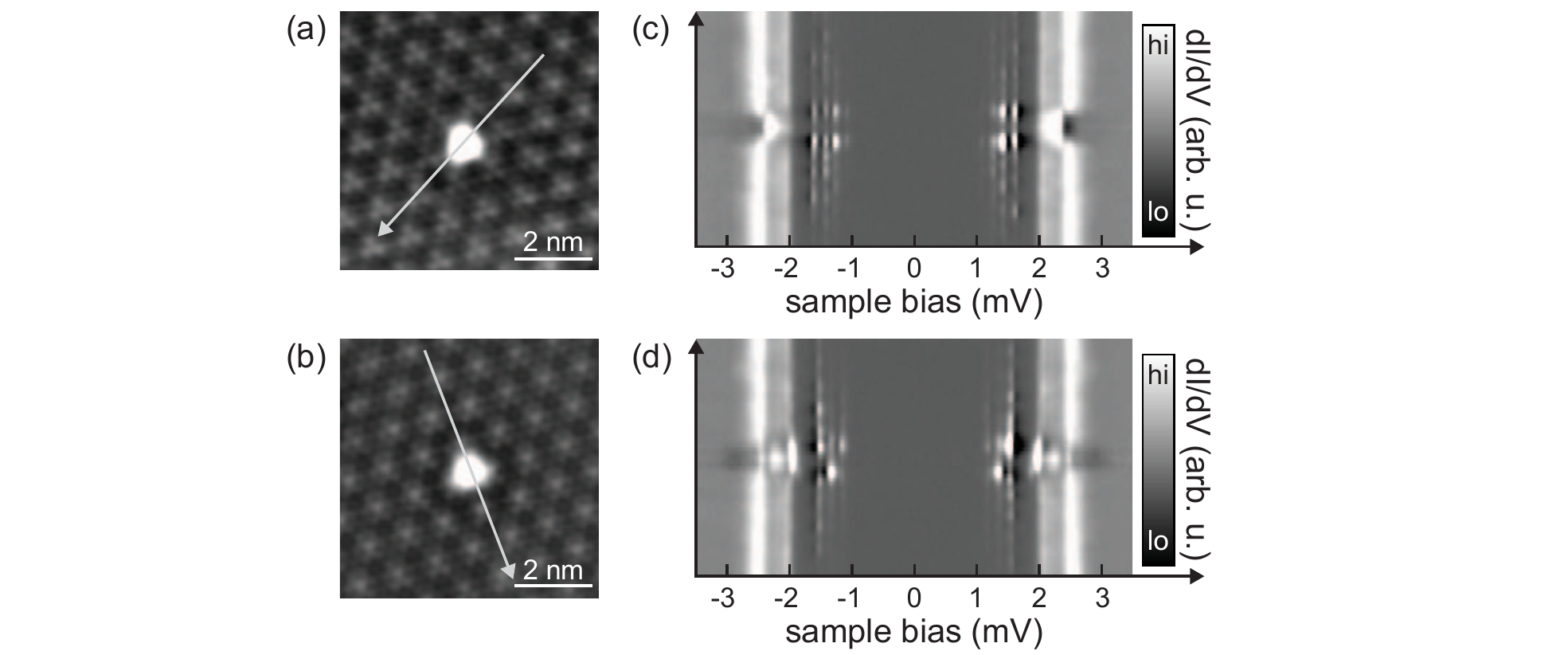}
\caption{
(a,b) Topography of atoms I and IV of the main text (set point:\ \spat). (c,d) Stacked \didv spectra (set point:\ \spat; $V_{\text{rms}}=15\umueV$) taken along the arrows indicated in (a,b).
}
\label{Fig:S13}
\end{figure}

Figure\ \ref{Fig:S13} shows false-color plots of the \didv spectra taken along high-symmetry directions of the YSR patterns across atoms I and IV of the main text. The YSR states exhibit long-range intensity oscillations with no detectable variations in energy within our experimental resolution. 

\begin{figure}[htbp]\centering
\includegraphics[width=0.95\textwidth]{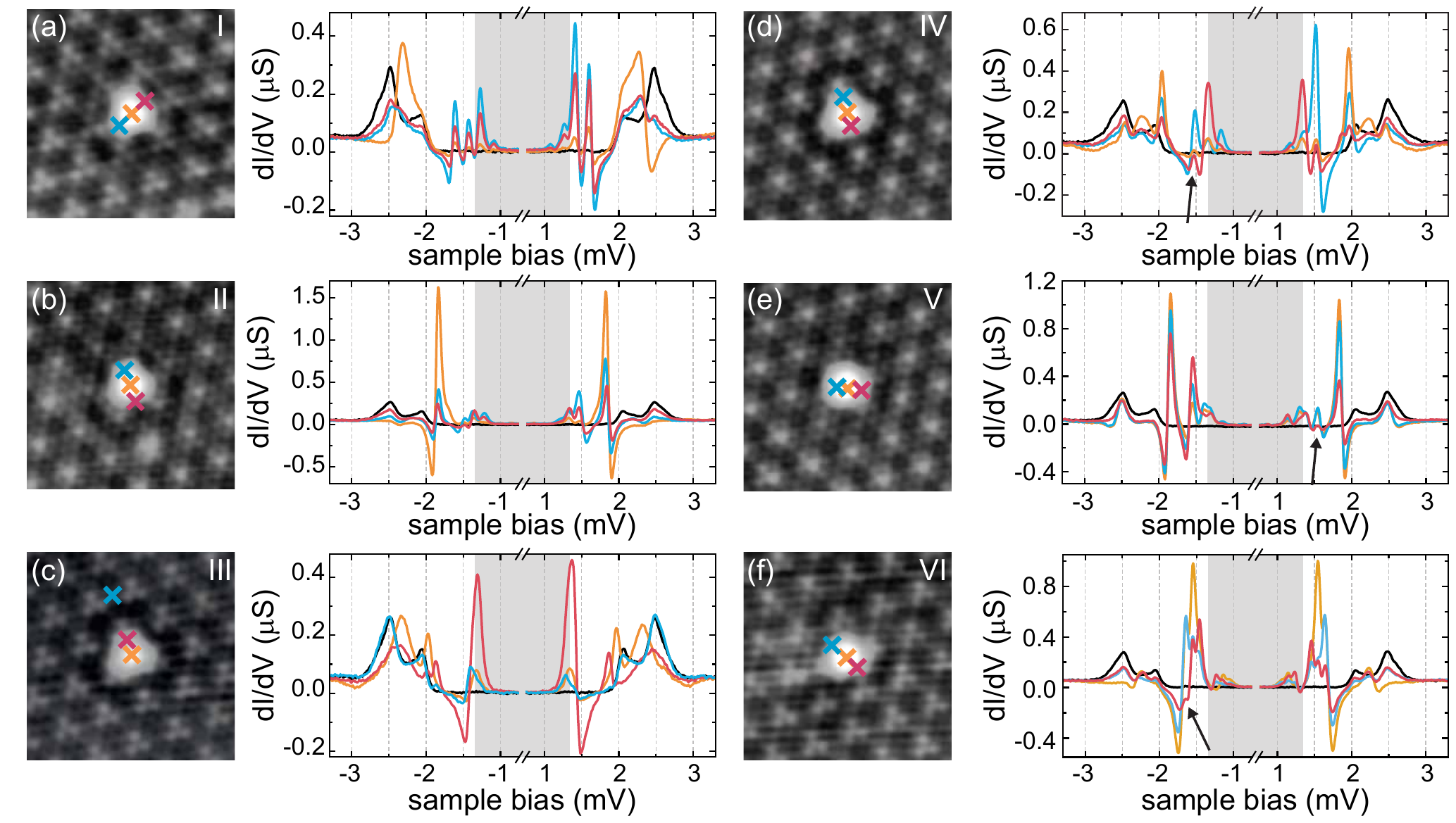}
\caption{
(a-f) Topography (left, $5\times 5\,\text{nm}^2$) and \didv spectra (right) taken on and in close vicinity of atoms I-VI of the main text. Colored crosses indicate the location of the spectra (set point:\ \spat; $V_{\text{rms}}=15\umueV$). Corresponding substrate data are shown in black for each atom. Black arrows highlight YSR states that are strongly affected by the NDC due to a close-by YSR resonance.
}
\label{Fig:S12}
\end{figure}

We note that in close vicinity to the atom (Figure\ \ref{Fig:S12}), the intensity variations are particularly strong and some YSR states are accompanied by a negative value of the differential conductance (NDC) at their high-bias flank. The NDC results from the convolution of YSR states with the tip density of states and can lead to a strong suppression of conductance. In an extreme case, an additional (weak) YSR resonance may be hidden in this part of the spectrum. Examples, where YSR resonances are suppressed below zero are shown by arrows for atom IV, V and VI in Fig.\ \ref{Fig:S12}. The local variation of the NDC also affects the spatial appearance of the YSR states in the \didv maps as is discussed in the following section.

%

%\newpage
\subsection{Additional \didv maps of YSR resonances (both bias polarities and thermal excitations)}

\begin{figure}[htbp]\centering
\includegraphics[width=\textwidth]{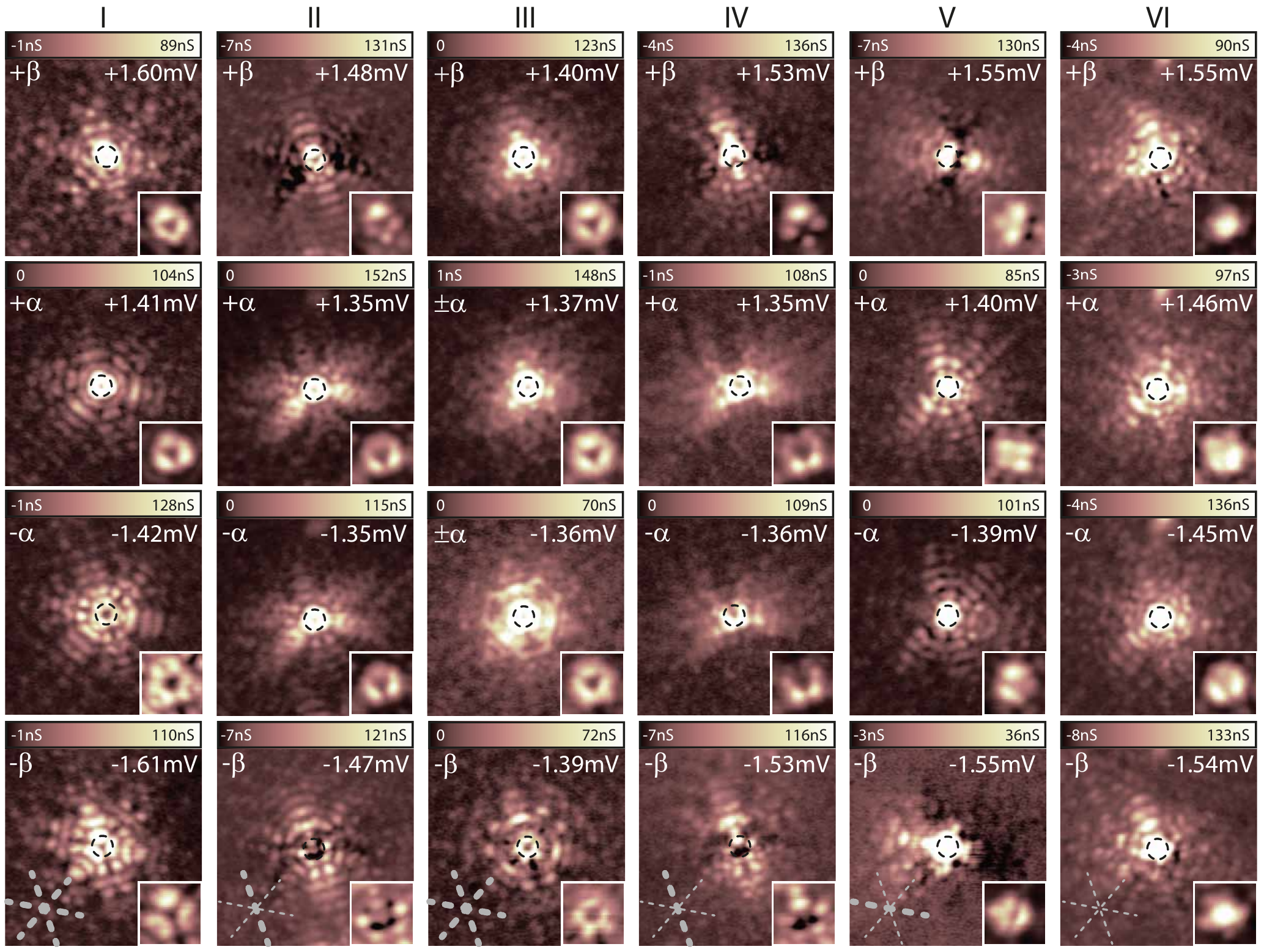}
\caption{
Constant-contour \didv maps ($9.5\times 9.5\unm^2$) of the YSR $(\pm\alpha,\pm\beta)$ states of HC atoms I-VI (set point:\ \spat; $V_{\text{rms}}=15$-$25\umueV$; $V_{\mathrm{bias}}$ as indicated in the images). The insets show a $2\times 2\unm^2$ close-up view around the center of the atoms. Black dashed circles (diameter 1 nm) outline the atoms' position. The grey dashed lines in the bottom map of each atom indicate the crystal's symmetry axes. Thick lines indicate mirror axes present in the \didv map.
}
\label{Fig:S14}
\end{figure}

Figure\ \ref{Fig:S14} shows the complete set of the \didv maps of atoms I-VI (the positive bias voltage was presented in the main manuscript). \didv maps at negative bias polarity are depicted in the lower two rows. Opposite bias polarities image the $\left|u\right|^2$ and $\left|v\right|^2$ components of the YSR states, respectively \cite{SRusinov1969}. 

As expected, the $\left|u\right|^2$ and $\left|v\right|^2$ components lead to distinct scattering patterns. As discussed above, the convolution of tip and sample density of states may lead to distortions of the intensity of individual YSR states, culminating in their suppression in the NDC region of a nearby state. Consequently, also the \didv maps at the energy of a specific YSR state may be affected by a close-lying YSR state. 

\begin{figure}[htbp]\centering
\includegraphics[width=\textwidth]{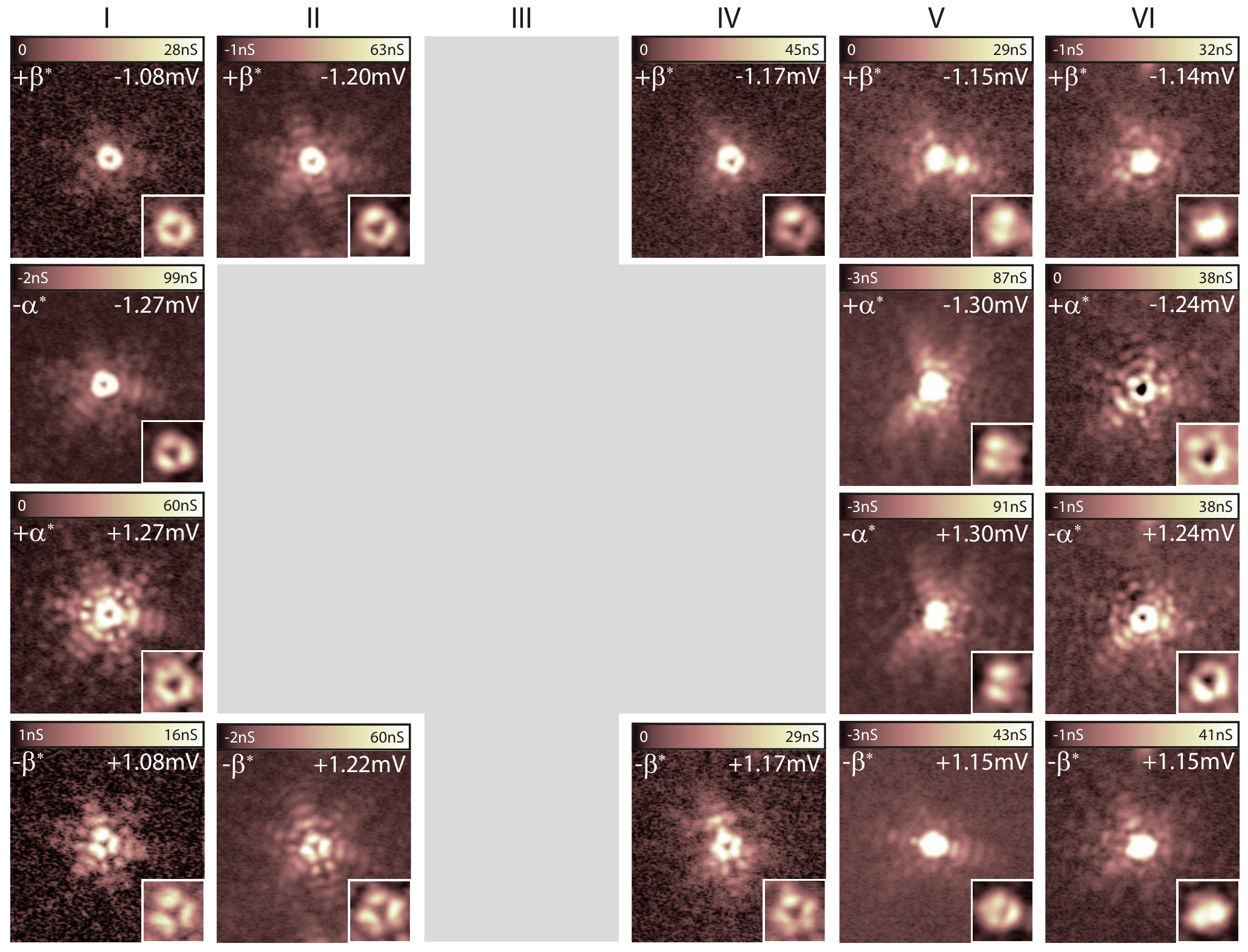}
\caption{
Constant-contour \didv maps ($9.5\times 9.5 \unm^2$) of the thermally excited YSR $(\pm\alpha,\pm\beta)$ states (set point:\ \spat; $V_{\text{rms}}=15$-$25\umueV$; $V_{\mathrm{bias}}$ as indicated in the images). The insets show a $2\times 2\unm^2$ close-up view around the center of the atoms. The images are arranged in the same way as in Fig.\ \ref{Fig:S14}, \textit{i.e.}\ the maps of the thermally excited YSR states can be found in the same position within the array as in Fig.\ \ref{Fig:S14}. For YSR resonances at zero energy (\textit{i.e.}\ $eV_{+\alpha}=eV_{-\alpha}=\Delta_{\text{t}}$) there are no maps of the corresponding thermal excitations (grey area).
}
\label{Fig:S15}
\end{figure}

\begin{figure}[htbp]\centering
\includegraphics[width=\textwidth]{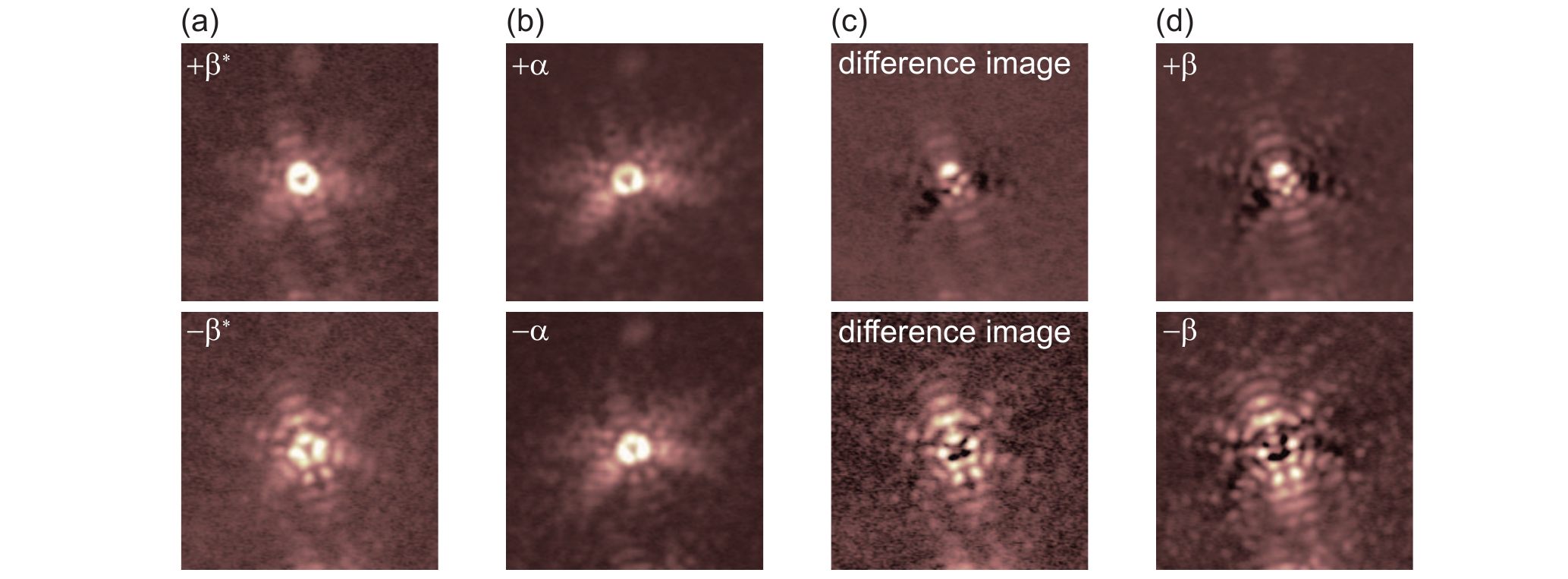}
\caption{Constant-contour \didv maps of
(a) $+\beta^\star$ (top) and $-\beta^\star$ (bottom) of atom II (same data can be found in Fig.\ \ref{Fig:S15}), and (b) $+\alpha$ (top) and $-\alpha$ (bottom) of atom II (same data can be found in Fig.\ \ref{Fig:S14}). (c) Difference images: $+\alpha$ ($-\alpha$) map subtracted with a factor of $0.7$ ($1.0$) from the map $+\beta^\star$ ($-\beta^\star$). (d) \didv map of the $+\beta$ (top) and $-\beta$ (bottom) resonance (same data can be found in Fig.\ \ref{Fig:S14}). (c) and (d) are very similar.
}
\label{Fig:S16}
\end{figure}

In particular, the $\pm\beta$-states of atom II, IV and V are affected by the NDC of the $\alpha$-resonances, which are close in energy (Fig.\ \ref{Fig:S14}). One trick to determine and eliminate the effect of NDC is to investigate the \didv maps of thermally excited YSR states. 
Thermal excitation of quasiparticles leads to additional resonances within the energy gap of the tip (\textit{cf.}\ grey shaded area in Fig.\ 2b,c in the main text). Whereas the original resonances are found at a bias of $eV_{\pm\alpha, \pm\beta}=\pm|\Delta_{\text{t}}+E_{\alpha, \beta}|$, the thermally excited states are found at $eV_{\pm\alpha^\star,\pm\beta^\star}=\mp|\Delta_{\text{t}}-E_{\alpha, \beta}|$ \cite{SRuby2015}. 
The thermally excited YSR states are of much less intensity and, therefore, exhibit a small or negligible region of NDC.
Figure\ \ref{Fig:S15} shows the \didv maps recorded at the corresponding voltages of the thermally excited YSR states for atoms I-VI. 
In these maps the similarity of the $\pm\beta^\star$ states of atoms I, II and IV is more striking than in the original maps of $\pm\beta$. 
To illustrate the impact of the NDC from the $\pm\alpha$-YSR resonances on the $\pm\beta$ state, we simulate the map of $\pm\beta$ by subtracting a fraction of the $\pm\alpha$ maps (mimicking the NDC) from the $\pm\beta^\star$ maps. The result is compared to the maps at the energies corresponding to $\pm\beta$ (Fig.\ \ref{Fig:S16}) and shows remarkable similarity, illustrating the impact of NDC. 

\newpage

\subsection{Determination of YSR energies}

To determine the YSR energies, we fit the deconvolved spectra (\textit{cf.} Fig.\ \ref{Fig:S6}) in the low-energy region with a set of 4 Gaussians (2 pairs of peaks):
\begin{align}
\text{LDOS}(E) = D_0+A_1 e^{-(E-E_\alpha)^2/(2\sigma^2)}+A_2 e^{-(E+E_\alpha)^2/(2 \sigma^2)} \nonumber \\
+B_1 e^{-(E-E_\beta)^2/(2\sigma^2)}+B_2 e^{-(E+E_\beta)^2/(2\sigma^2)}.
\label{Eq:current2}
\end{align}
Here, $D_0$ is an offset, $E_\alpha$ and $E_\beta$ are the YSR energies, $A_{1,2}$ and $B_{1,2}$ are four amplitudes and $\sigma$ is the width of the Gaussian peaks.

\begin{figure}[htbp]\centering
\includegraphics[width=\textwidth]{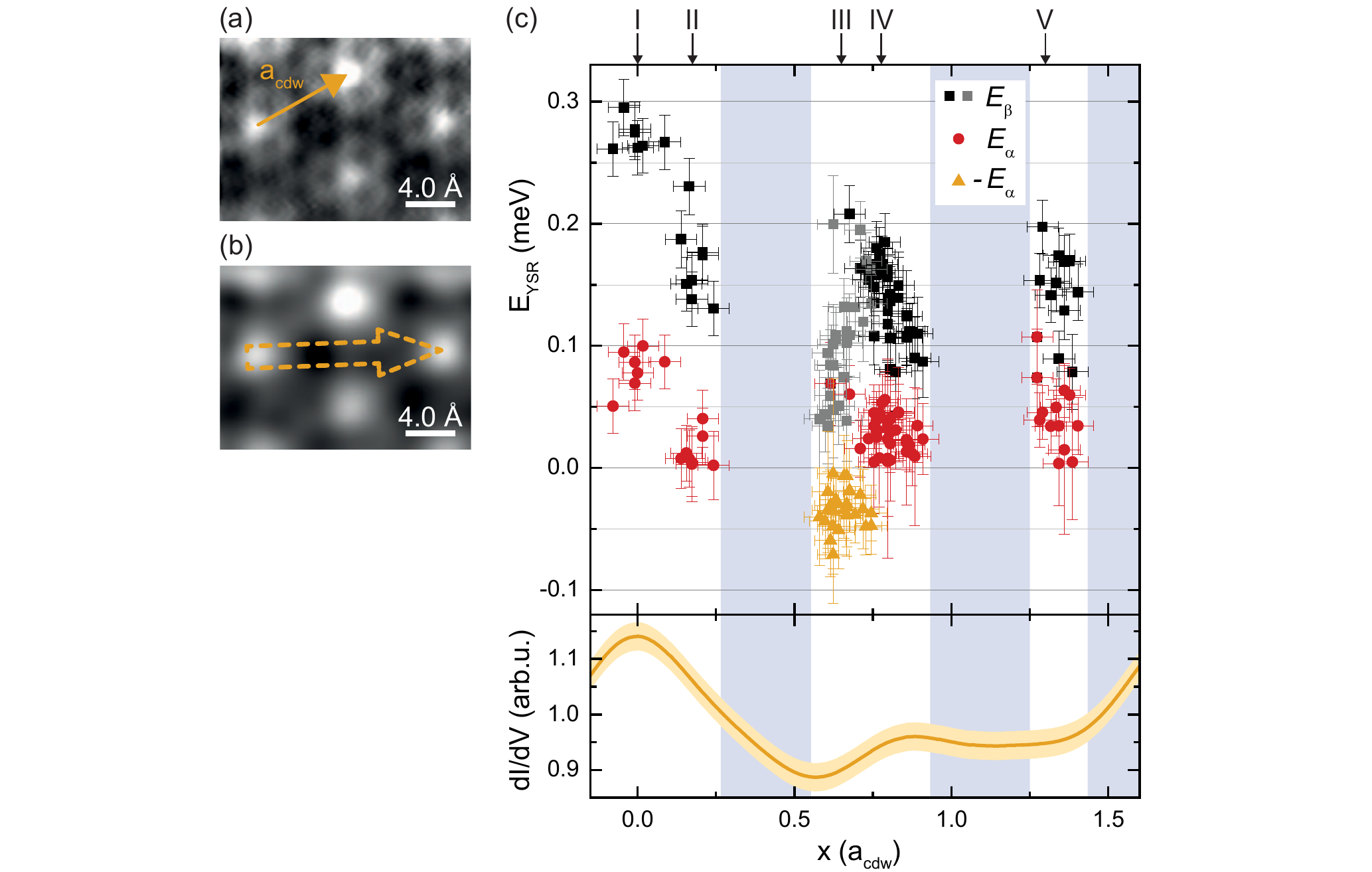}
\caption{
(a) Constant-height \didv map taken at $T=8\,$K and $V_{\mathrm{bias}}=0$ (set point:\ \spat; $V_{\mathrm{rms}}=100\umueV$; some z-drift is visible). (b) FFT-filtered data of (a) (already presented in Fig.\ 4 in the main text). The signal averaged over the radius of influence of $125\upm$, which is approx. the covalent radius of a Fe atom (dashed orange arrow) is plotted in the bottom panel of (c) for one complete period of the CDW. (c) Evaluation of YSR energies $E_\alpha,E_\beta$ of approximately 90 atoms measured on several samples with several Pb tips by fitting the deconvolved spectra as described in the text. Error bars are also discussed in the text. Black and grey show the evolution of $E_\beta$. Red and orange data points correspond to $E_\alpha$. For the orange data points a negative $E_\alpha$ was assumed. Corridors with no datapoints are marked in gray.
}
\label{Fig:S17}
\end{figure}

The top panel of Fig.\ \ref{Fig:S17}c shows the extracted energies of the YSR states of approximately 90 atoms with respect to their location relative to the CDW. The error bar in x-direction results from the inaccuracy in finding the exact position relative to the CDW as discussed along with Fig. \ref{Fig:S10}.
The error margins of the YSR energy include the uncertainty in the exact value of the tip's energy gap $\Delta_{\text{t}}$, the lock-in modulation of the bias voltage and the standard deviation of the fit routine. 
Additionally, an error of $\pm\sigma/2$ was added to the energy uncertainty if the peaks are not well separated (\textit{i.e.}\ $E_{\alpha}-E_{\beta}\leq 2\sigma$). 
This is mainly the case for atoms close to the CDW minimum, where we cannot distinguish the $\alpha$- and $\beta$-resonances as they are both very close to zero energy (\textit{e.g.}\ atom III in the main text). Some of these atoms can be fitted with only one pair of Gauss peaks as the YSR states overlap. In this case both energies have the same value ($E_\alpha=E_\beta$).

As discussed in the main manuscript, we find a correlation of the YSR energy with position along the CDW symmetry axis. $E_{\alpha}$ and $E_{\beta}$ are largest at the maximum of the CDW and decrease with distance from the maximum. 
By default, the fit outputs for $E_{\alpha}$ and $E_{\beta}$ are positive. Assuming that $E_{\alpha}$ and $E_{\beta}$ should follow a similar trend, we suggest that the $\alpha$-resonance crosses zero energy, \textit{i.e.}\ undergoes the quantum phase transition \cite{SFranke2011}. 
Therefore, $E_\alpha<0$ was assumed for the atoms in close vicinity to the CDW minimum (\textit{cf.}\ atom III, the set of atoms with $E_\alpha<0$ is marked in orange and grey). 
Having passed the minimum (\textit{cf.}\ atom IV) the $\alpha$- and $\beta$-states can be separated and identified again in the \didv maps (see Fig.\ 3 in the main manuscript and Figs.\ \ref{Fig:S14}, \ref{Fig:S15}, \ref{Fig:S16}), \textit{i.e.}\ $E_{\alpha}>0$.

The top panel of Fig.\ 4a in the main manuscript is obtained from the data in Fig.\ \ref{Fig:S17}c by averaging over all atoms that are found within an interval of $\pm 0.05$\ a$_{\text{cdw}}$. This corresponds to the uncertainty in the determination of the atoms' position with respect to the CDW (see above).
The error margins in energy are the standard deviations of the averaging including the average of the errors shown in  Fig.\ \ref{Fig:S17}c.

\begin{figure}[htbp]\centering
\includegraphics[width=0.9\textwidth]{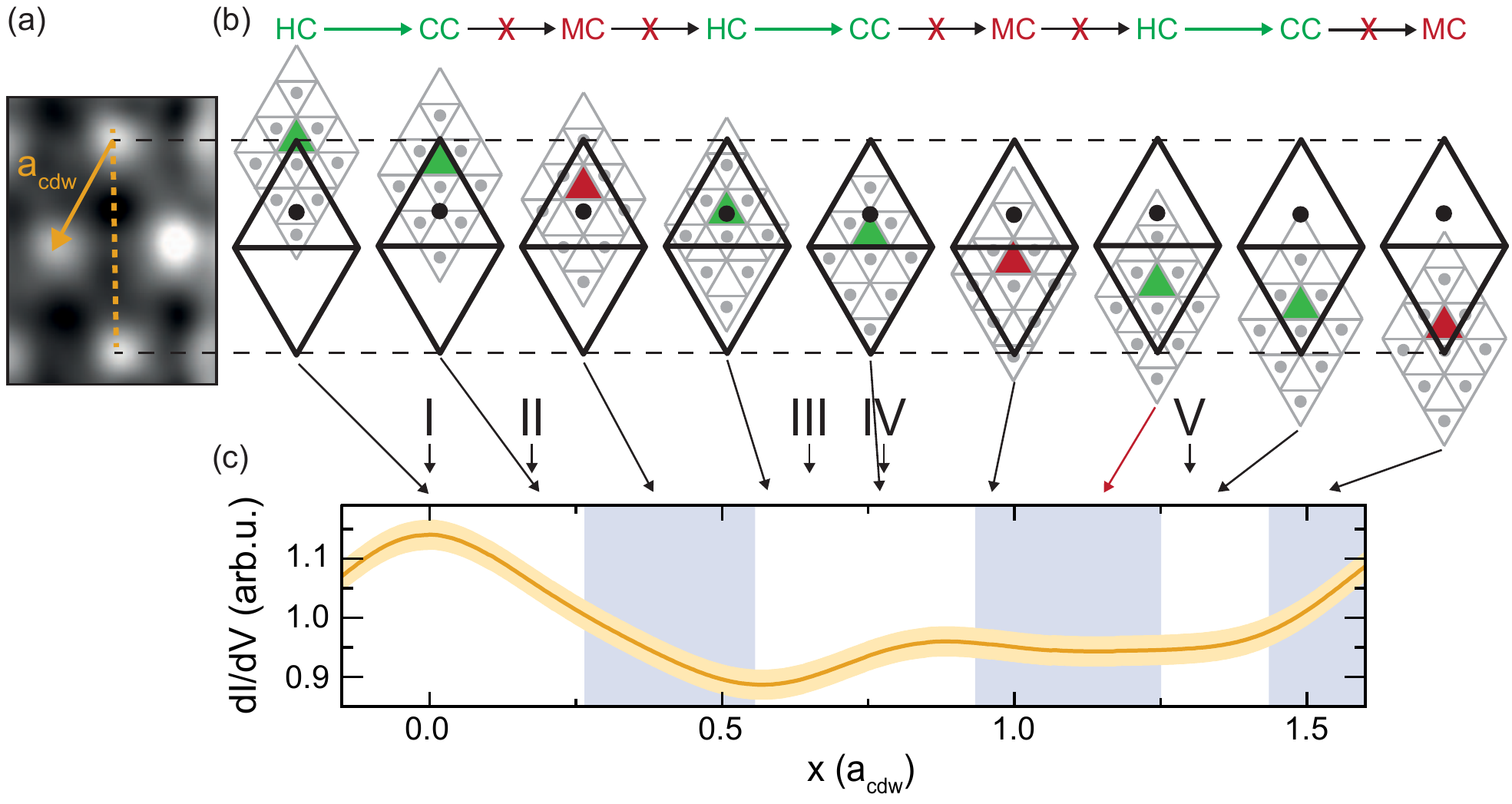}
\caption{
(a) FFT-filtered topography already presented in Fig.\ \ref{Fig:S17}b and inset of Fig.\ 4 in the main text. (b) Atomic lattice in grey and CDW lattice in black. Crossing of the atomic lattice represent Se atoms, Nb atoms of the second layer are indicated by the grey dots. Crossings of the CDW lattice correspond to CDW maxima, the black dot indicates the CDW minimum. One HC hollow site is marked in color. Green/red indicates if the relative orientation both lattices relative to each other can be found in experiment. (c) Connection to linecut along the symmetry axis of the CDW (see bottom panel of Fig.\ \ref{Fig:S17}b and Fig.\ 4b in the main text).
}
\label{Fig:S18}
\end{figure}

As seen in Fig.\ \ref{Fig:S17} and Fig.\ 4 of the main text, some regions along the CDW cannot be probed by Fe adatoms (grey areas). To understand the absence of data in these regions, we illustrate the phase change of the CDW (black rhombi) with respect to the atomic lattice (grey lattice) along the main symmetry axis in detail in Fig.\ \ref{Fig:S18}. A HC adsorption site of an Fe atom is marked by a colored triangle on the atomic lattice. The green triangles indicate observed lattice sites, whereas the red triangles indicate that we did not find these adsorption sites relative to the CDW. The latter cases correspond to configurations, where the maximum of the CDW is located on top of a Nb atom. These so-called MC configurations are energetically disfavored as shown by DFT calculations and do not occur on the surface \cite{SGye2019, SZheng2018, SGuster2019, SLian2018, SCossu2018}. Hence, the corresponding Fe adsorption sites cannot be found either. This explains the absence of data points in the grey areas with only one exception. The third high symmetry position, where an atomic HC adsorption site would sit exactly in the local minimum of the CDW (marked with the red arrow in Fig.\ \ref{Fig:S18}b), was also not found in experiment. We may speculate that the energy landscape of the CDW can be affected by the Fe adatoms for this specific adsorption site.

\bibliographystyle{apsrev4-1}
%\bibliography{paper}
%\bibliography{library}

\begin{thebibliography}{59}%
\makeatletter
\providecommand \@ifxundefined [1]{%
 \@ifx{#1\undefined}
}%
\providecommand \@ifnum [1]{%
 \ifnum #1\expandafter \@firstoftwo
 \else \expandafter \@secondoftwo
 \fi
}%
\providecommand \@ifx [1]{%
 \ifx #1\expandafter \@firstoftwo
 \else \expandafter \@secondoftwo
 \fi
}%
\providecommand \natexlab [1]{#1}%
\providecommand \enquote  [1]{``#1''}%
\providecommand \bibnamefont  [1]{#1}%
\providecommand \bibfnamefont [1]{#1}%
\providecommand \citenamefont [1]{#1}%
\providecommand \href@noop [0]{\@secondoftwo}%
\providecommand \href [0]{\begingroup \@sanitize@url \@href}%
\providecommand \@href[1]{\@@startlink{#1}\@@href}%
\providecommand \@@href[1]{\endgroup#1\@@endlink}%
\providecommand \@sanitize@url [0]{\catcode `\\12\catcode `\$12\catcode
  `\&12\catcode `\#12\catcode `\^12\catcode `\_12\catcode `\%12\relax}%
\providecommand \@@startlink[1]{}%
\providecommand \@@endlink[0]{}%
\providecommand \url  [0]{\begingroup\@sanitize@url \@url }%
\providecommand \@url [1]{\endgroup\@href {#1}{\urlprefix }}%
\providecommand \urlprefix  [0]{URL }%
\providecommand \Eprint [0]{\href }%
\providecommand \doibase [0]{http://dx.doi.org/}%
\providecommand \selectlanguage [0]{\@gobble}%
\providecommand \bibinfo  [0]{\@secondoftwo}%
\providecommand \bibfield  [0]{\@secondoftwo}%
\providecommand \translation [1]{[#1]}%
\providecommand \BibitemOpen [0]{}%
\providecommand \bibitemStop [0]{}%
\providecommand \bibitemNoStop [0]{.\EOS\space}%
\providecommand \EOS [0]{\spacefactor3000\relax}%
\providecommand \BibitemShut  [1]{\csname bibitem#1\endcsname}%
\let\auto@bib@innerbib\@empty
%</preamble>
\bibitem [{\citenamefont {Hess}\ \emph {et~al.}(1990)\citenamefont {Hess},
  \citenamefont {Robinson},\ and\ \citenamefont {Waszczak}}]{Hess1990}%
  \BibitemOpen
  \bibfield  {author} {\bibinfo {author} {\bibfnamefont {H.~F.}\ \bibnamefont
  {Hess}}, \bibinfo {author} {\bibfnamefont {R.~B.}\ \bibnamefont {Robinson}},
  \ and\ \bibinfo {author} {\bibfnamefont {J.~V.}\ \bibnamefont {Waszczak}},\
  }\href {\doibase 10.1103/PhysRevLett.64.2711} {\bibfield  {journal} {\bibinfo
   {journal} {Phys. Rev. Lett.}\ }\textbf {\bibinfo {volume} {64}},\ \bibinfo
  {pages} {2711} (\bibinfo {year} {1990})}\BibitemShut {NoStop}%
\bibitem [{\citenamefont {Yokoya}\ \emph {et~al.}(2001)\citenamefont {Yokoya},
  \citenamefont {Kiss}, \citenamefont {Chainani}, \citenamefont {Shin},
  \citenamefont {Nohara},\ and\ \citenamefont {Takagi}}]{Yokoya2001}%
  \BibitemOpen
  \bibfield  {author} {\bibinfo {author} {\bibfnamefont {T.}~\bibnamefont
  {Yokoya}}, \bibinfo {author} {\bibfnamefont {T.}~\bibnamefont {Kiss}},
  \bibinfo {author} {\bibfnamefont {A.}~\bibnamefont {Chainani}}, \bibinfo
  {author} {\bibfnamefont {S.}~\bibnamefont {Shin}}, \bibinfo {author}
  {\bibfnamefont {M.}~\bibnamefont {Nohara}}, \ and\ \bibinfo {author}
  {\bibfnamefont {H.}~\bibnamefont {Takagi}},\ }\href {\doibase
  10.1126/science.1065068} {\bibfield  {journal} {\bibinfo  {journal}
  {Science}\ }\textbf {\bibinfo {volume} {294}},\ \bibinfo {pages} {2518}
  (\bibinfo {year} {2001})}\BibitemShut {NoStop}%
\bibitem [{\citenamefont {Noat}\ \emph {et~al.}(2010)\citenamefont {Noat},
  \citenamefont {Cren}, \citenamefont {Debontridder}, \citenamefont
  {Roditchev}, \citenamefont {Sacks}, \citenamefont {Toulemonde},\ and\
  \citenamefont {San~Miguel}}]{Noat2010}%
  \BibitemOpen
  \bibfield  {author} {\bibinfo {author} {\bibfnamefont {Y.}~\bibnamefont
  {Noat}}, \bibinfo {author} {\bibfnamefont {T.}~\bibnamefont {Cren}}, \bibinfo
  {author} {\bibfnamefont {F.}~\bibnamefont {Debontridder}}, \bibinfo {author}
  {\bibfnamefont {D.}~\bibnamefont {Roditchev}}, \bibinfo {author}
  {\bibfnamefont {W.}~\bibnamefont {Sacks}}, \bibinfo {author} {\bibfnamefont
  {P.}~\bibnamefont {Toulemonde}}, \ and\ \bibinfo {author} {\bibfnamefont
  {A.}~\bibnamefont {San~Miguel}},\ }\href {\doibase
  10.1103/PhysRevB.82.014531} {\bibfield  {journal} {\bibinfo  {journal} {Phys.
  Rev. B}\ }\textbf {\bibinfo {volume} {82}},\ \bibinfo {pages} {014531}
  (\bibinfo {year} {2010})}\BibitemShut {NoStop}%
\bibitem [{\citenamefont {Noat}\ \emph {et~al.}(2015)\citenamefont {Noat},
  \citenamefont {Silva-Guill\'en}, \citenamefont {Cren}, \citenamefont
  {Cherkez}, \citenamefont {Brun}, \citenamefont {Pons}, \citenamefont
  {Debontridder}, \citenamefont {Roditchev}, \citenamefont {Sacks},
  \citenamefont {Cario}, \citenamefont {Ordej\'on}, \citenamefont
  {Garc\'{\i}a},\ and\ \citenamefont {Canadell}}]{Noat2015}%
  \BibitemOpen
  \bibfield  {author} {\bibinfo {author} {\bibfnamefont {Y.}~\bibnamefont
  {Noat}}, \bibinfo {author} {\bibfnamefont {J.~A.}\ \bibnamefont
  {Silva-Guill\'en}}, \bibinfo {author} {\bibfnamefont {T.}~\bibnamefont
  {Cren}}, \bibinfo {author} {\bibfnamefont {V.}~\bibnamefont {Cherkez}},
  \bibinfo {author} {\bibfnamefont {C.}~\bibnamefont {Brun}}, \bibinfo {author}
  {\bibfnamefont {S.}~\bibnamefont {Pons}}, \bibinfo {author} {\bibfnamefont
  {F.}~\bibnamefont {Debontridder}}, \bibinfo {author} {\bibfnamefont
  {D.}~\bibnamefont {Roditchev}}, \bibinfo {author} {\bibfnamefont
  {W.}~\bibnamefont {Sacks}}, \bibinfo {author} {\bibfnamefont
  {L.}~\bibnamefont {Cario}}, \bibinfo {author} {\bibfnamefont
  {P.}~\bibnamefont {Ordej\'on}}, \bibinfo {author} {\bibfnamefont
  {A.}~\bibnamefont {Garc\'{\i}a}}, \ and\ \bibinfo {author} {\bibfnamefont
  {E.}~\bibnamefont {Canadell}},\ }\href {\doibase 10.1103/PhysRevB.92.134510}
  {\bibfield  {journal} {\bibinfo  {journal} {Phys. Rev. B}\ }\textbf {\bibinfo
  {volume} {92}},\ \bibinfo {pages} {134510} (\bibinfo {year}
  {2015})}\BibitemShut {NoStop}%
\bibitem [{\citenamefont {Rodrigo}\ and\ \citenamefont
  {Vieira}(2004)}]{Rodrigo2004a}%
  \BibitemOpen
  \bibfield  {author} {\bibinfo {author} {\bibfnamefont {J.~G.}\ \bibnamefont
  {Rodrigo}}\ and\ \bibinfo {author} {\bibfnamefont {S.}~\bibnamefont
  {Vieira}},\ }\href {\doibase 10.1016/j.physc.2003.10.030} {\bibfield
  {journal} {\bibinfo  {journal} {Physica C}\ }\textbf {\bibinfo {volume}
  {404}},\ \bibinfo {pages} {306} (\bibinfo {year} {2004})}\BibitemShut
  {NoStop}%
\bibitem [{\citenamefont {Boaknin}\ \emph {et~al.}(2003)\citenamefont
  {Boaknin}, \citenamefont {Tanatar}, \citenamefont {Paglione}, \citenamefont
  {Hawthorn}, \citenamefont {Ronning}, \citenamefont {Hill}, \citenamefont
  {Sutherland}, \citenamefont {Taillefer}, \citenamefont {Sonier},
  \citenamefont {Hayden},\ and\ \citenamefont {Brill}}]{Boaknin2003}%
  \BibitemOpen
  \bibfield  {author} {\bibinfo {author} {\bibfnamefont {E.}~\bibnamefont
  {Boaknin}}, \bibinfo {author} {\bibfnamefont {M.~A.}\ \bibnamefont
  {Tanatar}}, \bibinfo {author} {\bibfnamefont {J.}~\bibnamefont {Paglione}},
  \bibinfo {author} {\bibfnamefont {D.}~\bibnamefont {Hawthorn}}, \bibinfo
  {author} {\bibfnamefont {F.}~\bibnamefont {Ronning}}, \bibinfo {author}
  {\bibfnamefont {R.~W.}\ \bibnamefont {Hill}}, \bibinfo {author}
  {\bibfnamefont {M.}~\bibnamefont {Sutherland}}, \bibinfo {author}
  {\bibfnamefont {L.}~\bibnamefont {Taillefer}}, \bibinfo {author}
  {\bibfnamefont {J.}~\bibnamefont {Sonier}}, \bibinfo {author} {\bibfnamefont
  {S.~M.}\ \bibnamefont {Hayden}}, \ and\ \bibinfo {author} {\bibfnamefont
  {J.~W.}\ \bibnamefont {Brill}},\ }\href {\doibase
  10.1103/PhysRevLett.90.117003} {\bibfield  {journal} {\bibinfo  {journal}
  {Phys. Rev. Lett.}\ }\textbf {\bibinfo {volume} {90}},\ \bibinfo {pages}
  {117003} (\bibinfo {year} {2003})}\BibitemShut {NoStop}%
\bibitem [{\citenamefont {Fletcher}\ \emph {et~al.}(2007)\citenamefont
  {Fletcher}, \citenamefont {Carrington}, \citenamefont {Diener}, \citenamefont
  {Rodi\`ere}, \citenamefont {Brison}, \citenamefont {Prozorov}, \citenamefont
  {Olheiser},\ and\ \citenamefont {Giannetta}}]{Fletcher2007}%
  \BibitemOpen
  \bibfield  {author} {\bibinfo {author} {\bibfnamefont {J.~D.}\ \bibnamefont
  {Fletcher}}, \bibinfo {author} {\bibfnamefont {A.}~\bibnamefont
  {Carrington}}, \bibinfo {author} {\bibfnamefont {P.}~\bibnamefont {Diener}},
  \bibinfo {author} {\bibfnamefont {P.}~\bibnamefont {Rodi\`ere}}, \bibinfo
  {author} {\bibfnamefont {J.~P.}\ \bibnamefont {Brison}}, \bibinfo {author}
  {\bibfnamefont {R.}~\bibnamefont {Prozorov}}, \bibinfo {author}
  {\bibfnamefont {T.}~\bibnamefont {Olheiser}}, \ and\ \bibinfo {author}
  {\bibfnamefont {R.~W.}\ \bibnamefont {Giannetta}},\ }\href {\doibase
  10.1103/PhysRevLett.98.057003} {\bibfield  {journal} {\bibinfo  {journal}
  {Phys. Rev. Lett.}\ }\textbf {\bibinfo {volume} {98}},\ \bibinfo {pages}
  {057003} (\bibinfo {year} {2007})}\BibitemShut {NoStop}%
\bibitem [{\citenamefont {Guillam{\'{o}}n}\ \emph
  {et~al.}(2008{\natexlab{a}})\citenamefont {Guillam{\'{o}}n}, \citenamefont
  {Suderow}, \citenamefont {Guinea},\ and\ \citenamefont
  {Vieira}}]{Guillamon2008}%
  \BibitemOpen
  \bibfield  {author} {\bibinfo {author} {\bibfnamefont {I.}~\bibnamefont
  {Guillam{\'{o}}n}}, \bibinfo {author} {\bibfnamefont {H.}~\bibnamefont
  {Suderow}}, \bibinfo {author} {\bibfnamefont {F.}~\bibnamefont {Guinea}}, \
  and\ \bibinfo {author} {\bibfnamefont {S.}~\bibnamefont {Vieira}},\ }\href
  {\doibase 10.1103/PhysRevB.77.134505} {\bibfield  {journal} {\bibinfo
  {journal} {Phys. Rev. B}\ }\textbf {\bibinfo {volume} {77}},\ \bibinfo
  {pages} {134505} (\bibinfo {year} {2008}{\natexlab{a}})}\BibitemShut
  {NoStop}%
\bibitem [{\citenamefont {Guillam{\'{o}}n}\ \emph
  {et~al.}(2008{\natexlab{b}})\citenamefont {Guillam{\'{o}}n}, \citenamefont
  {Suderow}, \citenamefont {Vieira}, \citenamefont {Cario}, \citenamefont
  {Diener},\ and\ \citenamefont {Rodi\`ere}}]{Guillamon2008a}%
  \BibitemOpen
  \bibfield  {author} {\bibinfo {author} {\bibfnamefont {I.}~\bibnamefont
  {Guillam{\'{o}}n}}, \bibinfo {author} {\bibfnamefont {H.}~\bibnamefont
  {Suderow}}, \bibinfo {author} {\bibfnamefont {S.}~\bibnamefont {Vieira}},
  \bibinfo {author} {\bibfnamefont {L.}~\bibnamefont {Cario}}, \bibinfo
  {author} {\bibfnamefont {P.}~\bibnamefont {Diener}}, \ and\ \bibinfo {author}
  {\bibfnamefont {P.}~\bibnamefont {Rodi\`ere}},\ }\href {\doibase
  10.1103/PhysRevLett.101.166407} {\bibfield  {journal} {\bibinfo  {journal}
  {Phys. Rev. Lett.}\ }\textbf {\bibinfo {volume} {101}},\ \bibinfo {pages}
  {166407} (\bibinfo {year} {2008}{\natexlab{b}})}\BibitemShut {NoStop}%
\bibitem [{\citenamefont {Xi}\ \emph {et~al.}(2016)\citenamefont {Xi},
  \citenamefont {Wang}, \citenamefont {Zhao}, \citenamefont {Park},
  \citenamefont {Law}, \citenamefont {Berger}, \citenamefont {Forr{\'{o}}},
  \citenamefont {Shan},\ and\ \citenamefont {Mak}}]{Xi2016}%
  \BibitemOpen
  \bibfield  {author} {\bibinfo {author} {\bibfnamefont {X.}~\bibnamefont
  {Xi}}, \bibinfo {author} {\bibfnamefont {Z.}~\bibnamefont {Wang}}, \bibinfo
  {author} {\bibfnamefont {W.}~\bibnamefont {Zhao}}, \bibinfo {author}
  {\bibfnamefont {J.-H.}\ \bibnamefont {Park}}, \bibinfo {author}
  {\bibfnamefont {K.~T.}\ \bibnamefont {Law}}, \bibinfo {author} {\bibfnamefont
  {H.}~\bibnamefont {Berger}}, \bibinfo {author} {\bibfnamefont
  {L.}~\bibnamefont {Forr{\'{o}}}}, \bibinfo {author} {\bibfnamefont
  {J.}~\bibnamefont {Shan}}, \ and\ \bibinfo {author} {\bibfnamefont {K.~F.}\
  \bibnamefont {Mak}},\ }\href {\doibase 10.1038/nphys3538} {\bibfield
  {journal} {\bibinfo  {journal} {Nature Phys.}\ }\textbf {\bibinfo {volume}
  {12}},\ \bibinfo {pages} {139} (\bibinfo {year} {2016})}\BibitemShut
  {NoStop}%
\bibitem [{\citenamefont {Ugeda}\ \emph {et~al.}(2016)\citenamefont {Ugeda},
  \citenamefont {Bradley}, \citenamefont {Zhang}, \citenamefont {Onishi},
  \citenamefont {Chen}, \citenamefont {Ruan}, \citenamefont
  {Ojeda-Aristizabal}, \citenamefont {Ryu}, \citenamefont {Edmonds},
  \citenamefont {Tsai},\ and\ \citenamefont {{et al.}}}]{Ugeda2016}%
  \BibitemOpen
  \bibfield  {author} {\bibinfo {author} {\bibfnamefont {M.~M.}\ \bibnamefont
  {Ugeda}}, \bibinfo {author} {\bibfnamefont {A.~J.}\ \bibnamefont {Bradley}},
  \bibinfo {author} {\bibfnamefont {Y.}~\bibnamefont {Zhang}}, \bibinfo
  {author} {\bibfnamefont {S.}~\bibnamefont {Onishi}}, \bibinfo {author}
  {\bibfnamefont {Y.}~\bibnamefont {Chen}}, \bibinfo {author} {\bibfnamefont
  {W.}~\bibnamefont {Ruan}}, \bibinfo {author} {\bibfnamefont {C.}~\bibnamefont
  {Ojeda-Aristizabal}}, \bibinfo {author} {\bibfnamefont {H.}~\bibnamefont
  {Ryu}}, \bibinfo {author} {\bibfnamefont {M.~T.}\ \bibnamefont {Edmonds}},
  \bibinfo {author} {\bibfnamefont {H.-Z.}\ \bibnamefont {Tsai}}, \ and\
  \bibinfo {author} {\bibnamefont {{et al.}}},\ }\href {\doibase
  10.1038/nphys3527} {\bibfield  {journal} {\bibinfo  {journal} {Nature Phys.}\
  }\textbf {\bibinfo {volume} {12}},\ \bibinfo {pages} {92} (\bibinfo {year}
  {2016})}\BibitemShut {NoStop}%
\bibitem [{\citenamefont {Rossnagel}\ \emph {et~al.}(2001)\citenamefont
  {Rossnagel}, \citenamefont {Seifarth}, \citenamefont {Kipp}, \citenamefont
  {Skibowski}, \citenamefont {Vo\ss{}}, \citenamefont {Kr\"uger}, \citenamefont
  {Mazur},\ and\ \citenamefont {Pollmann}}]{Rossnagel2001}%
  \BibitemOpen
  \bibfield  {author} {\bibinfo {author} {\bibfnamefont {K.}~\bibnamefont
  {Rossnagel}}, \bibinfo {author} {\bibfnamefont {O.}~\bibnamefont {Seifarth}},
  \bibinfo {author} {\bibfnamefont {L.}~\bibnamefont {Kipp}}, \bibinfo {author}
  {\bibfnamefont {M.}~\bibnamefont {Skibowski}}, \bibinfo {author}
  {\bibfnamefont {D.}~\bibnamefont {Vo\ss{}}}, \bibinfo {author} {\bibfnamefont
  {P.}~\bibnamefont {Kr\"uger}}, \bibinfo {author} {\bibfnamefont
  {A.}~\bibnamefont {Mazur}}, \ and\ \bibinfo {author} {\bibfnamefont
  {J.}~\bibnamefont {Pollmann}},\ }\href {\doibase 10.1103/PhysRevB.64.235119}
  {\bibfield  {journal} {\bibinfo  {journal} {Phys. Rev. B}\ }\textbf {\bibinfo
  {volume} {64}},\ \bibinfo {pages} {235119} (\bibinfo {year}
  {2001})}\BibitemShut {NoStop}%
\bibitem [{\citenamefont {Johannes}\ \emph {et~al.}(2006)\citenamefont
  {Johannes}, \citenamefont {Mazin},\ and\ \citenamefont
  {Howells}}]{Johannes2006}%
  \BibitemOpen
  \bibfield  {author} {\bibinfo {author} {\bibfnamefont {M.~D.}\ \bibnamefont
  {Johannes}}, \bibinfo {author} {\bibfnamefont {I.~I.}\ \bibnamefont {Mazin}},
  \ and\ \bibinfo {author} {\bibfnamefont {C.~A.}\ \bibnamefont {Howells}},\
  }\href {\doibase 10.1103/PhysRevB.73.205102} {\bibfield  {journal} {\bibinfo
  {journal} {Phys. Rev. B}\ }\textbf {\bibinfo {volume} {73}},\ \bibinfo
  {pages} {205102} (\bibinfo {year} {2006})}\BibitemShut {NoStop}%
\bibitem [{\citenamefont {Kiss}\ \emph {et~al.}(2007)\citenamefont {Kiss},
  \citenamefont {Yokoya}, \citenamefont {Chainani}, \citenamefont {Shin},
  \citenamefont {Hanaguri}, \citenamefont {Nohara},\ and\ \citenamefont
  {Takagi}}]{Kiss2007}%
  \BibitemOpen
  \bibfield  {author} {\bibinfo {author} {\bibfnamefont {T.}~\bibnamefont
  {Kiss}}, \bibinfo {author} {\bibfnamefont {T.}~\bibnamefont {Yokoya}},
  \bibinfo {author} {\bibfnamefont {A.}~\bibnamefont {Chainani}}, \bibinfo
  {author} {\bibfnamefont {S.}~\bibnamefont {Shin}}, \bibinfo {author}
  {\bibfnamefont {T.}~\bibnamefont {Hanaguri}}, \bibinfo {author}
  {\bibfnamefont {M.}~\bibnamefont {Nohara}}, \ and\ \bibinfo {author}
  {\bibfnamefont {H.}~\bibnamefont {Takagi}},\ }\href {\doibase
  10.1038/nphys699} {\bibfield  {journal} {\bibinfo  {journal} {Nature Phys.}\
  }\textbf {\bibinfo {volume} {3}},\ \bibinfo {pages} {720} (\bibinfo {year}
  {2007})}\BibitemShut {NoStop}%
\bibitem [{\citenamefont {Johannes}\ and\ \citenamefont
  {Mazin}(2008)}]{Johannes2008}%
  \BibitemOpen
  \bibfield  {author} {\bibinfo {author} {\bibfnamefont {M.~D.}\ \bibnamefont
  {Johannes}}\ and\ \bibinfo {author} {\bibfnamefont {I.~I.}\ \bibnamefont
  {Mazin}},\ }\href {\doibase 10.1103/PhysRevB.77.165135} {\bibfield  {journal}
  {\bibinfo  {journal} {Phys. Rev. B}\ }\textbf {\bibinfo {volume} {77}},\
  \bibinfo {pages} {165135} (\bibinfo {year} {2008})}\BibitemShut {NoStop}%
\bibitem [{\citenamefont {Rahn}\ \emph {et~al.}(2012)\citenamefont {Rahn},
  \citenamefont {Hellmann}, \citenamefont {Kall\"ane}, \citenamefont {Sohrt},
  \citenamefont {Kim}, \citenamefont {Kipp},\ and\ \citenamefont
  {Rossnagel}}]{Rahn2012}%
  \BibitemOpen
  \bibfield  {author} {\bibinfo {author} {\bibfnamefont {D.~J.}\ \bibnamefont
  {Rahn}}, \bibinfo {author} {\bibfnamefont {S.}~\bibnamefont {Hellmann}},
  \bibinfo {author} {\bibfnamefont {M.}~\bibnamefont {Kall\"ane}}, \bibinfo
  {author} {\bibfnamefont {C.}~\bibnamefont {Sohrt}}, \bibinfo {author}
  {\bibfnamefont {T.~K.}\ \bibnamefont {Kim}}, \bibinfo {author} {\bibfnamefont
  {L.}~\bibnamefont {Kipp}}, \ and\ \bibinfo {author} {\bibfnamefont
  {K.}~\bibnamefont {Rossnagel}},\ }\href {\doibase 10.1103/PhysRevB.85.224532}
  {\bibfield  {journal} {\bibinfo  {journal} {Phys. Rev. B}\ }\textbf {\bibinfo
  {volume} {85}},\ \bibinfo {pages} {224532} (\bibinfo {year}
  {2012})}\BibitemShut {NoStop}%
\bibitem [{\citenamefont {Malliakas}\ and\ \citenamefont
  {Kanatzidis}(2013)}]{Malliakas2013}%
  \BibitemOpen
  \bibfield  {author} {\bibinfo {author} {\bibfnamefont {C.~D.}\ \bibnamefont
  {Malliakas}}\ and\ \bibinfo {author} {\bibfnamefont {M.~G.}\ \bibnamefont
  {Kanatzidis}},\ }\href {\doibase 10.1021/ja3120554} {\bibfield  {journal}
  {\bibinfo  {journal} {J. Am. Chem. Soc.}\ }\textbf {\bibinfo {volume}
  {135}},\ \bibinfo {pages} {1719} (\bibinfo {year} {2013})}\BibitemShut
  {NoStop}%
\bibitem [{\citenamefont {Dai}\ \emph {et~al.}(2014)\citenamefont {Dai},
  \citenamefont {Calleja}, \citenamefont {Alldredge}, \citenamefont {Zhu},
  \citenamefont {Li}, \citenamefont {Lu}, \citenamefont {Sun}, \citenamefont
  {Wolf}, \citenamefont {Berger},\ and\ \citenamefont {McElroy}}]{Dai2014}%
  \BibitemOpen
  \bibfield  {author} {\bibinfo {author} {\bibfnamefont {J.}~\bibnamefont
  {Dai}}, \bibinfo {author} {\bibfnamefont {E.}~\bibnamefont {Calleja}},
  \bibinfo {author} {\bibfnamefont {J.}~\bibnamefont {Alldredge}}, \bibinfo
  {author} {\bibfnamefont {X.}~\bibnamefont {Zhu}}, \bibinfo {author}
  {\bibfnamefont {L.}~\bibnamefont {Li}}, \bibinfo {author} {\bibfnamefont
  {W.}~\bibnamefont {Lu}}, \bibinfo {author} {\bibfnamefont {Y.}~\bibnamefont
  {Sun}}, \bibinfo {author} {\bibfnamefont {T.}~\bibnamefont {Wolf}}, \bibinfo
  {author} {\bibfnamefont {H.}~\bibnamefont {Berger}}, \ and\ \bibinfo {author}
  {\bibfnamefont {K.}~\bibnamefont {McElroy}},\ }\href {\doibase
  10.1103/PhysRevB.89.165140} {\bibfield  {journal} {\bibinfo  {journal} {Phys.
  Rev. B}\ }\textbf {\bibinfo {volume} {89}},\ \bibinfo {pages} {165140}
  (\bibinfo {year} {2014})}\BibitemShut {NoStop}%
\bibitem [{\citenamefont {Arguello}\ \emph {et~al.}(2014)\citenamefont
  {Arguello}, \citenamefont {Chockalingam}, \citenamefont {Rosenthal},
  \citenamefont {Zhao}, \citenamefont {Guti{\'{e}}rrez}, \citenamefont {Kang},
  \citenamefont {Chung}, \citenamefont {Fernandes}, \citenamefont {Jia},
  \citenamefont {Millis}, \citenamefont {Cava},\ and\ \citenamefont
  {Pasupathy}}]{Arguello2014}%
  \BibitemOpen
  \bibfield  {author} {\bibinfo {author} {\bibfnamefont {C.~J.}\ \bibnamefont
  {Arguello}}, \bibinfo {author} {\bibfnamefont {S.~P.}\ \bibnamefont
  {Chockalingam}}, \bibinfo {author} {\bibfnamefont {E.~P.}\ \bibnamefont
  {Rosenthal}}, \bibinfo {author} {\bibfnamefont {L.}~\bibnamefont {Zhao}},
  \bibinfo {author} {\bibfnamefont {C.}~\bibnamefont {Guti{\'{e}}rrez}},
  \bibinfo {author} {\bibfnamefont {J.~H.}\ \bibnamefont {Kang}}, \bibinfo
  {author} {\bibfnamefont {W.~C.}\ \bibnamefont {Chung}}, \bibinfo {author}
  {\bibfnamefont {R.~M.}\ \bibnamefont {Fernandes}}, \bibinfo {author}
  {\bibfnamefont {S.}~\bibnamefont {Jia}}, \bibinfo {author} {\bibfnamefont
  {A.~J.}\ \bibnamefont {Millis}}, \bibinfo {author} {\bibfnamefont {R.~J.}\
  \bibnamefont {Cava}}, \ and\ \bibinfo {author} {\bibfnamefont {A.~N.}\
  \bibnamefont {Pasupathy}},\ }\href {\doibase 10.1103/PhysRevB.89.235115}
  {\bibfield  {journal} {\bibinfo  {journal} {Phys. Rev. B}\ }\textbf {\bibinfo
  {volume} {89}},\ \bibinfo {pages} {235115} (\bibinfo {year}
  {2014})}\BibitemShut {NoStop}%
\bibitem [{\citenamefont {Arguello}\ \emph {et~al.}(2015)\citenamefont
  {Arguello}, \citenamefont {Rosenthal}, \citenamefont {Andrade}, \citenamefont
  {Jin}, \citenamefont {Yeh}, \citenamefont {Zaki}, \citenamefont {Jia},
  \citenamefont {Cava}, \citenamefont {Fernandes}, \citenamefont {Millis},
  \citenamefont {Valla}, \citenamefont {Osgood},\ and\ \citenamefont
  {Pasupathy}}]{Arguello2015}%
  \BibitemOpen
  \bibfield  {author} {\bibinfo {author} {\bibfnamefont {C.~J.}\ \bibnamefont
  {Arguello}}, \bibinfo {author} {\bibfnamefont {E.~P.}\ \bibnamefont
  {Rosenthal}}, \bibinfo {author} {\bibfnamefont {E.~F.}\ \bibnamefont
  {Andrade}}, \bibinfo {author} {\bibfnamefont {W.}~\bibnamefont {Jin}},
  \bibinfo {author} {\bibfnamefont {P.~C.}\ \bibnamefont {Yeh}}, \bibinfo
  {author} {\bibfnamefont {N.}~\bibnamefont {Zaki}}, \bibinfo {author}
  {\bibfnamefont {S.}~\bibnamefont {Jia}}, \bibinfo {author} {\bibfnamefont
  {R.~J.}\ \bibnamefont {Cava}}, \bibinfo {author} {\bibfnamefont {R.~M.}\
  \bibnamefont {Fernandes}}, \bibinfo {author} {\bibfnamefont {A.~J.}\
  \bibnamefont {Millis}}, \bibinfo {author} {\bibfnamefont {T.}~\bibnamefont
  {Valla}}, \bibinfo {author} {\bibfnamefont {R.~M.}\ \bibnamefont {Osgood}}, \
  and\ \bibinfo {author} {\bibfnamefont {A.~N.}\ \bibnamefont {Pasupathy}},\
  }\href {\doibase 10.1103/PhysRevLett.114.037001} {\bibfield  {journal}
  {\bibinfo  {journal} {Phys. Rev. Lett.}\ }\textbf {\bibinfo {volume} {114}},\
  \bibinfo {pages} {037001} (\bibinfo {year} {2015})}\BibitemShut {NoStop}%
\bibitem [{\citenamefont {Flicker}\ and\ \citenamefont {van
  Wezel}(2015)}]{Flicker2015}%
  \BibitemOpen
  \bibfield  {author} {\bibinfo {author} {\bibfnamefont {F.}~\bibnamefont
  {Flicker}}\ and\ \bibinfo {author} {\bibfnamefont {J.}~\bibnamefont {van
  Wezel}},\ }\href {\doibase 10.1038/ncomms8034} {\bibfield  {journal}
  {\bibinfo  {journal} {Nature Commun.}\ }\textbf {\bibinfo {volume} {6}},\
  \bibinfo {pages} {7034} (\bibinfo {year} {2015})}\BibitemShut {NoStop}%
\bibitem [{\citenamefont {Soumyanarayanan}\ \emph {et~al.}(2013)\citenamefont
  {Soumyanarayanan}, \citenamefont {Yee}, \citenamefont {He}, \citenamefont
  {van Wezel}, \citenamefont {Rahn}, \citenamefont {Rossnagel}, \citenamefont
  {Hudson}, \citenamefont {Norman},\ and\ \citenamefont
  {Hoffman}}]{Soumyanarayanan2013}%
  \BibitemOpen
  \bibfield  {author} {\bibinfo {author} {\bibfnamefont {A.}~\bibnamefont
  {Soumyanarayanan}}, \bibinfo {author} {\bibfnamefont {M.~M.}\ \bibnamefont
  {Yee}}, \bibinfo {author} {\bibfnamefont {Y.}~\bibnamefont {He}}, \bibinfo
  {author} {\bibfnamefont {J.}~\bibnamefont {van Wezel}}, \bibinfo {author}
  {\bibfnamefont {D.~J.}\ \bibnamefont {Rahn}}, \bibinfo {author}
  {\bibfnamefont {K.}~\bibnamefont {Rossnagel}}, \bibinfo {author}
  {\bibfnamefont {E.~W.}\ \bibnamefont {Hudson}}, \bibinfo {author}
  {\bibfnamefont {M.~R.}\ \bibnamefont {Norman}}, \ and\ \bibinfo {author}
  {\bibfnamefont {J.~E.}\ \bibnamefont {Hoffman}},\ }\href {\doibase
  10.1073/pnas.1211387110} {\bibfield  {journal} {\bibinfo  {journal} {Proc.
  Natl. Acad. Sci.}\ }\textbf {\bibinfo {volume} {110}},\ \bibinfo {pages}
  {1623} (\bibinfo {year} {2013})}\BibitemShut {NoStop}%
\bibitem [{\citenamefont {Weber}\ \emph {et~al.}(2011)\citenamefont {Weber},
  \citenamefont {Rosenkranz}, \citenamefont {Castellan}, \citenamefont
  {Osborn}, \citenamefont {Hott}, \citenamefont {Heid}, \citenamefont {Bohnen},
  \citenamefont {Egami}, \citenamefont {Said},\ and\ \citenamefont
  {Reznik}}]{Weber2011}%
  \BibitemOpen
  \bibfield  {author} {\bibinfo {author} {\bibfnamefont {F.}~\bibnamefont
  {Weber}}, \bibinfo {author} {\bibfnamefont {S.}~\bibnamefont {Rosenkranz}},
  \bibinfo {author} {\bibfnamefont {J.-P.}\ \bibnamefont {Castellan}}, \bibinfo
  {author} {\bibfnamefont {R.}~\bibnamefont {Osborn}}, \bibinfo {author}
  {\bibfnamefont {R.}~\bibnamefont {Hott}}, \bibinfo {author} {\bibfnamefont
  {R.}~\bibnamefont {Heid}}, \bibinfo {author} {\bibfnamefont {K.-P.}\
  \bibnamefont {Bohnen}}, \bibinfo {author} {\bibfnamefont {T.}~\bibnamefont
  {Egami}}, \bibinfo {author} {\bibfnamefont {A.~H.}\ \bibnamefont {Said}}, \
  and\ \bibinfo {author} {\bibfnamefont {D.}~\bibnamefont {Reznik}},\ }\href
  {\doibase 10.1103/PhysRevLett.107.107403} {\bibfield  {journal} {\bibinfo
  {journal} {Phys. Rev. Lett.}\ }\textbf {\bibinfo {volume} {107}},\ \bibinfo
  {pages} {107403} (\bibinfo {year} {2011})}\BibitemShut {NoStop}%
\bibitem [{\citenamefont {Suderow}\ \emph {et~al.}(2005)\citenamefont
  {Suderow}, \citenamefont {Tissen}, \citenamefont {Brison}, \citenamefont
  {Mart\'{\i}nez},\ and\ \citenamefont {Vieira}}]{Suderow2005}%
  \BibitemOpen
  \bibfield  {author} {\bibinfo {author} {\bibfnamefont {H.}~\bibnamefont
  {Suderow}}, \bibinfo {author} {\bibfnamefont {V.~G.}\ \bibnamefont {Tissen}},
  \bibinfo {author} {\bibfnamefont {J.~P.}\ \bibnamefont {Brison}}, \bibinfo
  {author} {\bibfnamefont {J.~L.}\ \bibnamefont {Mart\'{\i}nez}}, \ and\
  \bibinfo {author} {\bibfnamefont {S.}~\bibnamefont {Vieira}},\ }\href
  {\doibase 10.1103/PhysRevLett.95.117006} {\bibfield  {journal} {\bibinfo
  {journal} {Phys. Rev. Lett.}\ }\textbf {\bibinfo {volume} {95}},\ \bibinfo
  {pages} {117006} (\bibinfo {year} {2005})}\BibitemShut {NoStop}%
\bibitem [{\citenamefont {Borisenko}\ \emph {et~al.}(2009)\citenamefont
  {Borisenko}, \citenamefont {Kordyuk}, \citenamefont {Zabolotnyy},
  \citenamefont {Inosov}, \citenamefont {Evtushinsky}, \citenamefont
  {B\"uchner}, \citenamefont {Yaresko}, \citenamefont {Varykhalov},
  \citenamefont {Follath}, \citenamefont {Eberhardt}, \citenamefont {Patthey},\
  and\ \citenamefont {Berger}}]{Borisenko2009}%
  \BibitemOpen
  \bibfield  {author} {\bibinfo {author} {\bibfnamefont {S.~V.}\ \bibnamefont
  {Borisenko}}, \bibinfo {author} {\bibfnamefont {A.~A.}\ \bibnamefont
  {Kordyuk}}, \bibinfo {author} {\bibfnamefont {V.~B.}\ \bibnamefont
  {Zabolotnyy}}, \bibinfo {author} {\bibfnamefont {D.~S.}\ \bibnamefont
  {Inosov}}, \bibinfo {author} {\bibfnamefont {D.}~\bibnamefont {Evtushinsky}},
  \bibinfo {author} {\bibfnamefont {B.}~\bibnamefont {B\"uchner}}, \bibinfo
  {author} {\bibfnamefont {A.~N.}\ \bibnamefont {Yaresko}}, \bibinfo {author}
  {\bibfnamefont {A.}~\bibnamefont {Varykhalov}}, \bibinfo {author}
  {\bibfnamefont {R.}~\bibnamefont {Follath}}, \bibinfo {author} {\bibfnamefont
  {W.}~\bibnamefont {Eberhardt}}, \bibinfo {author} {\bibfnamefont
  {L.}~\bibnamefont {Patthey}}, \ and\ \bibinfo {author} {\bibfnamefont
  {H.}~\bibnamefont {Berger}},\ }\href {\doibase
  10.1103/PhysRevLett.102.166402} {\bibfield  {journal} {\bibinfo  {journal}
  {Phys. Rev. Lett.}\ }\textbf {\bibinfo {volume} {102}},\ \bibinfo {pages}
  {166402} (\bibinfo {year} {2009})}\BibitemShut {NoStop}%
\bibitem [{\citenamefont {Cho}\ \emph {et~al.}(2018)\citenamefont {Cho},
  \citenamefont {Ko{\'n}czykowski}, \citenamefont {Teknowijoyo}, \citenamefont
  {Tanatar}, \citenamefont {Guss}, \citenamefont {Gartin}, \citenamefont
  {Wilde}, \citenamefont {Kreyssig}, \citenamefont {McQueeney}, \citenamefont
  {Goldman} \emph {et~al.}}]{Cho2018}%
  \BibitemOpen
  \bibfield  {author} {\bibinfo {author} {\bibfnamefont {K.}~\bibnamefont
  {Cho}}, \bibinfo {author} {\bibfnamefont {M.}~\bibnamefont
  {Ko{\'n}czykowski}}, \bibinfo {author} {\bibfnamefont {S.}~\bibnamefont
  {Teknowijoyo}}, \bibinfo {author} {\bibfnamefont {M.~A.}\ \bibnamefont
  {Tanatar}}, \bibinfo {author} {\bibfnamefont {J.}~\bibnamefont {Guss}},
  \bibinfo {author} {\bibfnamefont {P.}~\bibnamefont {Gartin}}, \bibinfo
  {author} {\bibfnamefont {J.~M.}\ \bibnamefont {Wilde}}, \bibinfo {author}
  {\bibfnamefont {A.}~\bibnamefont {Kreyssig}}, \bibinfo {author}
  {\bibfnamefont {R.}~\bibnamefont {McQueeney}}, \bibinfo {author}
  {\bibfnamefont {A.~I.}\ \bibnamefont {Goldman}},  \emph {et~al.},\
  }\href@noop {} {\bibfield  {journal} {\bibinfo  {journal} {Nature Commun.}\
  }\textbf {\bibinfo {volume} {9}},\ \bibinfo {pages} {2796} (\bibinfo {year}
  {2018})}\BibitemShut {NoStop}%
\bibitem [{\citenamefont {Sticlet}\ and\ \citenamefont
  {Morari}(2019)}]{Sticlet2019}%
  \BibitemOpen
  \bibfield  {author} {\bibinfo {author} {\bibfnamefont {D.}~\bibnamefont
  {Sticlet}}\ and\ \bibinfo {author} {\bibfnamefont {C.}~\bibnamefont
  {Morari}},\ }\href {\doibase 10.1103/PhysRevB.100.075420} {\bibfield
  {journal} {\bibinfo  {journal} {Phys. Rev. B}\ }\textbf {\bibinfo {volume}
  {100}},\ \bibinfo {pages} {075420} (\bibinfo {year} {2019})}\BibitemShut
  {NoStop}%
\bibitem [{\citenamefont {G{\l}odzik}\ and\ \citenamefont
  {Ojanen}(2019)}]{Glodzik2019}%
  \BibitemOpen
  \bibfield  {author} {\bibinfo {author} {\bibfnamefont {S.}~\bibnamefont
  {G{\l}odzik}}\ and\ \bibinfo {author} {\bibfnamefont {T.}~\bibnamefont
  {Ojanen}},\ }\href {https://arxiv.org/abs/1905.01063} {\bibfield  {journal}
  {\bibinfo  {journal} {arXiv}\ ,\ \bibinfo {pages} {1905.01063}} (\bibinfo
  {year} {2019})},\ \Eprint {http://arxiv.org/abs/arXiv:1905.01063}
  {arXiv:arXiv:1905.01063} \BibitemShut {NoStop}%
\bibitem [{\citenamefont {Nadj-Perge}\ \emph {et~al.}(2014)\citenamefont
  {Nadj-Perge}, \citenamefont {Drozdov}, \citenamefont {Li}, \citenamefont
  {Chen}, \citenamefont {Jeon}, \citenamefont {Seo}, \citenamefont {MacDonald},
  \citenamefont {Bernevig},\ and\ \citenamefont {Yazdani}}]{Nadj-Perge2014}%
  \BibitemOpen
  \bibfield  {author} {\bibinfo {author} {\bibfnamefont {S.}~\bibnamefont
  {Nadj-Perge}}, \bibinfo {author} {\bibfnamefont {I.~K.}\ \bibnamefont
  {Drozdov}}, \bibinfo {author} {\bibfnamefont {J.}~\bibnamefont {Li}},
  \bibinfo {author} {\bibfnamefont {H.}~\bibnamefont {Chen}}, \bibinfo {author}
  {\bibfnamefont {S.}~\bibnamefont {Jeon}}, \bibinfo {author} {\bibfnamefont
  {J.}~\bibnamefont {Seo}}, \bibinfo {author} {\bibfnamefont {A.~H.}\
  \bibnamefont {MacDonald}}, \bibinfo {author} {\bibfnamefont {B.~A.}\
  \bibnamefont {Bernevig}}, \ and\ \bibinfo {author} {\bibfnamefont
  {A.}~\bibnamefont {Yazdani}},\ }\href {\doibase 10.1126/science.1259327}
  {\bibfield  {journal} {\bibinfo  {journal} {Science}\ }\textbf {\bibinfo
  {volume} {346}},\ \bibinfo {pages} {602} (\bibinfo {year}
  {2014})}\BibitemShut {NoStop}%
\bibitem [{\citenamefont {Nadj-Perge}\ \emph {et~al.}(2013)\citenamefont
  {Nadj-Perge}, \citenamefont {Drozdov}, \citenamefont {Bernevig},\ and\
  \citenamefont {Yazdani}}]{NadjPerge2013}%
  \BibitemOpen
  \bibfield  {author} {\bibinfo {author} {\bibfnamefont {S.}~\bibnamefont
  {Nadj-Perge}}, \bibinfo {author} {\bibfnamefont {I.~K.}\ \bibnamefont
  {Drozdov}}, \bibinfo {author} {\bibfnamefont {B.~A.}\ \bibnamefont
  {Bernevig}}, \ and\ \bibinfo {author} {\bibfnamefont {A.}~\bibnamefont
  {Yazdani}},\ }\href {\doibase 10.1103/PhysRevB.88.020407} {\bibfield
  {journal} {\bibinfo  {journal} {Phys. Rev. B}\ }\textbf {\bibinfo {volume}
  {88}},\ \bibinfo {pages} {020407} (\bibinfo {year} {2013})}\BibitemShut
  {NoStop}%
\bibitem [{\citenamefont {Pientka}\ \emph {et~al.}(2013)\citenamefont
  {Pientka}, \citenamefont {Glazman},\ and\ \citenamefont {von
  Oppen}}]{Pientka2013}%
  \BibitemOpen
  \bibfield  {author} {\bibinfo {author} {\bibfnamefont {F.}~\bibnamefont
  {Pientka}}, \bibinfo {author} {\bibfnamefont {L.~I.}\ \bibnamefont
  {Glazman}}, \ and\ \bibinfo {author} {\bibfnamefont {F.}~\bibnamefont {von
  Oppen}},\ }\href {\doibase 10.1103/PhysRevB.88.155420} {\bibfield  {journal}
  {\bibinfo  {journal} {Phys. Rev. B}\ }\textbf {\bibinfo {volume} {88}},\
  \bibinfo {pages} {155420} (\bibinfo {year} {2013})}\BibitemShut {NoStop}%
\bibitem [{\citenamefont {Yu}(1965)}]{Yu1965}%
  \BibitemOpen
  \bibfield  {author} {\bibinfo {author} {\bibfnamefont {L.}~\bibnamefont
  {Yu}},\ }\href {\doibase 10.7498/aps.21.75} {\bibfield  {journal} {\bibinfo
  {journal} {Acta Phys. Sin.}\ }\textbf {\bibinfo {volume} {21}},\ \bibinfo
  {pages} {75} (\bibinfo {year} {1965})}\BibitemShut {NoStop}%
\bibitem [{\citenamefont {Shiba}(1968)}]{Shiba1968}%
  \BibitemOpen
  \bibfield  {author} {\bibinfo {author} {\bibfnamefont {H.}~\bibnamefont
  {Shiba}},\ }\href {\doibase 10.1143/PTP.40.435} {\bibfield  {journal}
  {\bibinfo  {journal} {Prog. Theor. Phys.}\ }\textbf {\bibinfo {volume}
  {40}},\ \bibinfo {pages} {435} (\bibinfo {year} {1968})}\BibitemShut
  {NoStop}%
\bibitem [{\citenamefont {Rusinov}(1969)}]{Rusinov1969}%
  \BibitemOpen
  \bibfield  {author} {\bibinfo {author} {\bibfnamefont {A.~I.}\ \bibnamefont
  {Rusinov}},\ }\href
  {http://www.jetpletters.ac.ru/ps/1658/article{\_}25295.shtml} {\bibfield
  {journal} {\bibinfo  {journal} {JETP Lett.}\ }\textbf {\bibinfo {volume}
  {9}},\ \bibinfo {pages} {85} (\bibinfo {year} {1969})}\BibitemShut {NoStop}%
\bibitem [{\citenamefont {Yazdani}\ \emph {et~al.}(1997)\citenamefont
  {Yazdani}, \citenamefont {Jones}, \citenamefont {Lutz}, \citenamefont
  {Crommie},\ and\ \citenamefont {Eigler}}]{Yazdani1997}%
  \BibitemOpen
  \bibfield  {author} {\bibinfo {author} {\bibfnamefont {A.}~\bibnamefont
  {Yazdani}}, \bibinfo {author} {\bibfnamefont {B.~A.}\ \bibnamefont {Jones}},
  \bibinfo {author} {\bibfnamefont {C.~P.}\ \bibnamefont {Lutz}}, \bibinfo
  {author} {\bibfnamefont {M.~F.}\ \bibnamefont {Crommie}}, \ and\ \bibinfo
  {author} {\bibfnamefont {D.~M.}\ \bibnamefont {Eigler}},\ }\href {\doibase
  10.1126/science.275.5307.1767} {\bibfield  {journal} {\bibinfo  {journal}
  {Science}\ }\textbf {\bibinfo {volume} {275}},\ \bibinfo {pages} {1767}
  (\bibinfo {year} {1997})}\BibitemShut {NoStop}%
\bibitem [{\citenamefont {M{\'{e}}nard}\ \emph {et~al.}(2015)\citenamefont
  {M{\'{e}}nard}, \citenamefont {Guissart}, \citenamefont {Brun}, \citenamefont
  {Pons}, \citenamefont {Stolyarov}, \citenamefont {Debontridder},
  \citenamefont {Leclerc}, \citenamefont {Janod}, \citenamefont {Cario},
  \citenamefont {Roditchev}, \citenamefont {Simon},\ and\ \citenamefont
  {Cren}}]{Menard2015}%
  \BibitemOpen
  \bibfield  {author} {\bibinfo {author} {\bibfnamefont {G.~C.}\ \bibnamefont
  {M{\'{e}}nard}}, \bibinfo {author} {\bibfnamefont {S.}~\bibnamefont
  {Guissart}}, \bibinfo {author} {\bibfnamefont {C.}~\bibnamefont {Brun}},
  \bibinfo {author} {\bibfnamefont {S.}~\bibnamefont {Pons}}, \bibinfo {author}
  {\bibfnamefont {V.~S.}\ \bibnamefont {Stolyarov}}, \bibinfo {author}
  {\bibfnamefont {F.}~\bibnamefont {Debontridder}}, \bibinfo {author}
  {\bibfnamefont {M.~V.}\ \bibnamefont {Leclerc}}, \bibinfo {author}
  {\bibfnamefont {E.}~\bibnamefont {Janod}}, \bibinfo {author} {\bibfnamefont
  {L.}~\bibnamefont {Cario}}, \bibinfo {author} {\bibfnamefont
  {D.}~\bibnamefont {Roditchev}}, \bibinfo {author} {\bibfnamefont
  {P.}~\bibnamefont {Simon}}, \ and\ \bibinfo {author} {\bibfnamefont
  {T.}~\bibnamefont {Cren}},\ }\href {\doibase 10.1038/nphys3508} {\bibfield
  {journal} {\bibinfo  {journal} {Nature Phys.}\ }\textbf {\bibinfo {volume}
  {11}},\ \bibinfo {pages} {1013} (\bibinfo {year} {2015})}\BibitemShut
  {NoStop}%
\bibitem [{\citenamefont {Kezilebieke}\ \emph {et~al.}(2018)\citenamefont
  {Kezilebieke}, \citenamefont {Dvorak}, \citenamefont {Ojanen},\ and\
  \citenamefont {Liljeroth}}]{Kezilebieke2018}%
  \BibitemOpen
  \bibfield  {author} {\bibinfo {author} {\bibfnamefont {S.}~\bibnamefont
  {Kezilebieke}}, \bibinfo {author} {\bibfnamefont {M.}~\bibnamefont {Dvorak}},
  \bibinfo {author} {\bibfnamefont {T.}~\bibnamefont {Ojanen}}, \ and\ \bibinfo
  {author} {\bibfnamefont {P.}~\bibnamefont {Liljeroth}},\ }\href {\doibase
  10.1021/acs.nanolett.7b05050} {\bibfield  {journal} {\bibinfo  {journal}
  {Nano Lett.}\ }\textbf {\bibinfo {volume} {18}},\ \bibinfo {pages} {2311}
  (\bibinfo {year} {2018})}\BibitemShut {NoStop}%
\bibitem [{\citenamefont {Ruby}\ \emph {et~al.}(2018)\citenamefont {Ruby},
  \citenamefont {Heinrich}, \citenamefont {Peng}, \citenamefont {von Oppen},\
  and\ \citenamefont {Franke}}]{Ruby2018}%
  \BibitemOpen
  \bibfield  {author} {\bibinfo {author} {\bibfnamefont {M.}~\bibnamefont
  {Ruby}}, \bibinfo {author} {\bibfnamefont {B.~W.}\ \bibnamefont {Heinrich}},
  \bibinfo {author} {\bibfnamefont {Y.}~\bibnamefont {Peng}}, \bibinfo {author}
  {\bibfnamefont {F.}~\bibnamefont {von Oppen}}, \ and\ \bibinfo {author}
  {\bibfnamefont {K.~J.}\ \bibnamefont {Franke}},\ }\href {\doibase
  10.1103/PhysRevLett.120.156803} {\bibfield  {journal} {\bibinfo  {journal}
  {Phys. Rev. Lett.}\ }\textbf {\bibinfo {volume} {120}},\ \bibinfo {pages}
  {156803} (\bibinfo {year} {2018})}\BibitemShut {NoStop}%
\bibitem [{\citenamefont {Choi}\ \emph {et~al.}(2018)\citenamefont {Choi},
  \citenamefont {Fern\'andez}, \citenamefont {Herrera}, \citenamefont
  {Rubio-Verd{\'{u}}}, \citenamefont {Ugeda}, \citenamefont {Guillam\'on},
  \citenamefont {Suderow}, \citenamefont {Pascual},\ and\ \citenamefont
  {Lorente}}]{Choi2018}%
  \BibitemOpen
  \bibfield  {author} {\bibinfo {author} {\bibfnamefont {D.-J.}\ \bibnamefont
  {Choi}}, \bibinfo {author} {\bibfnamefont {C.~G.}\ \bibnamefont
  {Fern\'andez}}, \bibinfo {author} {\bibfnamefont {E.}~\bibnamefont
  {Herrera}}, \bibinfo {author} {\bibfnamefont {C.}~\bibnamefont
  {Rubio-Verd{\'{u}}}}, \bibinfo {author} {\bibfnamefont {M.~M.}\ \bibnamefont
  {Ugeda}}, \bibinfo {author} {\bibfnamefont {I.}~\bibnamefont {Guillam\'on}},
  \bibinfo {author} {\bibfnamefont {H.}~\bibnamefont {Suderow}}, \bibinfo
  {author} {\bibfnamefont {J.~I.}\ \bibnamefont {Pascual}}, \ and\ \bibinfo
  {author} {\bibfnamefont {N.}~\bibnamefont {Lorente}},\ }\href {\doibase
  10.1103/PhysRevLett.120.167001} {\bibfield  {journal} {\bibinfo  {journal}
  {Phys. Rev. Lett.}\ }\textbf {\bibinfo {volume} {120}},\ \bibinfo {pages}
  {167001} (\bibinfo {year} {2018})}\BibitemShut {NoStop}%
\bibitem [{\citenamefont {Kamlapure}\ \emph {et~al.}(2018)\citenamefont
  {Kamlapure}, \citenamefont {Cornils}, \citenamefont {Wiebe},\ and\
  \citenamefont {Wiesendanger}}]{Kamlapure2018}%
  \BibitemOpen
  \bibfield  {author} {\bibinfo {author} {\bibfnamefont {A.}~\bibnamefont
  {Kamlapure}}, \bibinfo {author} {\bibfnamefont {L.}~\bibnamefont {Cornils}},
  \bibinfo {author} {\bibfnamefont {J.}~\bibnamefont {Wiebe}}, \ and\ \bibinfo
  {author} {\bibfnamefont {R.}~\bibnamefont {Wiesendanger}},\ }\href {\doibase
  10.1038/s41467-018-05701-8} {\bibfield  {journal} {\bibinfo  {journal}
  {Nature Commun.}\ }\textbf {\bibinfo {volume} {9}},\ \bibinfo {pages} {1}
  (\bibinfo {year} {2018})}\BibitemShut {NoStop}%
\bibitem [{\citenamefont {Kim}\ \emph {et~al.}(2018)\citenamefont {Kim},
  \citenamefont {Palacio-Morales}, \citenamefont {Posske}, \citenamefont
  {R{\'o}zsa}, \citenamefont {Palot{\'a}s}, \citenamefont {Szunyogh},
  \citenamefont {Thorwart},\ and\ \citenamefont {Wiesendanger}}]{Kim2019}%
  \BibitemOpen
  \bibfield  {author} {\bibinfo {author} {\bibfnamefont {H.}~\bibnamefont
  {Kim}}, \bibinfo {author} {\bibfnamefont {A.}~\bibnamefont
  {Palacio-Morales}}, \bibinfo {author} {\bibfnamefont {T.}~\bibnamefont
  {Posske}}, \bibinfo {author} {\bibfnamefont {L.}~\bibnamefont {R{\'o}zsa}},
  \bibinfo {author} {\bibfnamefont {K.}~\bibnamefont {Palot{\'a}s}}, \bibinfo
  {author} {\bibfnamefont {L.}~\bibnamefont {Szunyogh}}, \bibinfo {author}
  {\bibfnamefont {M.}~\bibnamefont {Thorwart}}, \ and\ \bibinfo {author}
  {\bibfnamefont {R.}~\bibnamefont {Wiesendanger}},\ }\href {\doibase
  10.1126/sciadv.aar5251} {\bibfield  {journal} {\bibinfo  {journal} {Science
  Advances}\ }\textbf {\bibinfo {volume} {4}},\ \bibinfo {pages} {eaar5251}
  (\bibinfo {year} {2018})}  \BibitemShut
  {NoStop}%
\bibitem [{\citenamefont {Senkpiel}\ \emph {et~al.}(2019)\citenamefont
  {Senkpiel}, \citenamefont {Rubio-Verd\'u}, \citenamefont {Etzkorn},
  \citenamefont {Drost}, \citenamefont {Schoop}, \citenamefont {Dambach},
  \citenamefont {Padurariu}, \citenamefont {Kubala}, \citenamefont {Ankerhold},
  \citenamefont {Ast},\ and\ \citenamefont {Kern}}]{Senkpiel2018}%
  \BibitemOpen
  \bibfield  {author} {\bibinfo {author} {\bibfnamefont {J.}~\bibnamefont
  {Senkpiel}}, \bibinfo {author} {\bibfnamefont {C.}~\bibnamefont
  {Rubio-Verd\'u}}, \bibinfo {author} {\bibfnamefont {M.}~\bibnamefont
  {Etzkorn}}, \bibinfo {author} {\bibfnamefont {R.}~\bibnamefont {Drost}},
  \bibinfo {author} {\bibfnamefont {L.~M.}\ \bibnamefont {Schoop}}, \bibinfo
  {author} {\bibfnamefont {S.}~\bibnamefont {Dambach}}, \bibinfo {author}
  {\bibfnamefont {C.}~\bibnamefont {Padurariu}}, \bibinfo {author}
  {\bibfnamefont {B.}~\bibnamefont {Kubala}}, \bibinfo {author} {\bibfnamefont
  {J.}~\bibnamefont {Ankerhold}}, \bibinfo {author} {\bibfnamefont {C.~R.}\
  \bibnamefont {Ast}}, \ and\ \bibinfo {author} {\bibfnamefont
  {K.}~\bibnamefont {Kern}},\ }\href {\doibase 10.1103/PhysRevB.100.014502}
  {\bibfield  {journal} {\bibinfo  {journal} {Phys. Rev. B}\ }\textbf {\bibinfo
  {volume} {100}},\ \bibinfo {pages} {014502} (\bibinfo {year}
  {2019})}\BibitemShut {NoStop}%
\bibitem [{\citenamefont {Ji}\ \emph {et~al.}(2008)\citenamefont {Ji},
  \citenamefont {Zhang}, \citenamefont {Fu}, \citenamefont {Chen},
  \citenamefont {Ma}, \citenamefont {Li}, \citenamefont {Duan}, \citenamefont
  {Jia},\ and\ \citenamefont {Xue}}]{Ji2008}%
  \BibitemOpen
  \bibfield  {author} {\bibinfo {author} {\bibfnamefont {S.-H.}\ \bibnamefont
  {Ji}}, \bibinfo {author} {\bibfnamefont {T.}~\bibnamefont {Zhang}}, \bibinfo
  {author} {\bibfnamefont {Y.-S.}\ \bibnamefont {Fu}}, \bibinfo {author}
  {\bibfnamefont {X.}~\bibnamefont {Chen}}, \bibinfo {author} {\bibfnamefont
  {X.-C.}\ \bibnamefont {Ma}}, \bibinfo {author} {\bibfnamefont
  {J.}~\bibnamefont {Li}}, \bibinfo {author} {\bibfnamefont {W.-H.}\
  \bibnamefont {Duan}}, \bibinfo {author} {\bibfnamefont {J.-F.}\ \bibnamefont
  {Jia}}, \ and\ \bibinfo {author} {\bibfnamefont {Q.-K.}\ \bibnamefont
  {Xue}},\ }\href {\doibase 10.1103/PhysRevLett.100.226801} {\bibfield
  {journal} {\bibinfo  {journal} {Phys. Rev. Lett.}\ }\textbf {\bibinfo
  {volume} {100}},\ \bibinfo {pages} {226801} (\bibinfo {year}
  {2008})}\BibitemShut {NoStop}%
\bibitem [{\citenamefont {Ruby}\ \emph {et~al.}(2015)\citenamefont {Ruby},
  \citenamefont {Pientka}, \citenamefont {Peng}, \citenamefont {von Oppen},
  \citenamefont {Heinrich},\ and\ \citenamefont {Franke}}]{Ruby2015}%
  \BibitemOpen
  \bibfield  {author} {\bibinfo {author} {\bibfnamefont {M.}~\bibnamefont
  {Ruby}}, \bibinfo {author} {\bibfnamefont {F.}~\bibnamefont {Pientka}},
  \bibinfo {author} {\bibfnamefont {Y.}~\bibnamefont {Peng}}, \bibinfo {author}
  {\bibfnamefont {F.}~\bibnamefont {von Oppen}}, \bibinfo {author}
  {\bibfnamefont {B.~W.}\ \bibnamefont {Heinrich}}, \ and\ \bibinfo {author}
  {\bibfnamefont {K.~J.}\ \bibnamefont {Franke}},\ }\href {\doibase
  10.1103/PhysRevLett.115.087001} {\bibfield  {journal} {\bibinfo  {journal}
  {Phys. Rev. Lett.}\ }\textbf {\bibinfo {volume} {115}},\ \bibinfo {pages}
  {087001} (\bibinfo {year} {2015})}\BibitemShut {NoStop}%
\bibitem [{\citenamefont {Ruby}\ \emph {et~al.}(2016)\citenamefont {Ruby},
  \citenamefont {Peng}, \citenamefont {von Oppen}, \citenamefont {Heinrich},\
  and\ \citenamefont {Franke}}]{Ruby2016}%
  \BibitemOpen
  \bibfield  {author} {\bibinfo {author} {\bibfnamefont {M.}~\bibnamefont
  {Ruby}}, \bibinfo {author} {\bibfnamefont {Y.}~\bibnamefont {Peng}}, \bibinfo
  {author} {\bibfnamefont {F.}~\bibnamefont {von Oppen}}, \bibinfo {author}
  {\bibfnamefont {B.~W.}\ \bibnamefont {Heinrich}}, \ and\ \bibinfo {author}
  {\bibfnamefont {K.~J.}\ \bibnamefont {Franke}},\ }\href {\doibase
  10.1103/PhysRevLett.117.186801} {\bibfield  {journal} {\bibinfo  {journal}
  {Phys. Rev. Lett.}\ }\textbf {\bibinfo {volume} {117}},\ \bibinfo {pages}
  {186801} (\bibinfo {year} {2016})}\BibitemShut {NoStop}%
\bibitem [{\citenamefont {Choi}\ \emph {et~al.}(2017)\citenamefont {Choi},
  \citenamefont {Rubio-Verd{\'{u}}}, \citenamefont {{De Bruijckere}},
  \citenamefont {Ugeda}, \citenamefont {Lorente},\ and\ \citenamefont
  {Pascual}}]{Choi2017}%
  \BibitemOpen
  \bibfield  {author} {\bibinfo {author} {\bibfnamefont {D.-J.}\ \bibnamefont
  {Choi}}, \bibinfo {author} {\bibfnamefont {C.}~\bibnamefont
  {Rubio-Verd{\'{u}}}}, \bibinfo {author} {\bibfnamefont {J.}~\bibnamefont {{De
  Bruijckere}}}, \bibinfo {author} {\bibfnamefont {M.~M.}\ \bibnamefont
  {Ugeda}}, \bibinfo {author} {\bibfnamefont {N.}~\bibnamefont {Lorente}}, \
  and\ \bibinfo {author} {\bibfnamefont {J.~I.}\ \bibnamefont {Pascual}},\
  }\href {\doibase 10.1038/ncomms15175} {\bibfield  {journal} {\bibinfo
  {journal} {Nature Commun.}\ }\textbf {\bibinfo {volume} {8}},\ \bibinfo
  {pages} {15175} (\bibinfo {year} {2017})}\BibitemShut {NoStop}%
\bibitem [{\citenamefont {Cornils}\ \emph {et~al.}(2017)\citenamefont
  {Cornils}, \citenamefont {Kamlapure}, \citenamefont {Zhou}, \citenamefont
  {Pradhan}, \citenamefont {Khajetoorians}, \citenamefont {Fransson},
  \citenamefont {Wiebe},\ and\ \citenamefont {Wiesendanger}}]{Cornils2017}%
  \BibitemOpen
  \bibfield  {author} {\bibinfo {author} {\bibfnamefont {L.}~\bibnamefont
  {Cornils}}, \bibinfo {author} {\bibfnamefont {A.}~\bibnamefont {Kamlapure}},
  \bibinfo {author} {\bibfnamefont {L.}~\bibnamefont {Zhou}}, \bibinfo {author}
  {\bibfnamefont {S.}~\bibnamefont {Pradhan}}, \bibinfo {author} {\bibfnamefont
  {A.~A.}\ \bibnamefont {Khajetoorians}}, \bibinfo {author} {\bibfnamefont
  {J.}~\bibnamefont {Fransson}}, \bibinfo {author} {\bibfnamefont
  {J.}~\bibnamefont {Wiebe}}, \ and\ \bibinfo {author} {\bibfnamefont
  {R.}~\bibnamefont {Wiesendanger}},\ }\href {\doibase
  10.1103/PhysRevLett.119.197002} {\bibfield  {journal} {\bibinfo  {journal}
  {Phys. Rev. Lett.}\ }\textbf {\bibinfo {volume} {119}},\ \bibinfo {pages}
  {197002} (\bibinfo {year} {2017})}\BibitemShut {NoStop}%
\bibitem [{SM()}]{SM}%
  \BibitemOpen
  \href@noop {} {}\bibinfo {note} {Supporting Information}\BibitemShut
  {NoStop}%
\bibitem [{\citenamefont {Gye}\ \emph {et~al.}(2019)\citenamefont {Gye},
  \citenamefont {Oh},\ and\ \citenamefont {Yeom}}]{Gye2019}%
  \BibitemOpen
  \bibfield  {author} {\bibinfo {author} {\bibfnamefont {G.}~\bibnamefont
  {Gye}}, \bibinfo {author} {\bibfnamefont {E.}~\bibnamefont {Oh}}, \ and\
  \bibinfo {author} {\bibfnamefont {H.~W.}\ \bibnamefont {Yeom}},\ }\href
  {\doibase 10.1103/PhysRevLett.122.016403} {\bibfield  {journal} {\bibinfo
  {journal} {Phys. Rev. Lett.}\ }\textbf {\bibinfo {volume} {122}},\ \bibinfo
  {pages} {016403} (\bibinfo {year} {2019})}\BibitemShut {NoStop}%
\bibitem [{\citenamefont {Guster}\ \emph {et~al.}(2019)\citenamefont {Guster},
  \citenamefont {Rubio-Verd{\'{u}}}, \citenamefont {Robles}, \citenamefont
  {Zald{\'\i}var}, \citenamefont {Dreher}, \citenamefont {Pruneda},
  \citenamefont {Silva-Guill{\'{e}}n}, \citenamefont {Choi}, \citenamefont
  {Pascual}, \citenamefont {Ugeda}, \citenamefont {Ordej\'{o}n},\ and\
  \citenamefont {Canadell}}]{Guster2019}%
  \BibitemOpen
  \bibfield  {author} {\bibinfo {author} {\bibfnamefont {B.}~\bibnamefont
  {Guster}}, \bibinfo {author} {\bibfnamefont {C.}~\bibnamefont
  {Rubio-Verd{\'{u}}}}, \bibinfo {author} {\bibfnamefont {R.}~\bibnamefont
  {Robles}}, \bibinfo {author} {\bibfnamefont {J.}~\bibnamefont
  {Zald{\'\i}var}}, \bibinfo {author} {\bibfnamefont {P.}~\bibnamefont
  {Dreher}}, \bibinfo {author} {\bibfnamefont {M.}~\bibnamefont {Pruneda}},
  \bibinfo {author} {\bibfnamefont {J.~{\'{A}}.}\ \bibnamefont
  {Silva-Guill{\'{e}}n}}, \bibinfo {author} {\bibfnamefont {D.-J.}\
  \bibnamefont {Choi}}, \bibinfo {author} {\bibfnamefont {J.~I.}\ \bibnamefont
  {Pascual}}, \bibinfo {author} {\bibfnamefont {M.~M.}\ \bibnamefont {Ugeda}},
  \bibinfo {author} {\bibfnamefont {P.}~\bibnamefont {Ordej\'{o}n}}, \ and\
  \bibinfo {author} {\bibfnamefont {E.}~\bibnamefont {Canadell}},\ }\href
  {\doibase 10.1021/acs.nanolett.9b00268} {\bibfield  {journal} {\bibinfo
  {journal} {Nano Lett.}\ }\textbf {\bibinfo {volume} {19}},\ \bibinfo {pages}
  {3027} (\bibinfo {year} {2019})}\BibitemShut {NoStop}%
\bibitem [{SC1()}]{SC1}%
  \BibitemOpen
  \href@noop {} {}\bibinfo {note} {Areas with Nb-centered CDW maxima
  (metal-centered, MC) do not appear in the STM images, as they have higher
  energy \cite{Gye2019, Guster2019}. See SI \cite{SM}, especially Fig.\ S18 for
  further details.}\BibitemShut {Stop}%
\bibitem [{SC3()}]{SC3}%
  \BibitemOpen
  \href@noop {} {}\bibinfo {note} {In addition to the apparent height, both
  species also differ in the energies of their $d$-level resonances. We note
  that the apparent heights and $d$-level resonances are only slightly affected
  by the CDW. See SI \cite{SM} for more details.}\BibitemShut {Stop}%
\bibitem [{SC2()}]{SC2}%
  \BibitemOpen
  \href@noop {} {}\bibinfo {note} {The long-range oscillations are expected to
  have a period of half the Fermi wavelength in simple models of YSR states
  \cite{Rusinov1969}. The Fermi surface of \twoH\, consists of several sheets
  \cite{Yokoya2001, Kiss2007, Rahn2012, Rossnagel2001} which can lead to a more
  complex structure in the long-range scattering pattern of the YSR
  wavefunctions.}\BibitemShut {Stop}%
\bibitem [{\citenamefont {Salkola}\ \emph {et~al.}(1997)\citenamefont
  {Salkola}, \citenamefont {Balatsky},\ and\ \citenamefont
  {Schrieffer}}]{Salkola1997}%
  \BibitemOpen
  \bibfield  {author} {\bibinfo {author} {\bibfnamefont {M.~I.}\ \bibnamefont
  {Salkola}}, \bibinfo {author} {\bibfnamefont {A.~V.}\ \bibnamefont
  {Balatsky}}, \ and\ \bibinfo {author} {\bibfnamefont {J.~R.}\ \bibnamefont
  {Schrieffer}},\ }\href {\doibase 10.1103/PhysRevB.55.12648} {\bibfield
  {journal} {\bibinfo  {journal} {Phys. Rev. B}\ }\textbf {\bibinfo {volume}
  {55}},\ \bibinfo {pages} {12648} (\bibinfo {year} {1997})}\BibitemShut
  {NoStop}%
\bibitem [{\citenamefont {Flatt{\'{e}}}\ and\ \citenamefont
  {Byers}(1997)}]{Flatte1997PRL}%
  \BibitemOpen
  \bibfield  {author} {\bibinfo {author} {\bibfnamefont {M.~E.}\ \bibnamefont
  {Flatt{\'{e}}}}\ and\ \bibinfo {author} {\bibfnamefont {J.~M.}\ \bibnamefont
  {Byers}},\ }\href {\doibase 10.1103/PhysRevLett.78.3761} {\bibfield
  {journal} {\bibinfo  {journal} {Phys. Rev. Lett.}\ }\textbf {\bibinfo
  {volume} {78}},\ \bibinfo {pages} {3761} (\bibinfo {year}
  {1997})}\BibitemShut {NoStop}%
\bibitem [{\citenamefont {Franke}\ \emph {et~al.}(2011)\citenamefont {Franke},
  \citenamefont {Schulze},\ and\ \citenamefont {Pascual}}]{Franke2011}%
  \BibitemOpen
  \bibfield  {author} {\bibinfo {author} {\bibfnamefont {K.~J.}\ \bibnamefont
  {Franke}}, \bibinfo {author} {\bibfnamefont {G.}~\bibnamefont {Schulze}}, \
  and\ \bibinfo {author} {\bibfnamefont {J.~I.}\ \bibnamefont {Pascual}},\
  }\href {\doibase 10.1126/science.1202204} {\bibfield  {journal} {\bibinfo
  {journal} {Science}\ }\textbf {\bibinfo {volume} {332}},\ \bibinfo {pages}
  {940} (\bibinfo {year} {2011})}\BibitemShut {NoStop}%
\bibitem [{\citenamefont {Farinacci}\ \emph {et~al.}(2018)\citenamefont
  {Farinacci}, \citenamefont {Ahmadi}, \citenamefont {Reecht}, \citenamefont
  {Ruby}, \citenamefont {Bogdanoff}, \citenamefont {Peters}, \citenamefont
  {Heinrich}, \citenamefont {von Oppen},\ and\ \citenamefont
  {Franke}}]{Farinacci2018}%
  \BibitemOpen
  \bibfield  {author} {\bibinfo {author} {\bibfnamefont {L.}~\bibnamefont
  {Farinacci}}, \bibinfo {author} {\bibfnamefont {G.}~\bibnamefont {Ahmadi}},
  \bibinfo {author} {\bibfnamefont {G.}~\bibnamefont {Reecht}}, \bibinfo
  {author} {\bibfnamefont {M.}~\bibnamefont {Ruby}}, \bibinfo {author}
  {\bibfnamefont {N.}~\bibnamefont {Bogdanoff}}, \bibinfo {author}
  {\bibfnamefont {O.}~\bibnamefont {Peters}}, \bibinfo {author} {\bibfnamefont
  {B.~W.}\ \bibnamefont {Heinrich}}, \bibinfo {author} {\bibfnamefont
  {F.}~\bibnamefont {von Oppen}}, \ and\ \bibinfo {author} {\bibfnamefont
  {K.~J.}\ \bibnamefont {Franke}},\ }\href {\doibase
  10.1103/PhysRevLett.121.196803} {\bibfield  {journal} {\bibinfo  {journal}
  {Phys. Rev. Lett.}\ }\textbf {\bibinfo {volume} {121}},\ \bibinfo {pages}
  {196803} (\bibinfo {year} {2018})}\BibitemShut {NoStop}%
\bibitem [{\citenamefont {Malavolti}\ \emph {et~al.}(2018)\citenamefont
  {Malavolti}, \citenamefont {Briganti}, \citenamefont {H{\"{a}}nze},
  \citenamefont {Serrano}, \citenamefont {Cimatti}, \citenamefont {McMurtrie},
  \citenamefont {Otero}, \citenamefont {Ohresser}, \citenamefont {Totti},
  \citenamefont {Mannini}, \citenamefont {Sessoli},\ and\ \citenamefont
  {Loth}}]{Malavolti2018}%
  \BibitemOpen
  \bibfield  {author} {\bibinfo {author} {\bibfnamefont {L.}~\bibnamefont
  {Malavolti}}, \bibinfo {author} {\bibfnamefont {M.}~\bibnamefont {Briganti}},
  \bibinfo {author} {\bibfnamefont {M.}~\bibnamefont {H{\"{a}}nze}}, \bibinfo
  {author} {\bibfnamefont {G.}~\bibnamefont {Serrano}}, \bibinfo {author}
  {\bibfnamefont {I.}~\bibnamefont {Cimatti}}, \bibinfo {author} {\bibfnamefont
  {G.}~\bibnamefont {McMurtrie}}, \bibinfo {author} {\bibfnamefont
  {E.}~\bibnamefont {Otero}}, \bibinfo {author} {\bibfnamefont
  {P.}~\bibnamefont {Ohresser}}, \bibinfo {author} {\bibfnamefont
  {F.}~\bibnamefont {Totti}}, \bibinfo {author} {\bibfnamefont
  {M.}~\bibnamefont {Mannini}}, \bibinfo {author} {\bibfnamefont
  {R.}~\bibnamefont {Sessoli}}, \ and\ \bibinfo {author} {\bibfnamefont
  {S.}~\bibnamefont {Loth}},\ }\href {\doibase 10.1021/acs.nanolett.8b03921}
  {\bibfield  {journal} {\bibinfo  {journal} {Nano Lett.}\ }\textbf {\bibinfo
  {volume} {18}},\ \bibinfo {pages} {7955} (\bibinfo {year}
  {2018})}\BibitemShut {NoStop}%
\bibitem [{Note1()}]{Note1}%
  \BibitemOpen
  \bibinfo {note} {Note that the experimental adsorption site falls in the
  middle of a Nb triangle. Such an impurity has equal couplings to the three
  neighboring Nb atoms. When modeling the impurity as a classical impurity,
  this easily leads to spurious YSR states which would not be present for a
  quantum impurity. In view of the fact that our treatment of the impurity is
  highly simplified in any case, we avoid this problem by placing the impurity
  on a Nb site.}\BibitemShut {Stop}%
\end{thebibliography}

\begin{thebibliography}{31}%
\makeatletter
\providecommand \@ifxundefined [1]{%
 \@ifx{#1\undefined}
}%
\providecommand \@ifnum [1]{%
 \ifnum #1\expandafter \@firstoftwo
 \else \expandafter \@secondoftwo
 \fi
}%
\providecommand \@ifx [1]{%
 \ifx #1\expandafter \@firstoftwo
 \else \expandafter \@secondoftwo
 \fi
}%
\providecommand \natexlab [1]{#1}%
\providecommand \enquote  [1]{``#1''}%
\providecommand \bibnamefont  [1]{#1}%
\providecommand \bibfnamefont [1]{#1}%
\providecommand \citenamefont [1]{#1}%
\providecommand \href@noop [0]{\@secondoftwo}%
\providecommand \href [0]{\begingroup \@sanitize@url \@href}%
\providecommand \@href[1]{\@@startlink{#1}\@@href}%
\providecommand \@@href[1]{\endgroup#1\@@endlink}%
\providecommand \@sanitize@url [0]{\catcode `\\12\catcode `\$12\catcode
  `\&12\catcode `\#12\catcode `\^12\catcode `\_12\catcode `\%12\relax}%
\providecommand \@@startlink[1]{}%
\providecommand \@@endlink[0]{}%
\providecommand \url  [0]{\begingroup\@sanitize@url \@url }%
\providecommand \@url [1]{\endgroup\@href {#1}{\urlprefix }}%
\providecommand \urlprefix  [0]{URL }%
\providecommand \Eprint [0]{\href }%
\providecommand \doibase [0]{http://dx.doi.org/}%
\providecommand \selectlanguage [0]{\@gobble}%
\providecommand \bibinfo  [0]{\@secondoftwo}%
\providecommand \bibfield  [0]{\@secondoftwo}%
\providecommand \translation [1]{[#1]}%
\providecommand \BibitemOpen [0]{}%
\providecommand \bibitemStop [0]{}%
\providecommand \bibitemNoStop [0]{.\EOS\space}%
\providecommand \EOS [0]{\spacefactor3000\relax}%
\providecommand \BibitemShut  [1]{\csname bibitem#1\endcsname}%
\let\auto@bib@innerbib\@empty
%</preamble>
\bibitem [{\citenamefont {Gr\"uner}(1994)}]{SGruner2018}%
  \BibitemOpen
  \bibfield  {author} {\bibinfo {author} {\bibfnamefont {G.}~\bibnamefont
  {Gr\"uner}},\ }\href {\doibase 10.1201/9780429501012} {\emph {\bibinfo
  {title} {Density waves in solids}}}\ (\bibinfo  {publisher} {CRC Press},\
  \bibinfo {year} {1994})\BibitemShut {NoStop}%
\bibitem [{\citenamefont {Smith}\ \emph {et~al.}(1985)\citenamefont {Smith},
  \citenamefont {Kevan},\ and\ \citenamefont {DiSalvo}}]{SSmith1985}%
  \BibitemOpen
  \bibfield  {author} {\bibinfo {author} {\bibfnamefont {N.~V.}\ \bibnamefont
  {Smith}}, \bibinfo {author} {\bibfnamefont {S.~D.}\ \bibnamefont {Kevan}}, \
  and\ \bibinfo {author} {\bibfnamefont {F.~J.}\ \bibnamefont {DiSalvo}},\
  }\href {\doibase 10.1088/0022-3719/18/16/013} {\bibfield  {journal} {\bibinfo
   {journal} {J. Phys. C: Solid State Phys.}\ }\textbf {\bibinfo {volume}
  {18}},\ \bibinfo {pages} {3175} (\bibinfo {year} {1985})}\BibitemShut
  {NoStop}%
\bibitem [{\citenamefont {Rossnagel}\ \emph {et~al.}(2005)\citenamefont
  {Rossnagel}, \citenamefont {Rotenberg}, \citenamefont {Koh}, \citenamefont
  {Smith},\ and\ \citenamefont {Kipp}}]{SRossnagel2005}%
  \BibitemOpen
  \bibfield  {author} {\bibinfo {author} {\bibfnamefont {K.}~\bibnamefont
  {Rossnagel}}, \bibinfo {author} {\bibfnamefont {E.}~\bibnamefont
  {Rotenberg}}, \bibinfo {author} {\bibfnamefont {H.}~\bibnamefont {Koh}},
  \bibinfo {author} {\bibfnamefont {N.~V.}\ \bibnamefont {Smith}}, \ and\
  \bibinfo {author} {\bibfnamefont {L.}~\bibnamefont {Kipp}},\ }\href {\doibase
  10.1103/PhysRevB.72.121103} {\bibfield  {journal} {\bibinfo  {journal} {Phys.
  Rev. B}\ }\textbf {\bibinfo {volume} {72}},\ \bibinfo {pages} {121103}
  (\bibinfo {year} {2005})}\BibitemShut {NoStop}%
\bibitem [{\citenamefont {Inosov}\ \emph {et~al.}(2008)\citenamefont {Inosov},
  \citenamefont {Zabolotnyy}, \citenamefont {Evtushinsky}, \citenamefont
  {Kordyuk}, \citenamefont {B\"uchner}, \citenamefont {Follath}, \citenamefont
  {Berger},\ and\ \citenamefont {Borisenko}}]{SInosov2008}%
  \BibitemOpen
  \bibfield  {author} {\bibinfo {author} {\bibfnamefont {D.~S.}\ \bibnamefont
  {Inosov}}, \bibinfo {author} {\bibfnamefont {V.~B.}\ \bibnamefont
  {Zabolotnyy}}, \bibinfo {author} {\bibfnamefont {D.~V.}\ \bibnamefont
  {Evtushinsky}}, \bibinfo {author} {\bibfnamefont {A.~A.}\ \bibnamefont
  {Kordyuk}}, \bibinfo {author} {\bibfnamefont {B.}~\bibnamefont {B\"uchner}},
  \bibinfo {author} {\bibfnamefont {R.}~\bibnamefont {Follath}}, \bibinfo
  {author} {\bibfnamefont {H.}~\bibnamefont {Berger}}, \ and\ \bibinfo {author}
  {\bibfnamefont {S.~V.}\ \bibnamefont {Borisenko}},\ }\href {\doibase
  10.1088/1367-2630/10/12/125027} {\bibfield  {journal} {\bibinfo  {journal}
  {New J. Phys.}\ }\textbf {\bibinfo {volume} {10}},\ \bibinfo {pages} {125027}
  (\bibinfo {year} {2008})}\BibitemShut {NoStop}%
\bibitem [{\citenamefont {Inosov}\ \emph {et~al.}(2009)\citenamefont {Inosov},
  \citenamefont {Evtushinsky}, \citenamefont {Zabolotnyy}, \citenamefont
  {Kordyuk}, \citenamefont {B\"uchner}, \citenamefont {Follath}, \citenamefont
  {Berger},\ and\ \citenamefont {Borisenko}}]{SInosov2009}%
  \BibitemOpen
  \bibfield  {author} {\bibinfo {author} {\bibfnamefont {D.~S.}\ \bibnamefont
  {Inosov}}, \bibinfo {author} {\bibfnamefont {D.~V.}\ \bibnamefont
  {Evtushinsky}}, \bibinfo {author} {\bibfnamefont {V.~B.}\ \bibnamefont
  {Zabolotnyy}}, \bibinfo {author} {\bibfnamefont {A.~A.}\ \bibnamefont
  {Kordyuk}}, \bibinfo {author} {\bibfnamefont {B.}~\bibnamefont {B\"uchner}},
  \bibinfo {author} {\bibfnamefont {R.}~\bibnamefont {Follath}}, \bibinfo
  {author} {\bibfnamefont {H.}~\bibnamefont {Berger}}, \ and\ \bibinfo {author}
  {\bibfnamefont {S.~V.}\ \bibnamefont {Borisenko}},\ }\href {\doibase
  10.1103/PhysRevB.79.125112} {\bibfield  {journal} {\bibinfo  {journal} {Phys.
  Rev. B}\ }\textbf {\bibinfo {volume} {79}},\ \bibinfo {pages} {125112}
  (\bibinfo {year} {2009})}\BibitemShut {NoStop}%
\bibitem [{\citenamefont {Rahn}\ \emph {et~al.}(2012)\citenamefont {Rahn},
  \citenamefont {Hellmann}, \citenamefont {Kall\"ane}, \citenamefont {Sohrt},
  \citenamefont {Kim}, \citenamefont {Kipp},\ and\ \citenamefont
  {Rossnagel}}]{SRahn2012}%
  \BibitemOpen
  \bibfield  {author} {\bibinfo {author} {\bibfnamefont {D.~J.}\ \bibnamefont
  {Rahn}}, \bibinfo {author} {\bibfnamefont {S.}~\bibnamefont {Hellmann}},
  \bibinfo {author} {\bibfnamefont {M.}~\bibnamefont {Kall\"ane}}, \bibinfo
  {author} {\bibfnamefont {C.}~\bibnamefont {Sohrt}}, \bibinfo {author}
  {\bibfnamefont {T.~K.}\ \bibnamefont {Kim}}, \bibinfo {author} {\bibfnamefont
  {L.}~\bibnamefont {Kipp}}, \ and\ \bibinfo {author} {\bibfnamefont
  {K.}~\bibnamefont {Rossnagel}},\ }\href {\doibase 10.1103/PhysRevB.85.224532}
  {\bibfield  {journal} {\bibinfo  {journal} {Phys. Rev. B}\ }\textbf {\bibinfo
  {volume} {85}},\ \bibinfo {pages} {224532} (\bibinfo {year}
  {2012})}\BibitemShut {NoStop}%
\bibitem [{\citenamefont {M{\'{e}}nard}\ \emph {et~al.}(2015)\citenamefont
  {M{\'{e}}nard}, \citenamefont {Guissart}, \citenamefont {Brun}, \citenamefont
  {Pons}, \citenamefont {Stolyarov}, \citenamefont {Debontridder},
  \citenamefont {Leclerc}, \citenamefont {Janod}, \citenamefont {Cario},
  \citenamefont {Roditchev}, \citenamefont {Simon},\ and\ \citenamefont
  {Cren}}]{SMenard2015}%
  \BibitemOpen
  \bibfield  {author} {\bibinfo {author} {\bibfnamefont {G.~C.}\ \bibnamefont
  {M{\'{e}}nard}}, \bibinfo {author} {\bibfnamefont {S.}~\bibnamefont
  {Guissart}}, \bibinfo {author} {\bibfnamefont {C.}~\bibnamefont {Brun}},
  \bibinfo {author} {\bibfnamefont {S.}~\bibnamefont {Pons}}, \bibinfo {author}
  {\bibfnamefont {V.~S.}\ \bibnamefont {Stolyarov}}, \bibinfo {author}
  {\bibfnamefont {F.}~\bibnamefont {Debontridder}}, \bibinfo {author}
  {\bibfnamefont {M.~V.}\ \bibnamefont {Leclerc}}, \bibinfo {author}
  {\bibfnamefont {E.}~\bibnamefont {Janod}}, \bibinfo {author} {\bibfnamefont
  {L.}~\bibnamefont {Cario}}, \bibinfo {author} {\bibfnamefont
  {D.}~\bibnamefont {Roditchev}}, \bibinfo {author} {\bibfnamefont
  {P.}~\bibnamefont {Simon}}, \ and\ \bibinfo {author} {\bibfnamefont
  {T.}~\bibnamefont {Cren}},\ }\href {\doibase 10.1038/nphys3508} {\bibfield
  {journal} {\bibinfo  {journal} {Nature Phys.}\ }\textbf {\bibinfo {volume}
  {11}},\ \bibinfo {pages} {1013} (\bibinfo {year} {2015})}\BibitemShut
  {NoStop}%
\bibitem [{\citenamefont {Bawden}\ \emph {et~al.}(2016)\citenamefont {Bawden},
  \citenamefont {Cooil}, \citenamefont {Mazzola}, \citenamefont {Riley},
  \citenamefont {Collins-McIntyre}, \citenamefont {Sunko}, \citenamefont
  {Hunvik}, \citenamefont {Leandersson}, \citenamefont {Polley}, \citenamefont
  {Balasubramanian}, \citenamefont {Kim}, \citenamefont {Hoesch}, \citenamefont
  {Wells}, \citenamefont {Balakrishnan}, \citenamefont {Bahramy},\ and\
  \citenamefont {King}}]{SBawden2016}%
  \BibitemOpen
  \bibfield  {author} {\bibinfo {author} {\bibfnamefont {L.}~\bibnamefont
  {Bawden}}, \bibinfo {author} {\bibfnamefont {S.~P.}\ \bibnamefont {Cooil}},
  \bibinfo {author} {\bibfnamefont {F.}~\bibnamefont {Mazzola}}, \bibinfo
  {author} {\bibfnamefont {J.~M.}\ \bibnamefont {Riley}}, \bibinfo {author}
  {\bibfnamefont {L.~J.}\ \bibnamefont {Collins-McIntyre}}, \bibinfo {author}
  {\bibfnamefont {V.}~\bibnamefont {Sunko}}, \bibinfo {author} {\bibfnamefont
  {K.~W.~B.}\ \bibnamefont {Hunvik}}, \bibinfo {author} {\bibfnamefont
  {M.}~\bibnamefont {Leandersson}}, \bibinfo {author} {\bibfnamefont {C.~M.}\
  \bibnamefont {Polley}}, \bibinfo {author} {\bibfnamefont {T.}~\bibnamefont
  {Balasubramanian}}, \bibinfo {author} {\bibfnamefont {T.~K.}\ \bibnamefont
  {Kim}}, \bibinfo {author} {\bibfnamefont {M.}~\bibnamefont {Hoesch}},
  \bibinfo {author} {\bibfnamefont {J.~W.}\ \bibnamefont {Wells}}, \bibinfo
  {author} {\bibfnamefont {G.}~\bibnamefont {Balakrishnan}}, \bibinfo {author}
  {\bibfnamefont {M.~S.}\ \bibnamefont {Bahramy}}, \ and\ \bibinfo {author}
  {\bibfnamefont {P.~D.~C.}\ \bibnamefont {King}},\ }\href {\doibase
  10.1038/ncomms11711} {\bibfield  {journal} {\bibinfo  {journal} {Nature
  Comm.}\ }\textbf {\bibinfo {volume} {7}},\ \bibinfo {pages} {11711} (\bibinfo
  {year} {2016})}\BibitemShut {NoStop}%
\bibitem [{\citenamefont {Xi}\ \emph {et~al.}(2016)\citenamefont {Xi},
  \citenamefont {Wang}, \citenamefont {Zhao}, \citenamefont {Park},
  \citenamefont {Law}, \citenamefont {Berger}, \citenamefont {Forr{\'{o}}},
  \citenamefont {Shan},\ and\ \citenamefont {Mak}}]{SXi2016}%
  \BibitemOpen
  \bibfield  {author} {\bibinfo {author} {\bibfnamefont {X.}~\bibnamefont
  {Xi}}, \bibinfo {author} {\bibfnamefont {Z.}~\bibnamefont {Wang}}, \bibinfo
  {author} {\bibfnamefont {W.}~\bibnamefont {Zhao}}, \bibinfo {author}
  {\bibfnamefont {J.-H.}\ \bibnamefont {Park}}, \bibinfo {author}
  {\bibfnamefont {K.~T.}\ \bibnamefont {Law}}, \bibinfo {author} {\bibfnamefont
  {H.}~\bibnamefont {Berger}}, \bibinfo {author} {\bibfnamefont
  {L.}~\bibnamefont {Forr{\'{o}}}}, \bibinfo {author} {\bibfnamefont
  {J.}~\bibnamefont {Shan}}, \ and\ \bibinfo {author} {\bibfnamefont {K.~F.}\
  \bibnamefont {Mak}},\ }\href {\doibase 10.1038/nphys3538} {\bibfield
  {journal} {\bibinfo  {journal} {Nature Phys.}\ }\textbf {\bibinfo {volume}
  {12}},\ \bibinfo {pages} {139} (\bibinfo {year} {2016})}\BibitemShut
  {NoStop}%
\bibitem [{\citenamefont {Hewson}(1993)}]{SHewson1993}%
  \BibitemOpen
  \bibfield  {author} {\bibinfo {author} {\bibfnamefont {A.~C.}\ \bibnamefont
  {Hewson}},\ }\href {\doibase 10.1201/9780429501012} {\emph {\bibinfo {title}
  {The Kondo Problem to Heavy Fermions}}}\ (\bibinfo  {publisher} {Cambridge
  University Press},\ \bibinfo {year} {1993})\BibitemShut {NoStop}%
\bibitem [{\citenamefont {Yu}(1965)}]{SYu1965}%
  \BibitemOpen
  \bibfield  {author} {\bibinfo {author} {\bibfnamefont {L.}~\bibnamefont
  {Yu}},\ }\href {\doibase 10.7498/aps.21.75} {\bibfield  {journal} {\bibinfo
  {journal} {Acta Phys. Sin.}\ }\textbf {\bibinfo {volume} {21}},\ \bibinfo
  {pages} {75} (\bibinfo {year} {1965})}\BibitemShut {NoStop}%
\bibitem [{\citenamefont {Shiba}(1968{\natexlab{a}})}]{SShiba1968}%
  \BibitemOpen
  \bibfield  {author} {\bibinfo {author} {\bibfnamefont {H.}~\bibnamefont
  {Shiba}},\ }\href {\doibase 10.1143/PTP.40.435} {\bibfield  {journal}
  {\bibinfo  {journal} {Prog. Theor. Phys.}\ }\textbf {\bibinfo {volume}
  {40}},\ \bibinfo {pages} {435} (\bibinfo {year}
  {1968}{\natexlab{a}})}\BibitemShut {NoStop}%
\bibitem [{\citenamefont {Rusinov}(1969)}]{SRusinov1969}%
  \BibitemOpen
  \bibfield  {author} {\bibinfo {author} {\bibfnamefont {A.~I.}\ \bibnamefont
  {Rusinov}},\ }\href
  {http://www.jetpletters.ac.ru/ps/1658/article{\_}25295.shtml} {\bibfield
  {journal} {\bibinfo  {journal} {JETP Lett.}\ }\textbf {\bibinfo {volume}
  {9}},\ \bibinfo {pages} {85} (\bibinfo {year} {1969})}\BibitemShut {NoStop}%
\bibitem [{\citenamefont {Moca}\ \emph {et~al.}(2008)\citenamefont {Moca},
  \citenamefont {Demler}, \citenamefont {Jank\'o},\ and\ \citenamefont
  {Zar\'and}}]{SMoca2008}%
  \BibitemOpen
  \bibfield  {author} {\bibinfo {author} {\bibfnamefont {C.~P.}\ \bibnamefont
  {Moca}}, \bibinfo {author} {\bibfnamefont {E.}~\bibnamefont {Demler}},
  \bibinfo {author} {\bibfnamefont {B.}~\bibnamefont {Jank\'o}}, \ and\
  \bibinfo {author} {\bibfnamefont {G.}~\bibnamefont {Zar\'and}},\ }\href
  {\doibase 10.1103/PhysRevB.77.174516} {\bibfield  {journal} {\bibinfo
  {journal} {Phys. Rev. B}\ }\textbf {\bibinfo {volume} {77}},\ \bibinfo
  {pages} {174516} (\bibinfo {year} {2008})}\BibitemShut {NoStop}%
\bibitem [{\citenamefont {Kir\u{s}anskas}\ \emph {et~al.}(2015)\citenamefont
  {Kir\u{s}anskas}, \citenamefont {Goldstein}, \citenamefont {Flensberg},
  \citenamefont {Glazman},\ and\ \citenamefont {Paaske}}]{SKirsanskas2015}%
  \BibitemOpen
  \bibfield  {author} {\bibinfo {author} {\bibfnamefont {G.}~\bibnamefont
  {Kir\u{s}anskas}}, \bibinfo {author} {\bibfnamefont {M.}~\bibnamefont
  {Goldstein}}, \bibinfo {author} {\bibfnamefont {K.}~\bibnamefont
  {Flensberg}}, \bibinfo {author} {\bibfnamefont {L.~I.}\ \bibnamefont
  {Glazman}}, \ and\ \bibinfo {author} {\bibfnamefont {J.}~\bibnamefont
  {Paaske}},\ }\href {\doibase 10.1103/PhysRevB.92.235422} {\bibfield
  {journal} {\bibinfo  {journal} {Phys. Rev. B}\ }\textbf {\bibinfo {volume}
  {92}},\ \bibinfo {pages} {235422} (\bibinfo {year} {2015})}\BibitemShut
  {NoStop}%
\bibitem [{\citenamefont {Shiba}(1968{\natexlab{b}})}]{SShiba1968a}%
  \BibitemOpen
  \bibfield  {author} {\bibinfo {author} {\bibfnamefont {H.}~\bibnamefont
  {Shiba}},\ }\href {\doibase 10.1143/PTP.40.435} {\bibfield  {journal}
  {\bibinfo  {journal} {Prog. Theor. Phys.}\ }\textbf {\bibinfo {volume}
  {40}},\ \bibinfo {pages} {435} (\bibinfo {year}
  {1968}{\natexlab{b}})}\BibitemShut {NoStop}%
\bibitem [{\citenamefont {Ruby}\ \emph
  {et~al.}(2015{\natexlab{a}})\citenamefont {Ruby}, \citenamefont {Pientka},
  \citenamefont {Peng}, \citenamefont {von Oppen}, \citenamefont {Heinrich},\
  and\ \citenamefont {Franke}}]{SRuby2015}%
  \BibitemOpen
  \bibfield  {author} {\bibinfo {author} {\bibfnamefont {M.}~\bibnamefont
  {Ruby}}, \bibinfo {author} {\bibfnamefont {F.}~\bibnamefont {Pientka}},
  \bibinfo {author} {\bibfnamefont {Y.}~\bibnamefont {Peng}}, \bibinfo {author}
  {\bibfnamefont {F.}~\bibnamefont {von Oppen}}, \bibinfo {author}
  {\bibfnamefont {B.~W.}\ \bibnamefont {Heinrich}}, \ and\ \bibinfo {author}
  {\bibfnamefont {K.~J.}\ \bibnamefont {Franke}},\ }\href {\doibase
  10.1103/PhysRevLett.115.087001} {\bibfield  {journal} {\bibinfo  {journal}
  {Phys. Rev. Lett.}\ }\textbf {\bibinfo {volume} {115}},\ \bibinfo {pages}
  {087001} (\bibinfo {year} {2015}{\natexlab{a}})}\BibitemShut {NoStop}%
\bibitem [{\citenamefont {Tersoff}\ and\ \citenamefont
  {Hamann}(1985)}]{STersoff1985}%
  \BibitemOpen
  \bibfield  {author} {\bibinfo {author} {\bibfnamefont {J.}~\bibnamefont
  {Tersoff}}\ and\ \bibinfo {author} {\bibfnamefont {D.~R.}\ \bibnamefont
  {Hamann}},\ }\href {\doibase 10.1103/PhysRevB.31.805} {\bibfield  {journal}
  {\bibinfo  {journal} {Phys. Rev. B}\ }\textbf {\bibinfo {volume} {31}},\
  \bibinfo {pages} {805} (\bibinfo {year} {1985})}\BibitemShut {NoStop}%
\bibitem [{\citenamefont {Ruby}\ \emph
  {et~al.}(2015{\natexlab{b}})\citenamefont {Ruby}, \citenamefont {Heinrich},
  \citenamefont {Pascual},\ and\ \citenamefont {Franke}}]{SRubyPb15}%
  \BibitemOpen
  \bibfield  {author} {\bibinfo {author} {\bibfnamefont {M.}~\bibnamefont
  {Ruby}}, \bibinfo {author} {\bibfnamefont {B.~W.}\ \bibnamefont {Heinrich}},
  \bibinfo {author} {\bibfnamefont {J.~I.}\ \bibnamefont {Pascual}}, \ and\
  \bibinfo {author} {\bibfnamefont {K.~J.}\ \bibnamefont {Franke}},\ }\href
  {\doibase 10.1103/PhysRevLett.114.157001} {\bibfield  {journal} {\bibinfo
  {journal} {Phys. Rev. Lett.}\ }\textbf {\bibinfo {volume} {114}},\ \bibinfo
  {pages} {157001} (\bibinfo {year} {2015}{\natexlab{b}})}\BibitemShut
  {NoStop}%
\bibitem [{\citenamefont {Choi}\ \emph {et~al.}(2017)\citenamefont {Choi},
  \citenamefont {Rubio-Verd{\'{u}}}, \citenamefont {{De Bruijckere}},
  \citenamefont {Ugeda}, \citenamefont {Lorente},\ and\ \citenamefont
  {Pascual}}]{SChoi2017}%
  \BibitemOpen
  \bibfield  {author} {\bibinfo {author} {\bibfnamefont {D.-J.}\ \bibnamefont
  {Choi}}, \bibinfo {author} {\bibfnamefont {C.}~\bibnamefont
  {Rubio-Verd{\'{u}}}}, \bibinfo {author} {\bibfnamefont {J.}~\bibnamefont {{De
  Bruijckere}}}, \bibinfo {author} {\bibfnamefont {M.~M.}\ \bibnamefont
  {Ugeda}}, \bibinfo {author} {\bibfnamefont {N.}~\bibnamefont {Lorente}}, \
  and\ \bibinfo {author} {\bibfnamefont {J.~I.}\ \bibnamefont {Pascual}},\
  }\href {\doibase 10.1038/ncomms15175} {\bibfield  {journal} {\bibinfo
  {journal} {Nature Commun.}\ }\textbf {\bibinfo {volume} {8}},\ \bibinfo
  {pages} {15175} (\bibinfo {year} {2017})}\BibitemShut {NoStop}%
\bibitem [{\citenamefont {Soumyanarayanan}\ \emph {et~al.}(2013)\citenamefont
  {Soumyanarayanan}, \citenamefont {Yee}, \citenamefont {He}, \citenamefont
  {van Wezel}, \citenamefont {Rahn}, \citenamefont {Rossnagel}, \citenamefont
  {Hudson}, \citenamefont {Norman},\ and\ \citenamefont
  {Hoffman}}]{SSoumyanarayanan2013}%
  \BibitemOpen
  \bibfield  {author} {\bibinfo {author} {\bibfnamefont {A.}~\bibnamefont
  {Soumyanarayanan}}, \bibinfo {author} {\bibfnamefont {M.~M.}\ \bibnamefont
  {Yee}}, \bibinfo {author} {\bibfnamefont {Y.}~\bibnamefont {He}}, \bibinfo
  {author} {\bibfnamefont {J.}~\bibnamefont {van Wezel}}, \bibinfo {author}
  {\bibfnamefont {D.~J.}\ \bibnamefont {Rahn}}, \bibinfo {author}
  {\bibfnamefont {K.}~\bibnamefont {Rossnagel}}, \bibinfo {author}
  {\bibfnamefont {E.~W.}\ \bibnamefont {Hudson}}, \bibinfo {author}
  {\bibfnamefont {M.~R.}\ \bibnamefont {Norman}}, \ and\ \bibinfo {author}
  {\bibfnamefont {J.~E.}\ \bibnamefont {Hoffman}},\ }\href {\doibase
  10.1073/pnas.1211387110} {\bibfield  {journal} {\bibinfo  {journal} {Proc.
  Natl. Acad. Sci.}\ }\textbf {\bibinfo {volume} {110}},\ \bibinfo {pages}
  {1623} (\bibinfo {year} {2013})}\BibitemShut {NoStop}%
\bibitem [{\citenamefont {Gye}\ \emph {et~al.}(2019)\citenamefont {Gye},
  \citenamefont {Oh},\ and\ \citenamefont {Yeom}}]{SGye2019}%
  \BibitemOpen
  \bibfield  {author} {\bibinfo {author} {\bibfnamefont {G.}~\bibnamefont
  {Gye}}, \bibinfo {author} {\bibfnamefont {E.}~\bibnamefont {Oh}}, \ and\
  \bibinfo {author} {\bibfnamefont {H.~W.}\ \bibnamefont {Yeom}},\ }\href
  {\doibase 10.1103/PhysRevLett.122.016403} {\bibfield  {journal} {\bibinfo
  {journal} {Phys. Rev. Lett.}\ }\textbf {\bibinfo {volume} {122}},\ \bibinfo
  {pages} {016403} (\bibinfo {year} {2019})}\BibitemShut {NoStop}%
\bibitem [{\citenamefont {Guster}\ \emph {et~al.}(2019)\citenamefont {Guster},
  \citenamefont {Rubio-Verd{\'{u}}}, \citenamefont {Robles}, \citenamefont
  {Zald{\'\i}var}, \citenamefont {Dreher}, \citenamefont {Pruneda},
  \citenamefont {Silva-Guill{\'{e}}n}, \citenamefont {Choi}, \citenamefont
  {Pascual}, \citenamefont {Ugeda}, \citenamefont {Ordej\'{o}n},\ and\
  \citenamefont {Canadell}}]{SGuster2019}%
  \BibitemOpen
  \bibfield  {author} {\bibinfo {author} {\bibfnamefont {B.}~\bibnamefont
  {Guster}}, \bibinfo {author} {\bibfnamefont {C.}~\bibnamefont
  {Rubio-Verd{\'{u}}}}, \bibinfo {author} {\bibfnamefont {R.}~\bibnamefont
  {Robles}}, \bibinfo {author} {\bibfnamefont {J.}~\bibnamefont
  {Zald{\'\i}var}}, \bibinfo {author} {\bibfnamefont {P.}~\bibnamefont
  {Dreher}}, \bibinfo {author} {\bibfnamefont {M.}~\bibnamefont {Pruneda}},
  \bibinfo {author} {\bibfnamefont {J.~{\'{A}}.}\ \bibnamefont
  {Silva-Guill{\'{e}}n}}, \bibinfo {author} {\bibfnamefont {D.-J.}\
  \bibnamefont {Choi}}, \bibinfo {author} {\bibfnamefont {J.~I.}\ \bibnamefont
  {Pascual}}, \bibinfo {author} {\bibfnamefont {M.~M.}\ \bibnamefont {Ugeda}},
  \bibinfo {author} {\bibfnamefont {P.}~\bibnamefont {Ordej\'{o}n}}, \ and\
  \bibinfo {author} {\bibfnamefont {E.}~\bibnamefont {Canadell}},\ }\href
  {\doibase 10.1021/acs.nanolett.9b00268} {\bibfield  {journal} {\bibinfo
  {journal} {Nano Lett.}\ }\textbf {\bibinfo {volume} {19}},\ \bibinfo {pages}
  {3027} (\bibinfo {year} {2019})}\BibitemShut {NoStop}%
\bibitem [{\citenamefont {Zheng}\ \emph {et~al.}(2018)\citenamefont {Zheng},
  \citenamefont {Zhou}, \citenamefont {Liu},\ and\ \citenamefont
  {Feng}}]{SZheng2018}%
  \BibitemOpen
  \bibfield  {author} {\bibinfo {author} {\bibfnamefont {F.}~\bibnamefont
  {Zheng}}, \bibinfo {author} {\bibfnamefont {Z.}~\bibnamefont {Zhou}},
  \bibinfo {author} {\bibfnamefont {X.}~\bibnamefont {Liu}}, \ and\ \bibinfo
  {author} {\bibfnamefont {J.}~\bibnamefont {Feng}},\ }\href {\doibase
  10.1103/PhysRevB.97.081101} {\bibfield  {journal} {\bibinfo  {journal} {Phys.
  Rev. B}\ }\textbf {\bibinfo {volume} {97}},\ \bibinfo {pages} {081101}
  (\bibinfo {year} {2018})}\BibitemShut {NoStop}%
\bibitem [{\citenamefont {Lian}\ \emph {et~al.}(2018)\citenamefont {Lian},
  \citenamefont {Si},\ and\ \citenamefont {Duan}}]{SLian2018}%
  \BibitemOpen
  \bibfield  {author} {\bibinfo {author} {\bibfnamefont {C.~S.}\ \bibnamefont
  {Lian}}, \bibinfo {author} {\bibfnamefont {C.}~\bibnamefont {Si}}, \ and\
  \bibinfo {author} {\bibfnamefont {W.}~\bibnamefont {Duan}},\ }\href {\doibase
  10.1021/acs.nanolett.8b00237} {\bibfield  {journal} {\bibinfo  {journal}
  {Nano Letters}\ }\textbf {\bibinfo {volume} {18}},\ \bibinfo {pages} {2924}
  (\bibinfo {year} {2018})}\BibitemShut {NoStop}%
\bibitem [{\citenamefont {Cossu}\ \emph {et~al.}(2018)\citenamefont {Cossu},
  \citenamefont {Moghaddam}, \citenamefont {Kim}, \citenamefont {Tahini},
  \citenamefont {Di~Marco}, \citenamefont {Yeom},\ and\ \citenamefont
  {Akbari}}]{SCossu2018}%
  \BibitemOpen
  \bibfield  {author} {\bibinfo {author} {\bibfnamefont {F.}~\bibnamefont
  {Cossu}}, \bibinfo {author} {\bibfnamefont {A.~G.}\ \bibnamefont
  {Moghaddam}}, \bibinfo {author} {\bibfnamefont {K.}~\bibnamefont {Kim}},
  \bibinfo {author} {\bibfnamefont {H.~A.}\ \bibnamefont {Tahini}}, \bibinfo
  {author} {\bibfnamefont {I.}~\bibnamefont {Di~Marco}}, \bibinfo {author}
  {\bibfnamefont {H.-W.}\ \bibnamefont {Yeom}}, \ and\ \bibinfo {author}
  {\bibfnamefont {A.}~\bibnamefont {Akbari}},\ }\href {\doibase
  10.1103/PhysRevB.98.195419} {\bibfield  {journal} {\bibinfo  {journal} {Phys.
  Rev. B}\ }\textbf {\bibinfo {volume} {98}},\ \bibinfo {pages} {195419}
  (\bibinfo {year} {2018})}\BibitemShut {NoStop}%
\bibitem [{\citenamefont {Langer}\ \emph {et~al.}(2014)\citenamefont {Langer},
  \citenamefont {Kisiel}, \citenamefont {Pawlak}, \citenamefont {Pellegrini},
  \citenamefont {Santoro}, \citenamefont {Buzio}, \citenamefont {Gerbi},
  \citenamefont {Balakrishnan}, \citenamefont {Baratoff}, \citenamefont
  {Tosatti},\ and\ \citenamefont {Meyer}}]{SLanger2014}%
  \BibitemOpen
  \bibfield  {author} {\bibinfo {author} {\bibfnamefont {M.}~\bibnamefont
  {Langer}}, \bibinfo {author} {\bibfnamefont {M.}~\bibnamefont {Kisiel}},
  \bibinfo {author} {\bibfnamefont {R.}~\bibnamefont {Pawlak}}, \bibinfo
  {author} {\bibfnamefont {F.}~\bibnamefont {Pellegrini}}, \bibinfo {author}
  {\bibfnamefont {G.~E.}\ \bibnamefont {Santoro}}, \bibinfo {author}
  {\bibfnamefont {R.}~\bibnamefont {Buzio}}, \bibinfo {author} {\bibfnamefont
  {A.}~\bibnamefont {Gerbi}}, \bibinfo {author} {\bibfnamefont
  {G.}~\bibnamefont {Balakrishnan}}, \bibinfo {author} {\bibfnamefont
  {A.}~\bibnamefont {Baratoff}}, \bibinfo {author} {\bibfnamefont
  {E.}~\bibnamefont {Tosatti}}, \ and\ \bibinfo {author} {\bibfnamefont
  {E.}~\bibnamefont {Meyer}},\ }\href {\doibase 10.1038/nmat3836} {\bibfield
  {journal} {\bibinfo  {journal} {Nat. Mater.}\ }\textbf {\bibinfo {volume}
  {13}},\ \bibinfo {pages} {173} (\bibinfo {year} {2014})}\BibitemShut
  {NoStop}%
\bibitem [{\citenamefont {Flicker}\ and\ \citenamefont {van
  Wezel}(2015)}]{SFlicker2015}%
  \BibitemOpen
  \bibfield  {author} {\bibinfo {author} {\bibfnamefont {F.}~\bibnamefont
  {Flicker}}\ and\ \bibinfo {author} {\bibfnamefont {J.}~\bibnamefont {van
  Wezel}},\ }\href {\doibase 10.1038/ncomms8034} {\bibfield  {journal}
  {\bibinfo  {journal} {Nature Commun.}\ }\textbf {\bibinfo {volume} {6}},\
  \bibinfo {pages} {7034} (\bibinfo {year} {2015})}\BibitemShut {NoStop}%
\bibitem [{\citenamefont {Chatterjee}\ \emph {et~al.}(2015)\citenamefont
  {Chatterjee}, \citenamefont {Zhao}, \citenamefont {Iavarone}, \citenamefont
  {{Di Capua}}, \citenamefont {Castellan}, \citenamefont {Karapetrov},
  \citenamefont {Malliakas}, \citenamefont {Kanatzidis}, \citenamefont {Claus},
  \citenamefont {Ruff}, \citenamefont {Weber}, \citenamefont {{Van Wezel}},
  \citenamefont {Campuzano}, \citenamefont {Osborn}, \citenamefont {Randeria},
  \citenamefont {Trivedi}, \citenamefont {Norman},\ and\ \citenamefont
  {Rosenkranz}}]{SChatterjee2015}%
  \BibitemOpen
  \bibfield  {author} {\bibinfo {author} {\bibfnamefont {U.}~\bibnamefont
  {Chatterjee}}, \bibinfo {author} {\bibfnamefont {J.}~\bibnamefont {Zhao}},
  \bibinfo {author} {\bibfnamefont {M.}~\bibnamefont {Iavarone}}, \bibinfo
  {author} {\bibfnamefont {R.}~\bibnamefont {{Di Capua}}}, \bibinfo {author}
  {\bibfnamefont {J.~P.}\ \bibnamefont {Castellan}}, \bibinfo {author}
  {\bibfnamefont {G.}~\bibnamefont {Karapetrov}}, \bibinfo {author}
  {\bibfnamefont {C.~D.}\ \bibnamefont {Malliakas}}, \bibinfo {author}
  {\bibfnamefont {M.~G.}\ \bibnamefont {Kanatzidis}}, \bibinfo {author}
  {\bibfnamefont {H.}~\bibnamefont {Claus}}, \bibinfo {author} {\bibfnamefont
  {J.~P.~C.}\ \bibnamefont {Ruff}}, \bibinfo {author} {\bibfnamefont
  {F.}~\bibnamefont {Weber}}, \bibinfo {author} {\bibfnamefont
  {J.}~\bibnamefont {{Van Wezel}}}, \bibinfo {author} {\bibfnamefont {J.~C.}\
  \bibnamefont {Campuzano}}, \bibinfo {author} {\bibfnamefont {R.}~\bibnamefont
  {Osborn}}, \bibinfo {author} {\bibfnamefont {M.}~\bibnamefont {Randeria}},
  \bibinfo {author} {\bibfnamefont {N.}~\bibnamefont {Trivedi}}, \bibinfo
  {author} {\bibfnamefont {M.~R.}\ \bibnamefont {Norman}}, \ and\ \bibinfo
  {author} {\bibfnamefont {S.}~\bibnamefont {Rosenkranz}},\ }\href {\doibase
  10.1038/ncomms7313} {\bibfield  {journal} {\bibinfo  {journal} {Nat.
  Commun.}\ }\textbf {\bibinfo {volume} {6}},\ \bibinfo {pages} {6313}
  (\bibinfo {year} {2015})}\BibitemShut {NoStop}%
\bibitem [{\citenamefont {Arguello}\ \emph {et~al.}(2014)\citenamefont
  {Arguello}, \citenamefont {Chockalingam}, \citenamefont {Rosenthal},
  \citenamefont {Zhao}, \citenamefont {Guti{\'{e}}rrez}, \citenamefont {Kang},
  \citenamefont {Chung}, \citenamefont {Fernandes}, \citenamefont {Jia},
  \citenamefont {Millis}, \citenamefont {Cava},\ and\ \citenamefont
  {Pasupathy}}]{SArguello2014}%
  \BibitemOpen
  \bibfield  {author} {\bibinfo {author} {\bibfnamefont {C.~J.}\ \bibnamefont
  {Arguello}}, \bibinfo {author} {\bibfnamefont {S.~P.}\ \bibnamefont
  {Chockalingam}}, \bibinfo {author} {\bibfnamefont {E.~P.}\ \bibnamefont
  {Rosenthal}}, \bibinfo {author} {\bibfnamefont {L.}~\bibnamefont {Zhao}},
  \bibinfo {author} {\bibfnamefont {C.}~\bibnamefont {Guti{\'{e}}rrez}},
  \bibinfo {author} {\bibfnamefont {J.~H.}\ \bibnamefont {Kang}}, \bibinfo
  {author} {\bibfnamefont {W.~C.}\ \bibnamefont {Chung}}, \bibinfo {author}
  {\bibfnamefont {R.~M.}\ \bibnamefont {Fernandes}}, \bibinfo {author}
  {\bibfnamefont {S.}~\bibnamefont {Jia}}, \bibinfo {author} {\bibfnamefont
  {A.~J.}\ \bibnamefont {Millis}}, \bibinfo {author} {\bibfnamefont {R.~J.}\
  \bibnamefont {Cava}}, \ and\ \bibinfo {author} {\bibfnamefont {A.~N.}\
  \bibnamefont {Pasupathy}},\ }\href {\doibase 10.1103/PhysRevB.89.235115}
  {\bibfield  {journal} {\bibinfo  {journal} {Phys. Rev. B}\ }\textbf {\bibinfo
  {volume} {89}},\ \bibinfo {pages} {235115} (\bibinfo {year}
  {2014})}\BibitemShut {NoStop}%
\bibitem [{\citenamefont {Franke}\ \emph {et~al.}(2011)\citenamefont {Franke},
  \citenamefont {Schulze},\ and\ \citenamefont {Pascual}}]{SFranke2011}%
  \BibitemOpen
  \bibfield  {author} {\bibinfo {author} {\bibfnamefont {K.~J.}\ \bibnamefont
  {Franke}}, \bibinfo {author} {\bibfnamefont {G.}~\bibnamefont {Schulze}}, \
  and\ \bibinfo {author} {\bibfnamefont {J.~I.}\ \bibnamefont {Pascual}},\
  }\href {\doibase 10.1126/science.1202204} {\bibfield  {journal} {\bibinfo
  {journal} {Science}\ }\textbf {\bibinfo {volume} {332}},\ \bibinfo {pages}
  {940} (\bibinfo {year} {2011})}\BibitemShut {NoStop}%
\end{thebibliography}
%
%
%
%

%
\end{document}